%% file: manuscript.tex
\PassOptionsToPackage{unicode}{hyperref}
\PassOptionsToPackage{hyphens}{url}
\PassOptionsToPackage{dvipsnames,svgnames,x11names}{xcolor}
\PassOptionsToPackage{unicode,hypertexnames=false}{hyperref}
\documentclass[12pt]{article}

\usepackage{amsmath,amssymb,amsthm}
\usepackage{bm}
\usepackage{needspace}

\usepackage{iftex}
\ifPDFTeX
  \usepackage[T1]{fontenc}
  \usepackage[utf8]{inputenc}
  \usepackage{textcomp} 
\else 
  \usepackage{unicode-math}
  \defaultfontfeatures{Scale=MatchLowercase}
  \defaultfontfeatures[\rmfamily]{Ligatures=TeX,Scale=1}
\fi
\usepackage{lmodern}
\ifPDFTeX\else  
\fi
\IfFileExists{upquote.sty}{\usepackage{upquote}}{}
\IfFileExists{microtype.sty}{
  \usepackage[]{microtype}
  \UseMicrotypeSet[protrusion]{basicmath} 
}{}
\makeatletter
\@ifundefined{KOMAClassName}{
  \IfFileExists{parskip.sty}{%
    \usepackage{parskip}
  }{
    \setlength{\parindent}{0pt}
    \setlength{\parskip}{6pt plus 2pt minus 1pt}}
}{
  \KOMAoptions{parskip=half}}
\makeatother
\usepackage{xcolor}
\makeatletter
\ifx\paragraph\undefined\else
  \let\oldparagraph\paragraph
  \renewcommand{\paragraph}{
    \@ifstar
      \xxxParagraphStar
      \xxxParagraphNoStar
  }
  \newcommand{\xxxParagraphStar}[1]{\oldparagraph*{#1}\mbox{}}
  \newcommand{\xxxParagraphNoStar}[1]{\oldparagraph{#1}\mbox{}}
\fi
\ifx\subparagraph\undefined\else
  \let\oldsubparagraph\subparagraph
  \renewcommand{\subparagraph}{
    \@ifstar
      \xxxSubParagraphStar
      \xxxSubParagraphNoStar
  }
  \newcommand{\xxxSubParagraphStar}[1]{\oldsubparagraph*{#1}\mbox{}}
  \newcommand{\xxxSubParagraphNoStar}[1]{\oldsubparagraph{#1}\mbox{}}
\fi
\makeatother

\usepackage{longtable,booktabs,array}
\usepackage{calc} 
\usepackage{etoolbox}
\makeatletter
\patchcmd\longtable{\par}{\if@noskipsec\mbox{}\fi\par}{}{}
\makeatother
\IfFileExists{footnotehyper.sty}{\usepackage{footnotehyper}}{\usepackage{footnote}}
\makesavenoteenv{longtable}
\usepackage{graphicx}
\makeatletter
\def\maxwidth{\ifdim\Gin@nat@width>\linewidth\linewidth\else\Gin@nat@width\fi}
\def\maxheight{\ifdim\Gin@nat@height>\textheight\textheight\else\Gin@nat@height\fi}
\makeatother
\setkeys{Gin}{width=\maxwidth,height=\maxheight,keepaspectratio}
\makeatletter
\def\fps@figure{htbp}
\makeatother

\usepackage[margin=1in]{geometry}

\makeatletter
\@ifpackageloaded{caption}{}{\usepackage{caption}}
\AtBeginDocument{%
\ifdefined\contentsname
  \renewcommand*\contentsname{Table of contents}
\else
  \newcommand\contentsname{Table of contents}
\fi
\ifdefined\listfigurename
  \renewcommand*\listfigurename{List of Figures}
\else
  \newcommand\listfigurename{List of Figures}
\fi
\ifdefined\listtablename
  \renewcommand*\listtablename{List of Tables}
\else
  \newcommand\listtablename{List of Tables}
\fi
\ifdefined\figurename
  \renewcommand*\figurename{Figure}
\else
  \newcommand\figurename{Figure}
\fi
\ifdefined\tablename
  \renewcommand*\tablename{Table}
\else
  \newcommand\tablename{Table}
\fi
}
\@ifpackageloaded{float}{}{\usepackage{float}}
\floatstyle{ruled}
\@ifundefined{c@chapter}{\newfloat{codelisting}{h}{lop}}{\newfloat{codelisting}{h}{lop}[chapter]}
\floatname{codelisting}{Listing}

\makeatother
\makeatletter
\@ifpackageloaded{caption}{}{\usepackage{caption}}
\@ifpackageloaded{subcaption}{}{\usepackage{subcaption}}
\makeatother

\ifLuaTeX
  \usepackage{selnolig}  
\fi
\usepackage[]{natbib}
\usepackage{bookmark}

\IfFileExists{xurl.sty}{\usepackage{xurl}}{} 
\hypersetup{
  pdftitle={Quantile-Only Regression: Inference and Partial Identification under Quantile Disclosure},
  pdfauthor={Masamune Iwasawa; Hayato Nakanishi},
  pdfkeywords={Group-level summary statistics; Quantiles; Partial identification; Simultaneous confidence regions; Disclosure design},
  colorlinks=true,
  linkcolor={blue},
  filecolor={Maroon},
  citecolor={Blue},
  urlcolor={Blue},
  pdfcreator={LaTeX via pandoc}}

\usepackage{booktabs}
\usepackage{threeparttable}
\usepackage{graphicx}

\newcommand{\anon}{1}

\newtheorem{theorem}{Theorem}
\newtheorem{proposition}{Proposition}
\newtheorem{lemma}{Lemma}
\newtheorem{corollary}{Corollary} 
\newtheorem{example}{Example} 
\newtheorem{assumption}{Assumption} 

\newtheorem{remark}{Remark}

\newcommand{\Var}{\mathrm{Var}}
\newcommand{\Cov}{\mathrm{Cov}}

\usepackage{setspace}

\newcommand{\TITLE}{Quantile-Only Regression: Inference and Partial Identification under Quantile Disclosure}
\makeatletter
\pretocmd{\@makefntext}{%
  \normalsize
  \doublespacing
}{}{%
  \PackageWarning{manuscript}{Could not patch footnote formatting}%
}
\makeatother

\begin{document}

\doublespacing


\if1\anon
{
  \title{\bf \TITLE}
  \author{Masamune Iwasawa\hspace{.2cm}\\
    Doshisha University, 602-8580 Karasuma-higashi-iru, Imadegawa-dori,\\ Kamigyo-ku, Kyoto-shi, Email: miwasawa@mail.doshisha.ac.jp\\
    and \\
    Hayato Nakanishi \\
    Kanagawa University, 3-27-1 Rokkakubashi Kanagawa-ku, Yokohama,\\ 
    Kanagawa 221-8686, Japan,
Email: hnakanishi@kanagawa-u.ac.jp}
  \maketitle
} \fi

\if0\anon
{
  \bigskip
  \bigskip
  \bigskip
  \begin{center}
    {\LARGE\bf Title}
\end{center}
  \medskip
} \fi

\bigskip
\begin{abstract}
Researchers often observe group quantiles but not group means, raising the question of whether these summaries alone support conventional regression analysis. We develop estimation and inference methods for linear models of group means under quantile-only and mixed disclosure. We establish conditions on disclosure design for consistent estimation and asymptotic normality at the parametric rate. Under fixed disclosure, we characterize sharp identified sets and construct simultaneous confidence regions, clarifying how informativeness depends on the number and placement of disclosed quantiles. Applications to education, social care, and medical expenditure illustrate the methods. The R package \texttt{qonlyreg} implements the proposed procedures.
\end{abstract}


\vfill
\newpage
\clubpenalty=10000 \widowpenalty=10000 \displaywidowpenalty=10000 

\section{Introduction}\label{sec_intro}

Researchers increasingly encounter data in which outcome distributions are disclosed only through group-level summary statistics rather than individual-level observations.
Official statistical publications, administrative reporting systems, and empirical studies used in meta-analysis often report group sizes and selected quantiles such as medians, quartiles, or deciles, while the underlying microdata and sometimes even the group means are unavailable.
Such disclosure practices arise for a variety of reasons, including privacy, confidentiality, and reporting conventions, and occur in substantively different settings ranging from education and public administration to medical research.

This creates a simple but consequential problem for regression analysis.
Researchers may wish to relate group-average outcomes to observable group characteristics, yet the dependent variable in such a regression---the group mean---is not observed.
For example, substituting reported medians for unobserved group means need not recover the regression coefficients of interest, because differences between group means and medians may vary systematically with group characteristics.
More generally, the disclosed quantiles contain information about the mean, but a finite collection need not determine it exactly.
It is therefore not immediate whether conventional regression analysis remains possible from the disclosed summaries alone.

This paper asks three questions.
First, can regression coefficients defined through group means be consistently estimated when only group-specific quantiles are observed?
Second, how do the number and placement of disclosed quantiles determine the accuracy of estimation and the feasibility of conventional inference?
Third, when the available quantile information is insufficient to point identify the regression coefficients, what can still be learned about them?

We study these questions in a linear model for group-level conditional means.
For each group, the researcher observes the group size, group-level covariates, and a collection of sample quantiles, but not the underlying individual-level outcomes.
We reconstruct the group mean by integrating a piecewise-linear interpolation of the disclosed quantiles and estimate the regression coefficients by a sample-size-weighted minimum-distance procedure.
The framework also accommodates mixed disclosure, in which sample means are directly reported for some groups while only quantile summaries are available for others.

Our results deliver three main messages.

First, sufficiently rich quantile disclosure can support conventional regression analysis.
Under increasing quantile disclosure, the error in reconstructing a group mean can be decomposed into a discretization component arising from the finite quantile grid and a sampling component arising from estimation of the disclosed quantiles.
As the quantile grid becomes denser, the discretization error vanishes, yielding consistent estimation of the regression coefficients. If the grid becomes sufficiently dense relative to the sample size, the resulting estimator attains the usual parametric rate and is asymptotically normal, allowing feasible inference based only on the disclosed summaries.
Importantly, refining the quantile grid reduces the discretization error without worsening the sampling-error rate in mean reconstruction.

Second, fixed quantile disclosure creates an identification problem, not merely a loss of precision.
With a fixed finite set of disclosed quantiles, even the population quantiles generally do not determine the group means.
Distinct conditional distributions can generate exactly the same disclosed population quantiles while implying different group means and regression coefficients.
Consequently, increasing the sample size alone cannot resolve the problem: non-identification persists even when sampling uncertainty vanishes.
For this setting, we derive sharp bounds for the group means, characterize the sharp identified set for the regression coefficients, and construct asymptotically valid simultaneous confidence regions.

Third, the informativeness of quantile disclosure depends on where quantiles are reported, not only on how many are reported.
For point estimation, the discretization-error bound is governed by the maximal spacing of the disclosed quantile grid.
Among designs with the same number of disclosed quantiles, evenly spaced grids minimize this maximal spacing and therefore minimize the grid-dependent component of the error bound.
The implication for partial identification is different.
The width of the identified interval for a group mean depends not only on grid spacing but also on the local variation of the quantile function.
A grid that is favorable for point estimation therefore need not be equally favorable for partial identification, and allocating more quantiles to regions where the quantile function changes rapidly may yield tighter identified intervals.

Taken together, these results show that disclosure design is itself a statistical design problem. The number and placement of disclosed quantiles determine how informative the released summaries are for recovering group means, estimating regression coefficients, and conducting statistical inference.
The framework therefore provides a way to assess not only whether regression analysis is possible from a given statistical release, but also how the design of that release affects estimation and identification.

The proposed methods are straightforward to implement using only the disclosed summaries. The reconstructed group means are available in closed form, and the accompanying R package, \texttt{qonlyreg}, implements the main estimation and inference procedures developed in the paper.\footnote{
The package is available at \url{https://github.com/masaiwasawa/qonlyreg}.
}

We examine the finite-sample performance of the procedures through Monte Carlo simulations and illustrate their scope in three substantively different empirical settings. 
The first application uses publicly released National Assessment of Educational Progress (NAEP) mathematics summaries for groups defined by U.S. jurisdiction and economically disadvantaged status. 
The point estimates are close to the benchmark based on published group means, and the projected confidence interval narrows substantially when disclosure expands from quartiles to five percentiles.

The second uses publicly released Children in Need assessment-duration summaries from English local authorities, where assessment counts and quartiles are observed but the corresponding means are not. We use a fixed-window comparison motivated by regression discontinuity (RD) to examine differences in mean assessment duration around the 50 percent Labour seat-share threshold. 
The point estimates are negative and have small conventional standard errors, but the projected confidence intervals accounting for identification uncertainty are wide and include zero under the support restrictions considered. 

The third considers a medical-expenditure meta-analysis with mixed disclosure, in which some study comparisons report sample means while others report only quartiles. 
Together, these applications illustrate that the same problem arises in official statistics, administrative data, and published study summaries, and that the framework can accommodate both quantile-only and mixed-disclosure environments.

A related literature studies regression and other statistical relationships when only aggregate rather than individual-level data are available, including work on ecological inference \citep{king97} and regression with grouped or aggregated data (e.g., \citealp{farebrother79,nicoletti12}).
These approaches typically seek to recover individual-level relationships or model parameters from aggregate observations.
Related work exploits aggregate sufficient statistics or moments to recover parameters of richer individual-level models.
\citet{kovalchik13}, for example, recovers parameters of an individual-participant random-effects model from aggregate study-level data, while \citet{mcculloch25} use tract-level means and within-tract second moments together with a small-variance approximation to recover parcel-level treatment-effect parameters.
Our problem is distinct: the target is a regression model for group means, while the group means themselves are unavailable and must be reconstructed or bounded from disclosed quantiles.

The literature on recovering distributional features from grouped or quantile-based summaries includes methods for estimating a single population mean from reported quantiles, primarily in the context of medical meta-analysis \citep{hozo05,wan14,luo18,mcgrath20,cai21}.
These methods are designed for a small number of pre-specified summaries such as medians, ranges, and quartiles.
Our focus is different.
Rather than recovering a single mean, we study regression coefficients defined through multiple group means and characterize how the number and placement of disclosed quantiles affect estimation, inference, and identification.

The paper is also related to the literature on partial identification under incomplete information; see \citet{manski03} and \citet{molinari20} for textbook and survey treatments.
Unlike settings in which identification failure arises from missing or interval-censored outcomes, the non-identification studied here is generated by the disclosure mechanism itself.
An early example in this spirit is \citet{cowell82}, who derive lower and upper bounds for income inequality measures from grouped income data and propose interpolation-based point estimators.
Their analysis combines bounding and interpolation arguments to study functionals of partially observed distributions.
To the best of our knowledge, however, the implications of quantile disclosure for regression identification and inference have received little attention.

Finally, the asymptotic analysis builds on classical results for sample quantiles.
For a fixed quantile index, sample quantiles converge at the usual square-root sample-size rate under standard regularity conditions and admit Bahadur representations; see \citet{bahadur66} and \citet{kiefer67}.
The increasing-disclosure setting considered here raises an additional issue because the number of disclosed quantiles may grow with the sample size and the extreme disclosed quantiles may approach the boundaries of the unit interval.
We therefore establish uniform error bounds that remain valid over such expanding grids and use them to derive the asymptotic behavior of the reconstructed-mean regression estimator.

The remainder of the paper is organized as follows.
Section~\ref{sec_model_estimation} introduces the framework and estimator.
Sections~\ref{sec_regression} and~\ref{sec_fixedDC} develop the theoretical results under increasing and fixed quantile disclosure.
Sections~\ref{sec_simulation} and~\ref{sec_empirical} present simulation and empirical illustrations.
Section~\ref{sec_conclude} concludes.

\section{Model and Estimation}\label{sec_model_estimation}
Let $(Y_i,\bm X_i)$ be independent and identically distributed (i.i.d.) copies of
$(Y,\bm X)$, where $Y_i\in\mathbb R$ is the outcome and
$\bm X_i\in\mathbb R^p$ is an underlying individual-level covariate vector.
The support $\mathcal X$ of $\bm X$ is partitioned into a finite collection of
measurable groups $\{\mathcal G_g\}_{g=1}^G$ satisfying
\[
    \mathcal G_g\cap\mathcal G_h=\varnothing\quad(g\ne h),
    \qquad \bigcup_{g=1}^G\mathcal G_g=\mathcal X.
\]
Write $\bm X_i\in g$ for $\bm X_i\in\mathcal G_g$.
Each observation belongs to exactly one group, so the total number of underlying individual observations is
$N=\sum_{g=1}^G n_g$, where $n_g$ is the group size.
For each group, define the conditional distribution and quantile functions by
\[
    F_g(y)=P(Y\le y\mid\bm X\in g),
    \qquad
    Q_g(\tau)=\inf\{y\in\mathbb R:F_g(y)\ge\tau\},
    \quad \tau\in(0,1).
\]

The researcher observes only the released summaries
\[
    \{n_g,\bm X_g,\widehat Q_g(\tau_1),\ldots,
      \widehat Q_g(\tau_K)\}_{g=1}^G,
    \qquad 0<\tau_1<\cdots<\tau_K<1,
\]
and not the individual-level observations $(Y_i,\bm X_i)$.
The partition is formed by the data provider and treated as given.
The group characteristics $\bm X_g$ are nonrandom and conceptually distinct
from $\bm X_i$: they may be aggregated or externally provided and need not
determine the partition.

Our target is a linear restriction on the group-level conditional means,
\[
    m_g=E(Y\mid\bm X\in g)=\bm X_g'\bm\beta_0.
\]
We focus on lower-dimensional, nonsaturated specifications that impose
cross-group restrictions, excluding a full set of group indicators.
This setting includes models obtained by aggregating an individual-level
linear conditional mean specification, \(E(Y\mid \bm X)=\bm X'\bm\beta_0\); 
Supplemental Material~G explains the connection.

The baseline setup above focuses on quantile-only disclosure.
The framework also accommodates mixed disclosure, in which sample means
are directly reported for some groups while only quantile summaries are
available for others.
The following examples relate the group-mean model to the applications
in Section~\ref{sec_empirical}.

\begin{example}[Educational achievement]
Let $g$ index groups defined by area $s_g$ and socioeconomic
status $D_g\in\{0,1\}$.
Suppose that the sample size $n_g$ and selected sample quantiles of
test scores are disclosed for each group, but the corresponding
sample mean is not.
We model the population mean test score for group $g$ as
$m_g=\alpha_{s_g}+\beta_0 D_g$, allowing area-specific intercepts
and imposing a common mean score difference between the two status groups.
\end{example}

\begin{example}[Administrative duration outcomes]
Let $g$ index groups defined by local authority and reporting period.
Suppose that the number of completed cases $n_g$ and selected sample
quantiles of case duration are disclosed for each group, but the
corresponding sample mean is not.
Within a fixed window of half-width $h$ around a threshold in an
observed authority characteristic, we model the population mean duration
as $m_g=\alpha_h+\delta_h D_g$, where $D_g\in\{0,1\}$ indicates whether
the characteristic is at or above the threshold.
The coefficient $\delta_h$ is the mean-duration difference between
the two sides within this window.
\end{example}

\begin{example}[Meta-analysis with mixed disclosure]
Let $g$ index the population groups in a meta-analysis, with $j_g$
denoting the study comparison to which group $g$ belongs and
$D_g\in\{0,1\}$ distinguishing the two populations being compared.
Suppose that sample sizes $n_g$ are disclosed for all groups,
while some groups report sample means and others only selected
sample quantiles.
We model the population mean outcome as
$m_g=\alpha_{j_g}+\beta_0 D_g$, allowing comparison-specific intercepts.
The coefficient $\beta_0$ is the mean difference between the $D_g=1$
and $D_g=0$ populations within each comparison.
\end{example}

When $E(|Y|\mid\bm X\in g)<\infty$,
$m_g=\int_0^1 Q_g(\tau)\,d\tau$.
We reconstruct the unobserved quantile function by piecewise linear
interpolation: for $\tau\in[\tau_k,\tau_{k+1}]$,
\[
    \widetilde Q_g(\tau)
    =\widehat Q_g(\tau_k)
     +\frac{\tau-\tau_k}{\tau_{k+1}-\tau_k}
      \{\widehat Q_g(\tau_{k+1})-\widehat Q_g(\tau_k)\}.
\]
Below $\tau_1$ and above $\tau_K$, use the constant values
$\widehat Q_g(\tau_1)$ and $\widehat Q_g(\tau_K)$, respectively.\footnote{
A known support endpoint can be included as a deterministic interpolation
node. Such endpoints do not contribute to sampling variance, and the
results below continue to hold with this modification.
}
The reconstructed mean is
\[
    \hat m_g=\int_0^1\widetilde Q_g(\tau)\,d\tau.
\]
For example, with only quartiles disclosed, the baseline constant-tail rule gives
\[
    \hat m_g
    =\frac38\widehat Q_g(0.25)
     +\frac14\widehat Q_g(0.50)
     +\frac38\widehat Q_g(0.75).
\]

We estimate the regression coefficients by
\[
    \hat{\bm\beta}
    =\arg\min_{\bm\beta}\sum_{g=1}^G
      n_g(\hat m_g-\bm X_g'\bm\beta)^2
    =\left(\sum_{g=1}^G n_g\bm X_g\bm X_g'\right)^{-1}
      \sum_{g=1}^G n_g\bm X_g\hat m_g.
\]
Piecewise linear interpolation preserves monotonicity, requires no tuning
parameters, and yields a closed-form reconstructed mean.
We use sample-size weights in the minimum-distance estimator to
account for differences in group size without requiring preliminary
variance estimates for point estimation.
To keep the paper focused on how quantile disclosure affects
estimation and inference, we leave the analysis of optimal
weighting to future research.
For mixed disclosure, replace $\hat m_g$ by the reported sample mean
where available.
The theory below focuses on quantile-only disclosure, while
Section~\ref{sec_empirical} illustrates the mixed-disclosure
implementation through a meta-analysis of medical expenditure.

\section{Asymptotic Theory: Increasing Quantile Disclosure}\label{sec_regression}
Throughout this section, $G$ is fixed, each $n_g\to\infty$, and
$K=K_N$ increases with $N=\sum_g n_g$.
We use a common quantile grid across groups to simplify notation
and exposition in the theoretical analysis.
The proposed estimator can also be implemented with
group-specific quantile grids.
Fixed disclosure is studied in Section~\ref{sec_fixedDC}.

\subsection{Assumptions}

\begin{assumption}[Sampling]
\label{ass_sample}
Let $(Y_i, \bm{X}_i)$ be i.i.d.\ random variables.
\end{assumption}

\begin{assumption}[Group-Level Linear Mean Restriction]
\label{ass_linear}
For each $g=1,\dots,G$,
$
    m_g = E(Y \mid \bm{X} \in g) = \bm{X}_g'\bm{\beta }_0.
$
\end{assumption}

\begin{assumption}[Quantile Function Smoothness]
\label{ass_smooth}
For each group $g$, the quantile function $Q_g(\tau)$ is nondecreasing and satisfies either:
\begin{itemize}
\item[(i)] Lipschitz continuity: $|Q_g(\tau) - Q_g(\tau')| \le L |\tau - \tau'|$, or
\item[(ii)] Twice continuously differentiable with bounded second derivative:
$
        \sup_{\tau\in(0,1)} |Q_g''(\tau)|
    \le
        M_g
$
for some finite constant $M_g$.
\end{itemize}
\end{assumption}

\begin{assumption}[Quantile Estimation]
\label{ass_quantile}
For each $g=1,\dots,G$,
\[
    		\sum_{k=1}^{K_N}
    		w_{k,N}
   		 \bigl|
        			\widehat Q_g(\tau_k)-Q_g(\tau_k)
    		\bigr|
    	=
    		O_p(r_{g,N}),
\]
where $r_{g,N}\to 0$, and $\{w_{k,N}\}_{k=1}^{K_N}$ are the reconstruction weights defined in the proof of Lemma~\ref{lem_mean}.
\end{assumption}

\begin{assumption}[Design]
\label{ass_design}
For each \(g=1,\ldots,G\),
$
		    \frac{n_g}{N}
	    \overset{p}{\rightarrow}
		    \pi_g
    	=
		    P(\bm X_i\in\mathcal G_g)
		    \in(0,1),
$
and \(\|\bm X_g\|<\infty\).
Moreover,
$
    \bm\Sigma_X
    =
    \sum_{g=1}^G
    \pi_g
    \bm X_g\bm X_g'
$
is positive definite.
\end{assumption}

Given the population group means, Assumptions~\ref{ass_linear} and
\ref{ass_design} imply that $\bm\beta_0$ uniquely minimizes
$\sum_g\pi_g(m_g-\bm X_g'\bm\beta)^2$.
The proposed estimator minimizes the sample analogue of this
criterion and is well-defined with probability approaching one
under Assumption~\ref{ass_design}.
Primitive conditions for the quantile-estimation error in
Assumption~\ref{ass_quantile} are given in Lemma~\ref{lem_uniform_quantile}.
Both alternatives in Assumption~\ref{ass_smooth} imply that the
outcome distribution within each group has bounded support,
although the endpoints need not be known for the estimation
and inference procedures developed in this section.
The Gaussian and Laplace simulations assess robustness outside these
sufficient conditions; see Section~\ref{sec_simulation} and Supplemental
Material~D.

\subsection{Reconstruction Error and Consistency}
Let $\mathcal T_{K_N}=\{\tau_1,\ldots,\tau_{K_N}\}$ denote the \emph{disclosure design}, 
with $\tau_0=0$, $\tau_{K_N+1}=1$, and maximal spacing
\[
    \Delta(\mathcal T_{K_N})
    =\max_{0\le k\le K_N}(\tau_{k+1}-\tau_k).
\]
This includes both boundary gaps. We consider designs satisfying
$\Delta(\mathcal T_{K_N})\to0$.

\begin{lemma}[Mean Reconstruction Error]
\label{lem_mean}
Under Assumptions \ref{ass_smooth} and \ref{ass_quantile}, and $\min_g n_g \to \infty$,
the reconstruction error of the group means satisfies
\[
        \max_{1\le g\le G} |\hat m_g-m_g|
    =
        O_p\!\left(
            \Delta(\mathcal T_{K_N})^a
            +
            r_N
        \right),
\]
where $r_N = \max_{1\le g\le G} r_{g,N}$, and
$a=1$ under Assumption \ref{ass_smooth}(i) and $a=2$ under Assumption \ref{ass_smooth}(ii).
\end{lemma}

Lemma~\ref{lem_mean} separates the reconstruction error into two components.
The term $\Delta(\mathcal T_{K_N})^a$ corresponds to the
\emph{discretization error} arising from the finite quantile grid,
while $r_N$ summarizes the contribution of \emph{sampling error}
in the disclosed quantiles.

The discretization bound improves from order
$\Delta(\mathcal T_{K_N})$ under Lipschitz continuity to order
$\Delta(\mathcal T_{K_N})^2$ under the stronger smoothness
condition in Assumption~\ref{ass_smooth}(ii).
For the latter case, integrating the linear-interpolation error
over the intervals between disclosed quantiles yields the
quadratic bound \citep[Chapter~3]{atkinson89}.
On the two boundary intervals, this assumption also implies
bounded $Q_g'$, so the integrated error from constant extrapolation
is bounded by a constant times the square of each interval's length.

The following theorem extends this error bound to the proposed
regression estimator.

\begin{theorem}[Regression Error Decomposition]
\label{thm_main}
Under Assumptions \ref{ass_linear}, \ref{ass_smooth}, \ref{ass_quantile}, and \ref{ass_design},
\[
    \hat{\bm{\beta}} - \bm{\beta }_0
    =
    O_p\!\left(
        \Delta(\mathcal{T}_{K_N})^a
        +
        r_N
    \right),
\]
where $a=1$ under Assumption~\ref{ass_smooth}(i) and $a=2$ under Assumption~\ref{ass_smooth}(ii).
\end{theorem}

Theorem~\ref{thm_main} implies that
$\hat{\bm\beta}\xrightarrow{p}\bm\beta_0$
if $\Delta(\mathcal T_{K_N})\to0$ and $r_N\to0$.
The discretization bound depends on the largest gap between
successive quantile indices, including the gaps to zero and one.
Increasing $K_N$ alone does not ensure that this spacing vanishes,
since additional quantiles may be concentrated in only part
of the unit interval.

\subsection{Sufficient Conditions for \texorpdfstring{$\sqrt{N}$}{root-N} Consistency}
\label{subsec_rootN}
Theorem~\ref{thm_main} gives sufficient conditions
$\Delta(\mathcal T_{K_N})^a=O(N^{-1/2})$ and $r_N=O(N^{-1/2})$
for $\sqrt N$-consistency.
The latter follows from a uniform quantile-error bound because the
reconstruction weights are nonnegative and sum to one.
Classical results for fixed interior quantiles \citep{bahadur66,kiefer67}
do not directly cover grids with $\tau_1\to0$ and $\tau_{K_N}\to1$.
The next lemma supplies primitive conditions that allow this boundary
behavior.\footnote{
The asymptotic analysis in this section focuses on continuously distributed outcomes. 
Discrete outcomes may have flat or discontinuous quantile functions and require separate treatment.
}
Here, sample quantiles are generalized inverses of the group empirical
distribution:
\[
    \widehat Q_g(\tau)=\inf\{y\in\mathbb R:F_{n,g}(y)\ge\tau\},
    \qquad
    F_{n,g}(y)=\frac{1}{n_g}\sum_{i:\bm X_i\in g}1\{Y_i\le y\}.
\]

\begin{lemma}[A sufficient condition for $\sqrt{n_g}$ quantile reconstruction]
\label{lem_uniform_quantile}
Fix a group $g$.
Suppose
Assumption \ref{ass_sample} and that:
(i) $F_g$ is strictly increasing on its support and admits a density \(f_g\) that is continuously differentiable on its support.
(ii) The support of $F_g$ is a compact interval $[\underline y_g,\overline y_g]$.
(iii) There exists a constant $c_g>0$ such that
$
    \inf_{y\in[\underline y_g,\overline y_g]} f_g(y) \ge c_g.
$
Then
\begin{equation}
\label{eq_uniform_bound}
    \sup_{1\le k\le K_N}
    \bigl|
        \widehat Q_g(\tau_k)-Q_g(\tau_k)
    \bigr|
    =
    O_p(n_g^{-1/2}).
\end{equation}
\end{lemma}

Under Assumption~\ref{ass_design}, $n_g^{-1/2}=O_p(N^{-1/2})$, so
\eqref{eq_uniform_bound} implies Assumption~\ref{ass_quantile} with
$r_N=N^{-1/2}$.
No additional rate restriction on $\tau_1$ or $1-\tau_{K_N}$
is needed for this uniform absolute-error bound.
Refining the grid therefore does not worsen the sampling-error
rate in mean reconstruction or in the proposed regression estimator.

Combining Lemma~\ref{lem_uniform_quantile} with
Theorem~\ref{thm_main} yields the following sufficient conditions for $\sqrt{N}$-consistency.

\begin{proposition}[Sufficient conditions for $\sqrt{N}$-consistency]
\label{prop_rootN}
Suppose Assumptions~\ref{ass_linear}, \ref{ass_smooth}, and \ref{ass_design} hold, and that the conditions of Lemma~\ref{lem_uniform_quantile} are satisfied for each $g=1,\ldots,G$.
If the quantile grid satisfies
$
    \Delta(\mathcal T_{K_N})^a = O(N^{-1/2}),
$
then
\[
    \hat{\bm{\beta}}-\bm{\beta}_0
    =
    O_p(N^{-1/2}),
\]
where $a=1$ under Assumption~\ref{ass_smooth}(i) and $a=2$ under Assumption~\ref{ass_smooth}(ii).
\end{proposition}

For the evenly spaced grid $\tau_k=k/(K_N+1)$,
$\Delta(\mathcal T_{K_N})=(K_N+1)^{-1}$.
Thus, a sufficient condition is that $K_N$ grow at least on the order
of $N^{1/(2a)}$.
Because $r_N=O(N^{-1/2})$ under the lemma's conditions regardless of
the design, minimizing maximal spacing also minimizes the
grid-dependent component of the error bound.

\begin{corollary}\label{cor_optimal_design}
Under the conditions of Lemma~\ref{lem_uniform_quantile},
 the evenly spaced grid minimizes the upper bound in Theorem~\ref{thm_main} by minimizing \(\Delta(\mathcal T_{K_N})\) among all disclosure designs of size \(K_N\).
\end{corollary}

\subsection{Asymptotic Expansion and Normality}\label{subsec_expansion}

To derive the asymptotic distribution of the proposed regression
estimator, we first establish an expansion of its estimation
error that separates discretization bias from sampling variation.
We impose Assumption~\ref{ass_smooth}(ii) throughout this subsection and Section~\ref{subsec_variance}.

Let $b_{g,N}$ denote group $g$'s discretization bias arising from
approximating the quantile function on a finite grid, and let
$\psi_{g,N}$ denote the influence function of the reconstructed mean $\hat m_g$.
Their formulas are given in the proof in Supplemental Material~A.

\begin{lemma}[Asymptotic expansion]
\label{lem_expansion}
Under the conditions of Proposition~\ref{prop_rootN} with Assumption \ref{ass_smooth}(ii),
\[
    		\hat{\bm{\beta}}-\bm{\beta}_0
    	=
    		\bm{B}_N
    		+
    		\frac{1}{N}\sum_{g=1}^G \sum_{i\in \mathcal{I}_g}
   		 \bm{\Psi}_{g, N}(Y_i)
    		+
    		o_p(N^{-1/2}),
\]
where
$
		\bm{B}_N
    	=
   		 \bm\Sigma_X^{-1}
    		\sum_{g=1}^G \pi_g \bm X_g b_{g,N}
	=
		O(\Delta(\mathcal T_{K_N})^2)
$ is a deterministic bias term,
$
		\bm{\Psi}_{g, N}(y)
	=
		\bm\Sigma_X^{-1}\bm X_g\psi_{g,N}(y)
$
is an influence function with
$
    E[
        \bm\Psi_{g,N}(Y_i)
        \mid
        \bm X_i\in g
    ]
    =
    \bm 0
$.
Moreover, $\bm{\Psi}_{g,N}(Y_i)$ is uniformly bounded in $N$ and hence has finite second moment.
Finally, $\mathcal{I}_g = \{i: \bm X_i \in g\}$ denotes the index set of observations in group $g$.
\end{lemma}

Lemma~\ref{lem_expansion} relies on the uniform Bahadur
representation established in Lemma~\ref{lem_bahadur_grid}.
Because the disclosed grid may approach the boundaries of the
unit interval, standard uniform Bahadur results on fixed compact
subsets of $(0,1)$ do not directly apply.
Under the maintained assumptions, Lemma~\ref{lem_bahadur_grid}
establishes an $o_p(n_g^{-1/2})$ bound on the linearization
remainder uniformly over the entire disclosed grid.
This uniform control allows us to aggregate quantile estimation
errors in mean reconstruction and derive the regression expansion.
Supplemental Material~C presents a standard fixed-interior
representation and its proof for comparison.

Although $\sqrt{N}$-consistency holds, the discretization bias
$\bm B_N$ may remain of first-order importance when
$\Delta(\mathcal T_{K_N})^2$ is of order $N^{-1/2}$.
Centering by this bias gives the following limit.

\begin{theorem}[Bias-corrected asymptotic normality]
\label{thm_bias_clt}
Suppose the conditions of Lemma \ref{lem_expansion} hold.
Then
\[
    		\sqrt{N}
    		\bigl(
        			\hat{\bm{\beta}}-\bm{\beta}_0-\bm B_N
    		\bigr)
    	\overset{d}{\longrightarrow}
   		\mathcal N(\bm 0,\bm V_\beta),
\]
where
\[
    		\bm V_\beta
    	=
    		\Var\!\left(
        			\sum_{g=1}^G
        			\mathbf 1\{\bm X_i\in g\}
        			\bm\Psi_g(Y_i)
   		 \right),
\]
and $\bm\Psi_g$ is the pointwise limit of $\bm\Psi_{g,N}$, given by
$
    		\bm\Psi_g(y)
    	=
    		\bm\Sigma_X^{-1}\bm X_g(y - m_{g}).
$
\end{theorem}

The limiting influence function coincides with that of the
estimator that would be obtained if the group sample means
were observed and used in place of the reconstructed means
$\hat m_g$.
The proposed estimator therefore has the same asymptotic
covariance matrix as this benchmark, although its discretization
bias may remain nonnegligible at the $N^{-1/2}$ scale.
The following stronger grid condition makes this bias negligible,
yielding the same centered normal approximation.

\Needspace{8\baselineskip}
\begin{corollary}[Standard centered asymptotic normality]
\label{cor_centered_clt}
Suppose the conditions of Theorem~\ref{thm_bias_clt} hold.
Assume further that
$
    \Delta(\mathcal T_{K_N})^2=o(N^{-1/2}).
$
Then
\[
    \sqrt{N}
    \bigl(
        \hat{\bm{\beta}}-\bm{\beta}_0
    \bigr)
    \overset{d}{\longrightarrow}
    \mathcal N(\bm 0,\bm V_\beta).
\]
In particular, for an evenly spaced grid, the above order conditions are satisfied if
$
    N^{1/4}=o(K_N)
$.
\end{corollary}

\subsection{Variance Estimation and Feasible Inference}\label{subsec_variance}
To implement inference based on Corollary~\ref{cor_centered_clt},
we estimate the unknown asymptotic covariance matrix
$\bm V_\beta$ using only the disclosed summaries.

Since the groups form a partition, each observation belongs to exactly one group.
Together with $E[\bm\Psi_g(Y_i)\mid\bm X_i\in g]=\bm0$,
this implies that the cross-group covariance terms in $\bm V_\beta$ vanish.
Hence
\[
    		\bm V_\beta
	    =
		    \sum_{g=1}^G
		    \pi_g
		    E\left[
		        \bm\Psi_g(Y_i)
		        \bm\Psi_g(Y_i)'
		        \mid
		        \bm X_i\in g
		    \right] 
	    =
		    \bm\Sigma_X^{-1}
		    \left[
		        \sum_{g=1}^G
		        \pi_g
		        \bm X_g\bm X_g'
		        \Var(Y_i-m_g\mid\bm X_i\in g)
		    \right]
		    \bm\Sigma_X^{-1}.
\]

Feasible estimation of $\bm V_\beta$ therefore requires
estimating the group-specific variance components.\footnote{
For overlapping groups, observations may contribute to several summaries,
generating cross-group covariance terms. Estimating these terms generally
requires joint information unavailable from the marginal summaries alone
\citep[see, e.g.,][]{farebrother79}. The role of the partition in sharp
identification is discussed in Section~\ref{sec_partial_identification}.
}
Write $\hat m_g=\sum_{k=1}^{K_N}\alpha_{k,N}\widehat Q_g(\tau_k)$,
where the reconstruction weights are
\begin{align*}
    \alpha_{1,N}&=\tau_1+\frac{\tau_2-\tau_1}{2},\\
    \alpha_{k,N}&=\frac{\tau_k-\tau_{k-1}}{2}
                  +\frac{\tau_{k+1}-\tau_k}{2},
                  \quad k=2,\ldots,K_N-1,\\
    \alpha_{K_N,N}&=\frac{\tau_{K_N}-\tau_{K_N-1}}{2}
                    +(1-\tau_{K_N}).
\end{align*}
They are nonnegative and sum to one.

Because $\hat m_g$ is a weighted sum of the disclosed sample
quantiles, we estimate its variance using an asymptotic
approximation to their covariances\footnote{Supplemental Material~L provides the derivation of this variance estimator.}:
\[
    \widehat{\Var}(\hat m_g)
    =\frac{1}{n_g}\sum_{k=1}^{K_N}\sum_{l=1}^{K_N}
      \alpha_{k,N}\alpha_{l,N}
      \frac{\min(\tau_k,\tau_l)-\tau_k\tau_l}
      {\hat f_g(\widehat Q_g(\tau_k))\,
       \hat f_g(\widehat Q_g(\tau_l))},
\]
where the density estimates are constructed using the identity
$f_g(Q_g(\tau))=1/Q_g'(\tau)$:
\[
    \hat f_g(\widehat Q_g(\tau_k))
    =\frac{1}{\widehat{Q_g'}(\tau_k)},
    \qquad
    \widehat{Q_g'}(\tau_k)
    =\frac{\widehat Q_g(\tau_{k+1})-\widehat Q_g(\tau_{k-1})}
           {\tau_{k+1}-\tau_{k-1}}
\]
for $k=2,\ldots,K_N-1$.
At $\tau_1$ and $\tau_{K_N}$, we use forward and backward
differences, respectively.

Combining these group-level variance estimates, we estimate
the covariance matrix of $\hat{\bm\beta}$ by
\[
    \widehat{\bm V}_\beta
    =\left(\sum_{g=1}^G n_g\bm X_g\bm X_g'\right)^{-1}
      \left(\sum_{g=1}^G n_g^2\bm X_g\bm X_g'
        \widehat{\Var}(\hat m_g)\right)
      \left(\sum_{g=1}^G n_g\bm X_g\bm X_g'\right)^{-1}.
\]
All components in $\widehat{\bm V}_\beta$ use disclosed summaries only.
Under the conditions stated below,
$n_g\widehat{\Var}(\hat m_g)$ consistently estimates
$\Var(Y_i-m_g\mid\bm X_i\in g)$, the group-specific variance
component appearing in $\bm V_\beta$. 
Accordingly, $N\widehat{\bm V}_\beta$ consistently estimates
the asymptotic covariance matrix $\bm V_\beta$.

The finite-difference estimator $ \widehat{Q_g'}(\tau_k)$ divides differences of sample
quantiles by the corresponding differences in quantile indices.
Small spacings can therefore amplify sampling errors in the
derivative estimates.
To control this effect, define the minimum spacing
\[
    \underline\Delta(\mathcal T_{K_N})
    =
    \min_{1\le k\le K_N-1}(\tau_{k+1}-\tau_k).
\]
The additional condition
$N^{-1/2}=o(\underline\Delta(\mathcal T_{K_N}))$
in the following proposition is sufficient to make this
sampling-error contribution vanish.

\begin{proposition}[Consistency of the variance estimator]
\label{prop_var}
Under the conditions of Proposition~\ref{prop_rootN} with Assumption \ref{ass_smooth}(ii), suppose that
$
	N^{-1/2} = o(\underline\Delta(\mathcal T_{K_N}))
$.
Then
\[
    N \widehat{\bm V}_\beta
    \xrightarrow{p}
    \bm V_\beta.
\]
\end{proposition}

\begin{remark}
Under an evenly spaced grid, $\tau_k = k/(K_N+1)$, we have
$
    \underline\Delta(\mathcal T_{K_N})
    =
    \frac{1}{K_N+1}.
$
Hence, the condition
$
    N^{-1/2} = o(\underline\Delta(\mathcal T_{K_N}))
$
reduces to
$
    K_N = o(N^{1/2}).
$
\end{remark}

Together, Corollary~\ref{cor_centered_clt} and Proposition~\ref{prop_var}
give a feasible centered normal approximation. For an evenly spaced grid,
their conditions require $N^{1/4}=o(K_N)$ and $K_N=o(N^{1/2})$.
The finite-difference implementation requires at least two disclosed
quantiles.

The minimum-spacing condition is specific to the finite-difference
variance estimator.
Supplemental Material~H gives an alternative density-free variance
estimator based on reconstruction of the conditional second moment.
That estimator does not require the minimum-spacing condition;
its consistency is established for quantile grids becoming dense.
We retain the finite-difference procedure as the baseline in the
simulations, applications, and \texttt{qonlyreg} package because
the same approach to estimating sample-quantile covariances is
also used to implement inference under partial identification
in Section~\ref{sec_fixedDC}.

\section{Fixed Quantile Disclosure: Non-identification and Partial Identification}\label{sec_fixedDC}
The preceding asymptotic results require increasingly dense quantile disclosure.
We now study identification and inference when only a fixed finite set of quantiles is disclosed.

\subsection{Non-identification}\label{subsec_nonID}
Under fixed quantile disclosure, the regression coefficients are not generally point identified.

\begin{theorem}[Non-identification under fixed quantile disclosure]
\label{thm_nonidentification}
Suppose that $\bm\Sigma_X=\sum_{g=1}^G\pi_g\bm X_g\bm X_g'$ is positive definite, and let $\mathcal T_K=\{\tau_1,\ldots,\tau_K\}$ be any fixed finite disclosure design.
Then there exist two distinct joint distributions of $(Y,\bm X)$ with the same marginal distribution of $\bm X$ such that, for $\ell=1,2$, the group-specific conditional means satisfy $m_g^{(\ell)}=\bm X_g'\bm\beta_0^{(\ell)}$ for all $g$, and the conditional quantile functions satisfy Assumption~\ref{ass_smooth}(ii), with
\[
    Q_g^{(1)}(\tau_k)=Q_g^{(2)}(\tau_k)
    \quad(g=1,\ldots,G;\ k=1,\ldots,K),
    \qquad
    \bm\beta_0^{(1)}\ne\bm\beta_0^{(2)}.
\]
Consequently, $\bm\beta_0$ is not point identified from any fixed finite collection of disclosed quantiles.
\end{theorem}

Theorem~\ref{thm_nonidentification} isolates an identification problem caused by disclosure: even population quantiles need not determine the group means or regression coefficients.
Increasing the number of disclosed quantiles alone need not resolve it.
If $\liminf_{N\to\infty}\Delta(\mathcal T_{K_N})>0$, distinct observationally equivalent distributions with different coefficients exist for every sufficiently large $N$; these distributions may depend on $N$ (Supplemental Material~K).

\subsection{Partial Identification}\label{sec_partial_identification}
The non-identification result above motivates a partial-identification
analysis under fixed quantile disclosure.
We first derive sharp bounds for the group means and then use
these bounds to characterize the identified set for the regression
coefficients.

For partial identification, our aim is to derive bounds on group
means using only the probability restrictions implied by the
population quantiles and known support bounds.
We do not impose the smoothness or positive-density conditions
used in the asymptotic analysis on candidate distributions.

For the remainder of this section, assume that known bounds
$\underline y_g$ and $\overline y_g$ satisfy
$
    P(\underline y_g\le Y\le\overline y_g
      \mid\bm X\in g)=1.
$
Let $\mathcal F_g(\mathcal T_K)$ denote the class of candidate
conditional distribution functions $F$ having support contained
in $[\underline y_g,\overline y_g]$ and satisfying
\[
    F(Q_g(\tau_k)-)
    \le \tau_k
    \le F(Q_g(\tau_k)),
    \qquad k=1,\ldots,K,
\]
where $F(y-)=\lim_{z\uparrow y}F(z)$.
In defining $\mathcal F_g(\mathcal T_K)$, we hold fixed the
population values $Q_g(\tau_k)$ defined under the
generalized-inverse convention in
Section~\ref{sec_model_estimation}, but do not require them to
satisfy
$
    Q_g(\tau_k)=\inf\{y:F(y)\ge\tau_k\}
$
for each candidate distribution $F$.

To characterize the means attainable under
$\mathcal F_g(\mathcal T_K)$, we introduce a class of
nondecreasing functions.
Augment the fixed grid by $\tau_0=0$ and $\tau_{K+1}=1$,
and set $Q_g(\tau_0)=\underline y_g$ and
$Q_g(\tau_{K+1})=\overline y_g$ for notational convenience.
Let $\mathcal Q_g(\mathcal T_K)$ be the set of nondecreasing
functions $Q:[0,1]\to\mathbb R$ satisfying
$Q(\tau_k)=Q_g(\tau_k)$ for $k=0,\ldots,K+1$.
No left- or right-continuity is required for these functions.
Define
\[
    \mathcal M_g(\mathcal T_K)
    =
    \left\{
        \int_0^1 Q(\tau)\,d\tau:
        Q\in\mathcal Q_g(\mathcal T_K)
    \right\}.
\]
The following theorem shows that $\mathcal M_g(\mathcal T_K)$
is precisely the set of means attainable by distributions in
$\mathcal F_g(\mathcal T_K)$ and gives its sharp lower and
upper bounds.

By monotonicity, any $Q\in\mathcal Q_g(\mathcal T_K)$ satisfies
$Q_g(\tau_k)\le Q(\tau)\le Q_g(\tau_{k+1})$
for $\tau\in(\tau_k,\tau_{k+1})$.
Integrating the step functions that take the left- and
right-endpoint quantile values on each interval gives the lower and upper bounds for the mean.

\begin{theorem}[Sharp bounds for the mean]
\label{thm_mean_bounds}
The set $\mathcal M_g(\mathcal T_K)$ coincides with the
identified set for the group mean $m_g$ and is given by
\[
    \mathcal M_g(\mathcal T_K)
    =
    \left\{
        \int_{\mathbb R} y\,dF(y):
        F\in\mathcal F_g(\mathcal T_K)
    \right\}
    =
    [\underline m_g,\overline m_g],
\]
where
$
    \underline m_g
    =
    \sum_{k=0}^{K}
    (\tau_{k+1}-\tau_k)Q_g(\tau_k)
$
and
$
    \overline m_g
    =
    \sum_{k=0}^{K}
    (\tau_{k+1}-\tau_k)Q_g(\tau_{k+1}).
$
These bounds are sharp: every value in
$[\underline m_g,\overline m_g]$, including both endpoints,
is attainable as the mean of a distribution in
$\mathcal F_g(\mathcal T_K)$.
\end{theorem}

The following corollary characterizes the width of this identified
interval and provides an upper bound in terms of the maximal spacing $\Delta(\mathcal T_K)$.

\begin{corollary}[Identification Width]
\label{cor_width}
The width of the identified interval satisfies
\[
    \overline m_g-\underline m_g
    =\sum_{k=0}^K(\tau_{k+1}-\tau_k)
      \bigl(Q_g(\tau_{k+1})-Q_g(\tau_k)\bigr)
    \le(\overline y_g-\underline y_g)\Delta(\mathcal T_K).
\]
\end{corollary}

For a fixed number of disclosed quantiles, evenly spaced grids
minimize the maximal-spacing bound on discretization error
(Corollary~\ref{cor_optimal_design}) and the upper bound
$(\overline y_g-\underline y_g)\Delta(\mathcal T_K)$ on
the width of the identified interval.
They need not minimize the width itself, which also depends on
the increments of $Q_g$ between adjacent quantile indices.
A nonuniform grid that places more quantiles where $Q_g$ changes
rapidly may therefore yield a narrower identified interval.

The bounds on the group means translate into restrictions on
the regression coefficients through the relation
$m_g=\bm X_g'\bm\beta_0$.
Over a parameter space $\Theta\subset\mathbb R^p$, we therefore define
\[
    \mathcal B(\mathcal T_K)
    =\left\{\bm\beta\in\Theta:
      \underline m_g\le\bm X_g'\bm\beta\le\overline m_g
      \text{ for all }g=1,\ldots,G\right\}.
\]

The following proposition shows that every
$\bm\beta\in\mathcal B(\mathcal T_K)$ can be attained by
group-specific distributions that match the disclosed quantiles
and have means $\bm X_g'\bm\beta$.

\begin{proposition}[Sharp Partial Identification of $\bm\beta_0$]
\label{prop_beta}
Suppose $\Theta$ is compact and convex.
Then $\mathcal B(\mathcal T_K)$ is the sharp identified set
for $\bm\beta_0$ and is compact and convex.
\end{proposition}

Sharpness relies on the partition structure, which permits the group-specific conditional distributions to be specified separately and combined into a joint distribution.
With overlapping groups, cross-group compatibility imposes additional restrictions, so the same group-by-group bounds need not characterize the sharp set.

\subsection{Estimation and Inference}\label{subsec_inference}
Replace $Q_g(\tau_k)$ by $\widehat Q_g(\tau_k)$ in Theorem~\ref{thm_mean_bounds}, retaining the known support bounds as the boundary-node values, to obtain $\widehat{\underline m}_g$ and $\widehat{\overline m}_g$.
The estimated identified set is
\[
    \widehat{\mathcal B}(\mathcal T_K)
    =\left\{\bm\beta\in\Theta:
      \widehat{\underline m}_g\le\bm X_g'\bm\beta
      \le\widehat{\overline m}_g
      \text{ for all }g=1,\ldots,G\right\}.
\]
Proposition~\ref{prop_set_consistency} in Supplemental Material~E gives sufficient conditions for its Hausdorff consistency.

The estimated group-mean bounds inherit sampling error from the
disclosed sample quantiles.
Consequently, the plug-in set $\widehat{\mathcal B}(\mathcal T_K)$
need not contain the population identified set
$\mathcal B(\mathcal T_K)$.
We construct a confidence region by adjusting the estimated
lower bounds downward and the estimated upper bounds upward.
To calibrate these adjustments simultaneously across all groups,
we use the joint asymptotic distribution of the estimated bounds.

Specifically, let
\[
    \bm s
    =
    (\underline m_1,\overline m_1,\ldots,
     \underline m_G,\overline m_G)',
\]
and let $\widehat{\bm s}$ be its sample analog.
Under the conditions of Lemma~\ref{lem_joint_bound_normality},
\[
    \sqrt N(\widehat{\bm s}-\bm s)
    \xrightarrow{d}\mathcal N(\bm0,\bm\Omega).
\]
The partition structure makes $\bm\Omega$ block diagonal; its formula in terms of group shares and quantile densities is given in Supplemental Material~E.
A plug-in estimator $\widehat{\bm\Omega}$ replaces these quantities by $n_g/N$ and 
$\hat f_g(\widehat Q_g(\tau_k))$.

The asymptotic coverage guarantee requires
$\widehat{\bm\Omega}\xrightarrow{p}\bm\Omega$ as $N \to \infty$.
With fixed quantile disclosure, the finite-difference density
estimates $\hat f_g(\widehat Q_g(\tau_k))$ need not consistently
estimate $f_g(Q_g(\tau_k))$, so consistency of the resulting
plug-in estimator $\widehat{\bm\Omega}$ is not generally guaranteed.
Consistency holds in special cases, such as when each $Q_g$
is linear on all intervals used for differencing.
Supplemental Material~E discusses consistency using additional
disclosed covariance or density estimates, and possible
extensions using quantiles in shrinking neighborhoods.
In the partial-identification simulations and applications,
we construct $\widehat{\bm\Omega}$ as a plug-in approximation
using finite-difference density estimates.
Exact support endpoints, when known, are used as boundary
nodes, whereas assumed support bounds do not enter the
finite-difference estimation.

Let $\widehat\omega_j^2$ be the $j$th diagonal element of $\widehat{\bm\Omega}$.
Compute the $1-\alpha$ quantile $\hat q_{1-\alpha}$ of $\max_{1\le j\le2G}|Z_j/\widehat\omega_j|$ for $\bm Z\sim\mathcal N(\bm0,\widehat{\bm\Omega})$ by Monte Carlo simulation.
Enlarging the estimated constraints gives a feasible simultaneous confidence region for $\bm\beta_0$:
\[
    \widehat{\mathcal C}_{1-\alpha}
    =\left\{\bm\beta\in\Theta:
      \widehat{\underline m}_g
      -\hat q_{1-\alpha}\frac{\widehat\omega_{2g-1}}{\sqrt N}
      \le\bm X_g'\bm\beta\le
      \widehat{\overline m}_g
      +\hat q_{1-\alpha}\frac{\widehat\omega_{2g}}{\sqrt N}
      \text{ for all }g\right\}.
\]
Under the conditions of Theorem~\ref{thm_confidence_region_coverage}, including covariance consistency,
\[
    \liminf_{N\to\infty}
    P\bigl(\mathcal B(\mathcal T_K)
      \subseteq\widehat{\mathcal C}_{1-\alpha}\bigr)
    \ge1-\alpha.
\]
Thus the region covers the entire identified set with asymptotic probability at least $1-\alpha$.

A confidence interval for the $j$th coefficient is obtained by projecting
$\widehat{\mathcal C}_{1-\alpha}$ onto the $j$th coordinate axis.
Its lower and upper endpoints are
\[
    \hat\beta_j^L
    =\min_{\bm\beta\in\widehat{\mathcal C}_{1-\alpha}}\beta_j,
    \qquad
    \hat\beta_j^U
    =\max_{\bm\beta\in\widehat{\mathcal C}_{1-\alpha}}\beta_j.
\]
The lower and upper endpoints are obtained by solving convex optimization problems.
If $\Theta$ is defined by finitely many linear equality and inequality constraints, these problems are linear programs.

\section{Simulation Study}\label{sec_simulation}

We examine finite-sample estimation accuracy and confidence coverage
under increasing and fixed quantile disclosure.
All simulations and empirical analyses use the \texttt{qonlyreg}
R package developed for this paper.

\subsection{Point Estimation}
For each replication, we generate data according to the model
\[
Y_i=\beta_0+\beta_1X_i+u_i,
\qquad
u_i\mid X_i\in g\sim\mathcal N(0,\sigma_g^2),
\qquad
\sigma_g=\exp(0.5X_g),
\]
with $(\beta_0,\beta_1)=(1,2)$ and $G=10$ equally sized groups.
For each observation in group $g$, we set $X_i=X_g$, where
the group-level values $X_g$ are equally spaced over $[-1,1]$.
We consider $N\in\{2000,20000\}$ and $K\in\{3,5,7,10,15,20,50\}$.
For each of $M=10{,}000$ replications, the same sample is used across all values of $K$ and three disclosure designs: an evenly spaced grid and two progressively more right-tail-coarse grids.
The latter have larger maximal spacing $\Delta(\mathcal T_K)$ for a given $K$; their definitions appear in Supplemental Material~D.

The Gaussian design assesses robustness outside Assumption~\ref{ass_smooth}: its support is unbounded and its quantile derivatives diverge at the endpoints.
Supplemental Material~D also reports a bounded-support design satisfying the maintained conditions, additional sample sizes, and results indexed by $\Delta(\mathcal T_K)$.
Under the evenly spaced grid, symmetry makes the population mean-reconstruction error zero for every finite $K$.
Its finite-sample performance therefore reflects sample-quantile and feasible variance-estimation error, whereas the asymmetric grids generally induce population reconstruction bias.

[Figure~\ref{fig_simulation} here ]

Figure~\ref{fig_simulation} reports results for $\beta_1$.
Bias and MSE fall as $K$ increases, and grids with larger gaps perform worse for a given $K$.
Evenly spaced grids attain coverage close to 95\% with relatively few quantiles; right-tail-coarse grids require substantially more.
For the asymmetric grids, increasing $N$ alone cannot remove fixed-$K$ reconstruction bias: intervals can concentrate around a biased estimator and substantially undercover.
Supplemental Material~D provides numerical coverage results for evenly spaced grids (Table~\ref{tab:coverage_even}) and a Laplace-error experiment that violates Assumption~\ref{ass_smooth}(ii).
The grid ranking persists under Laplace errors, but more quantiles are needed for coverage to approach its nominal level.

\subsection{Partial Identification}

For fixed-disclosure inference, we retain the preceding design but replace the errors by
$u_i\mid X_i\in g\sim\mathrm{Uniform}(-\sigma_g,\sigma_g)$ to ensure bounded support.
The true group-specific support endpoints are treated as known
and are used both to construct the group-mean bounds and as
boundary nodes in finite-difference density estimation.
Because each $Q_g$ is linear under the uniform design, the
finite-difference density estimates
$\hat f_g(\widehat Q_g(\tau_k))$ used to construct
$\widehat{\bm\Omega}$ are consistent even with fixed $K$.
With evenly spaced grids and fixed $K\in\{3,9,19\}$, we examine coverage of the entire true identified set and the contraction of projection intervals.

[Table~\ref{tab:partial_identification_region_coverage_even} here ]

Table~\ref{tab:partial_identification_region_coverage_even} shows slight undercoverage when $N=1{,}000$, with coverage approaching its nominal level as $N$ increases.
Projection intervals become shorter with larger $N$, reflecting contraction toward the identified set, and with larger $K$, reflecting sharper identification.
Supplemental Material~D illustrates this geometry (Figure~\ref{fig_simulation_partial}) and reports alternative grid designs.

\section{Empirical Illustration}\label{sec_empirical}

Three applications illustrate quantile-only and mixed disclosure.
Conventional intervals account for sampling uncertainty but not finite-grid mean-reconstruction bias.
The simultaneous regions instead allow all group means consistent with the disclosed summaries and support restrictions.

\subsection{Educational Achievement}

We use publicly released NAEP summaries for the 2024 Grade~8
mathematics assessment.\footnote{All NAEP summary statistics used in
this empirical illustration were retrieved from the NAEP Data Explorer
on May 19, 2026:
\url{https://www.nationsreportcard.gov/ndecore/xplore/NDE}.}
The 50 states, District of Columbia, and Puerto Rico each contribute economically disadvantaged and non-economically disadvantaged cells, giving $G=104$ groups.
The release provides group sizes, means, and selected percentiles rather than individual-level data in this setting.
We reserve the published means for validation: point estimation uses only the disclosed quantiles, while partial identification additionally uses the known score bounds $[0,500]$.

For the group mean mathematics score $m_g$, we specify
\[
    m_g=\alpha_{s_g}+\beta\,\mathrm{nonEDS}_g,
\]
where $s_g$ denotes the jurisdiction to which group $g$ belongs,
$\alpha_{s_g}$ is a jurisdiction-specific intercept, and
$\mathrm{nonEDS}_g$ indicates non-economically disadvantaged status.
Thus, $\beta$ measures the score difference after controlling for state fixed effects.
We compare median-only, quartile, and five-percentile (10th, 25th, 50th, 75th, and 90th) disclosure.
With a median alone, however, the finite-difference covariance estimators used for conventional and partial-identification inference are unavailable.

[Table~\ref{tab:naep_qonlyreg} here ]

Table~\ref{tab:naep_qonlyreg} shows that all three point estimates are close to the published-mean benchmark of 28.4.
The five-percentile conventional interval contains the benchmark, whereas the quartile interval does not.
Increasing disclosure from quartiles to five percentiles reduces the width of the projection interval by about $54\%$.
Both projected regions nevertheless remain much wider than the conventional intervals, include zero, and contain the published-mean benchmark.

\subsection{Political Control and Children in Need Assessment Duration}

English local authorities assess children's needs following referrals to social care \citep{dfe_cin_methodology}; statutory guidance calls for completion within 45 working days \citep{dfe_working_together_2026}.
The Department for Education (DfE) publishes council-year assessment counts and duration quartiles, but not mean duration.
We combine these summaries for reporting years 2017--2025 with annual council-composition records (Supplemental Material~F).
Completed assessments are the underlying units.
Let $g$ index council-year groups, with group size $n_g$ equal to the corresponding assessment count.

We compare mean assessment duration around the 50\% Labour seat-share threshold.\footnote{
This is an RD-motivated fixed-window illustration.
A conventional local-polynomial RD analysis would require a
many-group extension with a shrinking bandwidth.
}
Following \citet{alonso20}, define $R_g$ as Labour's percentage share of council seats minus 50, and let
$D_g=\mathbf 1\{R_g\geq0\}$.
For a fixed window $h$, define
$\mathcal G_h=\{g:|R_g|\leq h\}$.
Within this window, we impose
\[
    m_g=\alpha_h+\delta_hD_g,
    \qquad g\in\mathcal G_h,
\]
where $m_g$ is the population mean assessment duration in group $g$
and $\delta_h$ is the coefficient of interest.
The primary window is $h=4$ percentage points, with $h=3$ and $h=5$
as sensitivity specifications; estimation weights groups by
assessment counts $n_g$.

Using the known lower endpoint $Q_g(0)=0$, the disclosed quartiles,
and constant extrapolation above the 75th percentile,
reconstruction gives
\[
    \hat m_g
    =0.125\times0
    +0.250\widehat Q_g(0.25)
    +0.250\widehat Q_g(0.50)
    +0.375\widehat Q_g(0.75).
\]
For partial identification, we retain the known lower support endpoint $Q_g(0)=0$ and impose the upper support bound $Y_g\leq U$, with $U\in\{150,175,200\}$ days.
These upper bounds are sensitivity assumptions, not administrative maximum durations.
We project the 95\% simultaneous region from Section~\ref{sec_fixedDC} onto $\delta_h$.

[Table~\ref{tab:qonlyreg_main} here ]

\begin{figure}[ht]
\centering
\includegraphics[width=\linewidth]
{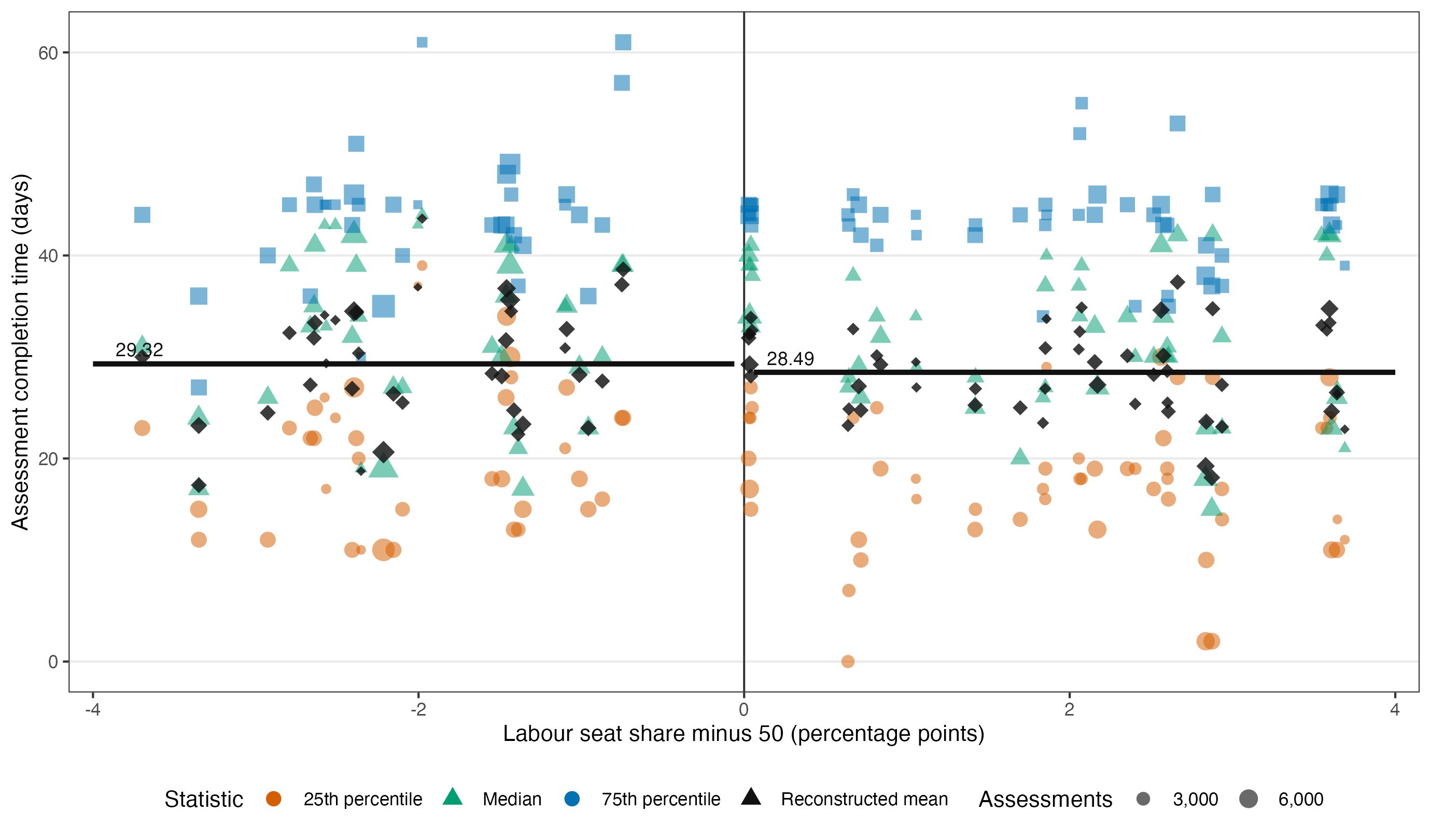}

\caption[Disclosed quantiles and reconstructed means]{\setstretch{2}
Disclosed assessment-duration quantiles and reconstructed means
around the Labour seat-share threshold.
Notes:
Horizontal segments show assessment-count-weighted averages
of reconstructed means on each side of the threshold.
Horizontal jitter is for display only.
Observations exactly at the threshold are assigned to the treated
group and shifted slightly to the right for display.
}
\label{fig_cin_main}
\end{figure}

Table~\ref{tab:qonlyreg_main} reports a main-window estimate of $-0.830$ days with a conventional standard error of $0.061$.
Point estimates are negative in all three windows, but every projected interval is wide and contains zero across window and upper-support specifications.
Figure~\ref{fig_cin_main} displays the disclosed quartiles and reconstructed means.
The negative point estimate is driven primarily by the lower disclosed quartile;
Supplemental Material~F reports the quantile decomposition, reporting-year
fixed effects, alternative political-control definitions, and
leave-one-council-out results. The latter retain a negative estimate in 35 of
36 exclusions; excluding Bristol yields $0.107$ days.

[Figure~\ref{fig_cin_main} here ]
%

\subsection{Healthcare-Associated Infections and Medical Expenditure}
\label{subsec_hai_expenditure}

To illustrate mixed disclosure, we use Table~IV of \citet{liu22_hai}, a meta-analysis of healthcare-associated infections (HAIs) in Chinese general hospitals.
Seven study--year comparisons from five publications report expenditure in Chinese yuan (CNY) for HAI and non-HAI inpatients.
Four comparisons disclose sample means and standard deviations, including three annual comparisons for 2013--2015 from \citet{li17_hai}; three disclose quartiles.

Let $Y_g^a$ denote total medical expenditure per inpatient in comparison
$g$ and HAI status $a\in\{H,C\}$, and write $\mu_g^a=E(Y_g^a)$.
Each comparison-by-status pair $(g,a)$ constitutes a group in the general
framework. With a study--year-specific intercept $\alpha_g$, we specify
\[
\mu_g^a=\alpha_g+\beta_0D_g^a,
\qquad g=1,\ldots,7,
\qquad D_g^H=1,\quad D_g^C=0,
\]
so that $\beta_0$ is the common within-comparison mean difference, $\mu_g^H-\mu_g^C$.
We use reported means directly, reconstruct quartile-based means as in Section~\ref{sec_model_estimation}, and apply sample-size-weighted minimum distance.
Inference incorporates the reported paired-difference uncertainty for the three \citet{li17_hai} comparisons.
For the others, the unavailable covariance between HAI- and non-HAI-specific summaries is set to zero as a working assumption.

Fixed-disclosure inference imposes $Y_g^a\in[0,U]$, with $U=200{,}000$ CNY in the main specification.
This upper bound is an assumption, not an established expenditure limit.
Supplemental Material~I provides details on the elimination of
study--year-specific intercepts, covariance construction,
simultaneous inference, and support sensitivity.

[Table~\ref{tab_hai_expenditure} here ]

Table~\ref{tab_hai_expenditure} shows a positive point estimate whose conventional 95\% confidence interval excludes zero.
The estimated identified set is empty: directly reported sample mean differences impose incompatible singleton constraints under the common-effect specification.
Allowing for sampling uncertainty yields a nonempty projected interval entirely above zero, but well above the point estimate.
This discrepancy may reflect finite-grid reconstruction bias in the point estimate, as well as tension with the common-effect specification, for example due to between-study heterogeneity.
The discrepancy persists across all upper support bounds considered: the confidence region remains above the point estimate and excludes zero throughout.
The region is unchanged at \(U=200{,}000\) and \(U=500{,}000\) CNY: once support is sufficiently loose, mean-reporting comparisons bind and keep the region informative.\footnote{Supplemental Material~J gives a vitamin D--tuberculosis illustration in which mean-reporting studies similarly anchor the region, which contains the point estimate.}

\section{Conclusion}\label{sec_conclude}

What can be learned about group-mean regression coefficients depends
on the information released, particularly the number and placement
of the disclosed quantiles.
Our framework accommodates quantile-only and mixed disclosure.
Increasing disclosure permits consistent, asymptotically normal estimation under the stated conditions; fixed disclosure generally yields partial identification, for which we derive sharp sets and simultaneous confidence regions.
The number and placement of quantiles affect reconstruction bias,
the feasibility of conventional inference, and identification
uncertainty, making disclosure design consequential for regression analysis.

Extensions include shape restrictions between disclosed quantiles to sharpen identification, panel and nonlinear models, more efficient weighting, and random-effects meta-analysis.
A framework with diverging $G$ would also permit local-polynomial RD methods with a shrinking bandwidth and an increasing number of groups near the cutoff.

\textbf{Acknowledgements: } 
We would like to thank Yoshiyasu Rai 
and
the participants of the Summer Workshop on Economic Theory 2026
for their useful comments.
We used ChatGPT (OpenAI, GPT-5) to assist with English-language editing,
the development and debugging of R code, and the exposition and
organization of mathematical arguments.
All AI-assisted output was reviewed by the authors, who take full responsibility for the content.

\textbf{Funding: } 
This work was supported by JSPS KAKENHI Grant Number 24K04826.

\textbf{Disclosure Statement: } 
The authors report there are no competing interests to declare.

\textbf{Data and Code Availability:}
The empirical applications use publicly available aggregate data as
well as study-level summary statistics transcribed from published articles.
Detailed information on data sources and, where applicable, data
retrieval, construction, and licensing is provided in the
application-specific README files in the replication repository.
The replication package, including the analysis data, data-retrieval
and construction scripts where applicable, simulation and empirical
analysis code, generated outputs, and detailed README files, is
available at
\url{https://github.com/masaiwasawa/QOR_Replication}.

\bibliography{bibliography.bib}

\clearpage
\section*{Figures}
\begin{figure}[!htbp]
\centering
\includegraphics[width=\linewidth]{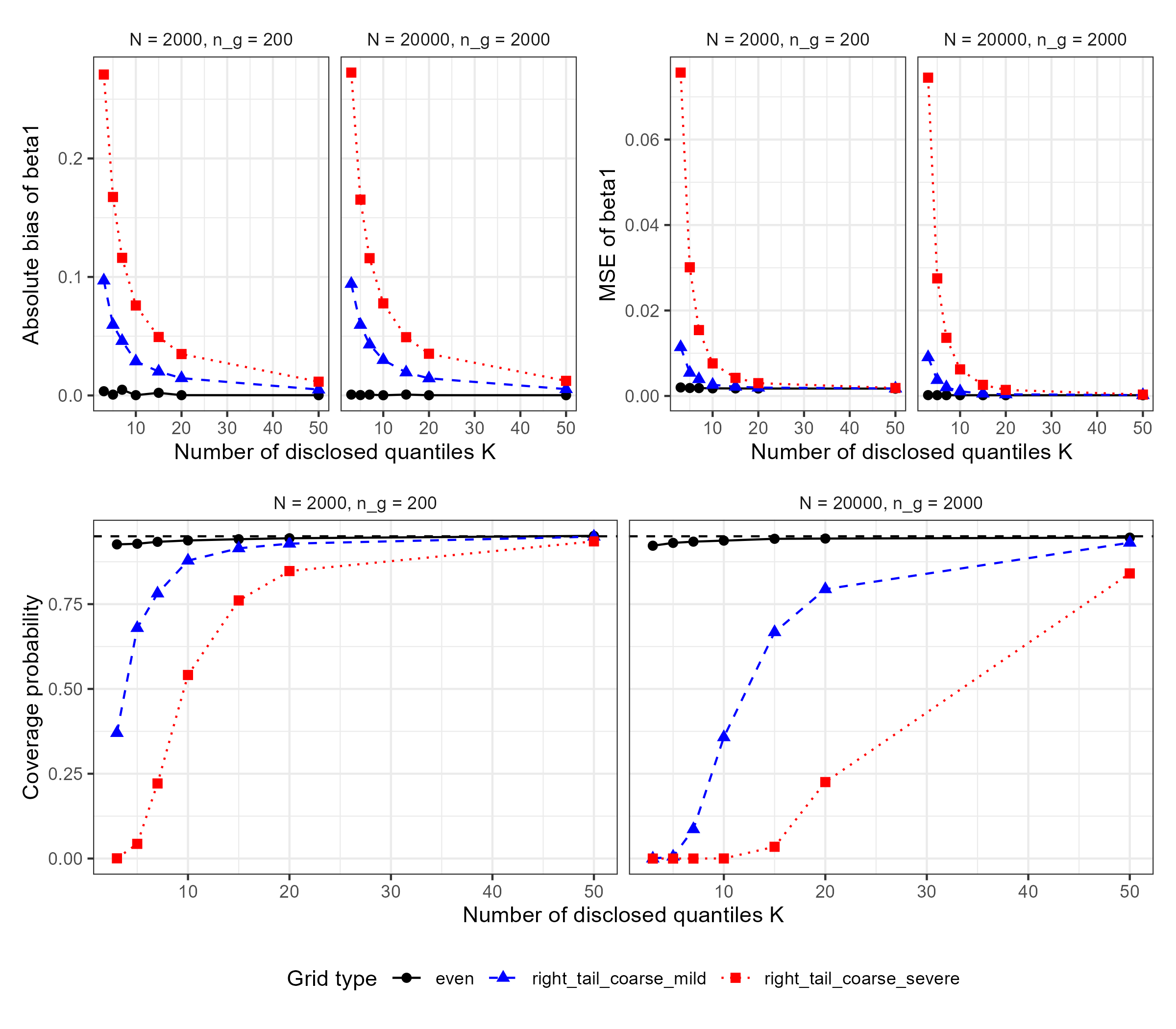}
\caption{\label{fig_simulation}\setstretch{2}
Absolute bias, MSE, and coverage probability of nominal 95\% confidence intervals under different quantile designs.
}
\end{figure}

\clearpage

\clearpage
\section*{Tables}

\input{QOR_Replication/simulation/results_partial_qlm/table_region_coverage_even.tex}
\clearpage

\input{QOR_Replication/empirical_illustration/NAEP_data/table_naep_qonlyreg.tex}
\clearpage

\input{QOR_Replication/empirical_illustration/UK_election/output/paper_tables/table_main_results.tex}
\clearpage

\begin{table}[htbp]
\centering
\begin{threeparttable}
\caption{HAIs and medical expenditure in CNY per inpatient}
\label{tab_hai_expenditure}
{
\renewcommand{\arraystretch}{1.8}
\begin{tabular}{@{}lrr@{}}
\toprule
Method & Estimate & Interval \\
\midrule
Point estimator
    & 21,250.23
    & [19,435.27, 23,065.20] \\
Estimated identified set
    &
    & Empty set \\
Feasible confidence region
    &
    & [28,894.94, 38,163.10] \\
\bottomrule
\end{tabular}
}
\begin{tablenotes}[flushleft]
\setstretch{2}
\item \textit{Notes:}
Positive values indicate higher expenditure among HAI patients.
The point-estimator interval is a conventional 95\% confidence interval.
The estimated identified set and the 95 percent simultaneous confidence region are based on the support restriction
\(Y_g^a\in[0,\,200\,000]\) CNY.
\end{tablenotes}
\end{threeparttable}
\end{table}

\clearpage
 \begin{center} {\Large 
 Supplemental material for the paper entitled \\
 "\TITLE"
 }
 \end{center}

\clubpenalty=150 \widowpenalty=150 \displaywidowpenalty=50
\appendix

\def\thesection{}
\renewcommand{\thesubsection}{A-\arabic{subsection}} 
 \def\theequation{A.\arabic{equation}}
 \def\theassumption{A.\arabic{assumption}}	
 \makeatletter
 \@addtoreset{equation}{section}
 \@addtoreset{figure}{section}
 \@addtoreset{table}{section}
 \@addtoreset{assumption}{section}
 \makeatother
 \def\thelemma{A.\arabic{lemma}}	  

\setcounter{equation}{0} 
\setcounter{subsection}{0} 
\setcounter{lemma}{0}
\setcounter{page}{1}
\section{Supplemental Material A: Proofs}

\subsection{Proof of Lemma \ref{lem_mean}}

\begin{proof}
For each group $g$, let $Q_g^*(\tau)$ denote the piecewise linear interpolation constructed from the population quantiles $\{Q_g(\tau_k): k=1,\dots,K_{N}\}$. That is, for $\tau \in [\tau_k,\tau_{k+1}]$ with $k=1,\dots,K_N-1$,
\[
    Q_g^*(\tau)
    =
    Q_g(\tau_k)
    +
    \frac{\tau-\tau_k}{\tau_{k+1}-\tau_k}
    \bigl[
        Q_g(\tau_{k+1})-Q_g(\tau_k)
    \bigr].
\]
For $\tau<\tau_1$ and $\tau>\tau_K$, define
\[
    Q_g^*(\tau)=Q_g(\tau_1),
    \qquad
    Q_g^*(\tau)=Q_g(\tau_K),
\]
respectively.

Since
$
    m_g=\int_0^1 Q_g(\tau)\,d\tau
$
and
$
    \hat m_g=\int_0^1 \widetilde Q_g(\tau)\,d\tau,
$
we have
\[
    \hat m_g-m_g
    =
    \int_0^1 \bigl[
        \widetilde Q_g(\tau)-Q_g(\tau)
    \bigr]\,d\tau
    =
    \int_0^1 \bigl[
        \widetilde Q_g(\tau)-Q_g^*(\tau)
    \bigr]\,d\tau
    +
    \int_0^1 \bigl[
        Q_g^*(\tau)-Q_g(\tau)
    \bigr]\,d\tau.
\]
Hence,
\[
    |\hat m_g-m_g|
    \le
    \int_0^1
        \bigl|
            \widetilde Q_g(\tau)-Q_g^*(\tau)
        \bigr|
    \,d\tau
    +
    \int_0^1
        \bigl|
            Q_g^*(\tau)-Q_g(\tau)
        \bigr|
    \,d\tau.
\]
We bound the two terms separately.

\medskip

\noindent
\textbf{Step 1: sampling error.}
For $\tau\in[\tau_k,\tau_{k+1}]$ with $k=1,\dots,K_N-1$, since both $\widetilde Q_g(\tau)$ and $Q_g^*(\tau)$ are linear interpolants,
\begin{equation}
    \widetilde Q_g(\tau)-Q_g^*(\tau)
    =
    \lambda_k(\tau)
    \bigl[
        \widehat Q_g(\tau_k)-Q_g(\tau_k)
    \bigr]
    +
    \bigl[
        1-\lambda_k(\tau)
    \bigr]
    \bigl[
        \widehat Q_g(\tau_{k+1})-Q_g(\tau_{k+1})
    \bigr],
    \label{eq_linearityInterpolants}
\end{equation}
where
\[
    \lambda_k(\tau)
    =
    \frac{\tau_{k+1}-\tau}{\tau_{k+1}-\tau_k}.
\]
Hence, by the triangle inequality,
\[
    \bigl|
        \widetilde Q_g(\tau)-Q_g^*(\tau)
    \bigr|
    \le
    \lambda_k(\tau)
    \bigl|
        \widehat Q_g(\tau_k)-Q_g(\tau_k)
    \bigr|
    +
    \bigl[
        1-\lambda_k(\tau)
    \bigr]
    \bigl|
        \widehat Q_g(\tau_{k+1})-Q_g(\tau_{k+1})
    \bigr|.
\]
Since 
$
        \int_{\tau_k}^{\tau_{k+1}}\lambda_k(\tau)\,d\tau
    =
        \int_{\tau_k}^{\tau_{k+1}}[1-\lambda_k(\tau)]\,d\tau
    =
        \frac{\tau_{k+1}-\tau_k}{2}
$,
integrating over $[\tau_k,\tau_{k+1}]$ gives
\[
    \int_{\tau_k}^{\tau_{k+1}}
    \bigl|
        \widetilde Q_g(\tau)-Q_g^*(\tau)
    \bigr|
    d\tau
    \le
    \frac{\tau_{k+1}-\tau_k}{2}
    \bigl(
        |\widehat Q_g(\tau_k)-Q_g(\tau_k)|
        +
        |\widehat Q_g(\tau_{k+1})-Q_g(\tau_{k+1})|
    \bigr).
\]

On the lower extrapolation region,
\[
        \widetilde Q_g(\tau)-Q_g^*(\tau)
    =
        \widehat Q_g(\tau_1)-Q_g(\tau_1),
        \qquad \tau\in[0,\tau_1),
\]
so that
\[
        \int_0^{\tau_1}
        \bigl|
            \widetilde Q_g(\tau)-Q_g^*(\tau)
        \bigr|
         d\tau
    =
        \tau_1
        |\widehat Q_g(\tau_1)-Q_g(\tau_1)|.
\]
Similarly, on the upper extrapolation region,
\[
        \int_{\tau_{K_N}}^1
        \bigl|
           \widetilde Q_g(\tau)-Q_g^*(\tau)
        \bigr|
        d\tau
    =
        (1-\tau_{K_N})
        |\widehat Q_g(\tau_{K_N})-Q_g(\tau_{K_N})|.
\]

Therefore,
\begin{align*}
        \int_0^1
        \bigl|
            \widetilde Q_g(\tau)-Q_g^*(\tau)
        \bigr|
        d\tau
    &\le
        \tau_1 |\widehat Q_g(\tau_1)-Q_g(\tau_1)| \\
        &+
        \sum_{k=1}^{K_N-1}
        \frac{\tau_{k+1}-\tau_k}{2}
        \bigl(
            |\widehat Q_g(\tau_k)-Q_g(\tau_k)|
            +
            |\widehat Q_g(\tau_{k+1})-Q_g(\tau_{k+1})|
        \bigr) \\
        &+
        (1-\tau_{K_N})
        |\widehat Q_g(\tau_{K_N})-Q_g(\tau_{K_N})|.
\end{align*}
Define weights $\{w_{k,N}\}_{k=1}^{K_N}$ by collecting the coefficients in the above bound. 
That is,
\begin{align*}
        w_{1,N}
    &=
        \tau_1+\frac{\tau_2-\tau_1}{2},
        \\
        w_{k,N}
    &=
        \frac{\tau_k-\tau_{k-1}}{2}
        +
        \frac{\tau_{k+1}-\tau_k}{2},
        \qquad 
        k=2,\dots,K_N-1,
    \\
        w_{K_N,N}
    &=
        \frac{\tau_{K_N}-\tau_{K_N-1}}{2}
        +
        (1-\tau_{K_N}).
\end{align*}
Then \(w_{k,N}\ge 0\) for all \(k\), and
$
    \sum_{k=1}^{K_N} w_{k,N}=1.
$
Hence
\[
        \int_0^1
        \bigl|
            \widetilde Q_g(\tau)-Q_g^*(\tau)
        \bigr|
        d\tau
    \le
        \sum_{k=1}^{K_N}
        w_{k,N}
        \bigl|
            \widehat Q_g(\tau_k)-Q_g(\tau_k)
        \bigr|
    =
      O_p(r_{g,N})
\]
by Assumption~\ref{ass_quantile}.
Therefore,
\[
    \max_{1\le g\le G}
    \int_0^1
    \bigl|
        \widetilde Q_g(\tau)-Q_g^*(\tau)
    \bigr|
    d\tau
    =
    O_p(r_N).
\]

\medskip

\noindent
\textbf{Step 2: discretization error under Assumption~\ref{ass_smooth}(i).}
Suppose that each $Q_g$ is Lipschitz continuous with constant $L$. 
For any $\tau \in [\tau_k,\tau_{k+1}]$ with $k=1,\dots,K_N-1$, since $Q_g^*(\tau)$ is a convex combination of $Q_g(\tau_k)$ and $Q_g(\tau_{k+1})$,
\[
    \bigl|
        Q_g^*(\tau)-Q_g(\tau)
    \bigr|
    \le
    \bigl|
        Q_g(\tau_k)-Q_g(\tau)
    \bigr|
    +
    \bigl|
        Q_g(\tau_{k+1})-Q_g(\tau_k)
    \bigr|
    \le
    2L|\tau_{k+1}-\tau_k|.
\]
On the lower extrapolation region,
\[
    |Q_g^*(\tau)-Q_g(\tau)|
    =
    |Q_g(\tau_1)-Q_g(\tau)|
    \le
    L|\tau_1-\tau|,
    \qquad \tau \in [0,\tau_1),
\]
and similarly on the upper extrapolation region,
\[
    |Q_g^*(\tau)-Q_g(\tau)|
    =
    |Q_g(\tau_{K_N})-Q_g(\tau)|
    \le
    L|\tau-\tau_{K_N}|,
    \qquad \tau \in (\tau_{K_N},1].
\]
Therefore,
\[
    \int_0^1
        \bigl|
            Q_g^*(\tau)-Q_g(\tau)
        \bigr|
    \,d\tau
    \le
    C
    \sum_{k=0}^{K_N}
    (\tau_{k+1}-\tau_k)^2
\]
for some constant $C>0$ independent of $g$ and $N$. Since
\[
    \sum_{k=0}^{K_N}
    (\tau_{k+1}-\tau_k)^2
    \le
    \Delta(\mathcal T_{K_N})
    \sum_{k=0}^{K_N}
    (\tau_{k+1}-\tau_k)
    =
    \Delta(\mathcal T_{K_N}),
\]
it follows that
\[
    \max_{1\le g\le G}
    \int_0^1
        \bigl|
            Q_g^*(\tau)-Q_g(\tau)
        \bigr|
    \,d\tau
    =
    O\bigl(
        \Delta(\mathcal T_{K_N})
    \bigr).
\]

\medskip

\noindent
\textbf{Step 3: discretization error under Assumption~\ref{ass_smooth}(ii).}
Suppose instead that each $Q_g$ is twice continuously differentiable with bounded second derivative. 
Since $G$ is fixed, there exists a finite constant $M$ such that
\[
    \sup_{\tau\in(0,1)}
    |Q_g''(\tau)|
    \leq M,
    \qquad
    g=1,\ldots,G.
\]

Assumption~\ref{ass_smooth}(ii) also implies that the first derivatives are uniformly
bounded. Indeed, fixing $\tau'$, we have
$
    |Q_g'(\tau)|
    \leq
    |Q_g'(\tau')|
    +
    M|\tau-\tau'|,
$
for any
$\tau\in(0,1)$.
Hence, since $G$ is fixed, there exists a finite constant $L$ such that
\[
    \sup_{1\leq g\leq G}
    \sup_{\tau\in(0,1)}
    |Q_g'(\tau)|
    \leq L.
\]

We first consider the interior interpolation intervals. 
For $k=1,\ldots,K_N-1$, the standard linear-interpolation error formula (see, e.g., \citealp[Chapter~3]{atkinson89}) implies that, for each $\tau\in[\tau_k,\tau_{k+1}]$, there exists $\xi_{g,\tau}\in(\tau_k,\tau_{k+1})$ such that
\[
    Q_g(\tau)-Q_g^*(\tau)
    =
    \frac{Q_g''(\xi_{g,\tau})}{2}
    (\tau-\tau_k)(\tau-\tau_{k+1}).
\]
Therefore,
\[
\begin{aligned}
		    \int_{\tau_k}^{\tau_{k+1}}
		    |Q_g^*(\tau)-Q_g(\tau)|
		    \,d\tau
	    &\leq
		    \frac{M}{2}
		    \int_{\tau_k}^{\tau_{k+1}}
		    (\tau-\tau_k)(\tau_{k+1}-\tau)
		    \,d\tau
		    \\
	    &=
		    \frac{M}{12}
		    (\tau_{k+1}-\tau_k)^3.
\end{aligned}
\]

Second, on the lower boundary,
\[
\begin{aligned}
		    \int_0^{\tau_1}
		    |Q_g^*(\tau)-Q_g(\tau)|
		    \,d\tau
	    &=
		    \int_0^{\tau_1}
		    |Q_g(\tau_1)-Q_g(\tau)|
		    \,d\tau
	\leq
		    L\int_0^{\tau_1}(\tau_1-\tau)\,d\tau
		    =
		    \frac{L}{2}\tau_1^2.
\end{aligned}
\]
Finally, on the upper boundary,
\[
\begin{aligned}
		    \int_{\tau_{K_N}}^1
		    |Q_g^*(\tau)-Q_g(\tau)|
		    \,d\tau
	    &=
		    \int_{\tau_{K_N}}^1
		    |Q_g(\tau)-Q_g(\tau_{K_N})|
		    \,d\tau
		    \\
	    &\leq
		    L\int_{\tau_{K_N}}^1
		    (\tau-\tau_{K_N})\,d\tau
		    =
		    \frac{L}{2}(1-\tau_{K_N})^2.
\end{aligned}
\]

Combining the interior and boundary bounds gives
\[
\begin{aligned}
		    \int_0^1
		    |Q_g^*(\tau)-Q_g(\tau)|
		    \,d\tau
	    &\leq
		    \frac{L}{2}
		    \left\{
		        \tau_1^2+(1-\tau_{K_N})^2
		    \right\}
		    +
		    \frac{M}{12}
		    \sum_{k=1}^{K_N-1}
		    (\tau_{k+1}-\tau_k)^3.
\end{aligned}
\]
Since
$
    \tau_1
    \leq
    \Delta(\mathcal T_{K_N}),
$
$
    1-\tau_{K_N}
    \leq
    \Delta(\mathcal T_{K_N}),
$
and
\[
		    \sum_{k=1}^{K_N-1}
		    (\tau_{k+1}-\tau_k)^3
    \leq
		    \Delta(\mathcal T_{K_N})^2
		    \sum_{k=1}^{K_N-1}
		    (\tau_{k+1}-\tau_k)
    \leq
		    \Delta(\mathcal T_{K_N})^2,
\]
we obtain
\[
		    \max_{1\leq g\leq G}
		    \int_0^1
		    |Q_g^*(\tau)-Q_g(\tau)|
		    \,d\tau
	    =
		    O\!\left(
	        		\Delta(\mathcal T_{K_N})^2
		    \right).
\]

\medskip

\noindent
\textbf{Step 4: conclusion.}
Combining the bounds in Steps 1--3, we conclude that
\[
    \max_{1\le g\le G}
    |\hat m_g-m_g|
    =
    O_p\!\left(
        \Delta(\mathcal T_{K_N})^a
        +
        r_N
    \right),
\]
where $a=1$ under Assumption~\ref{ass_smooth}(i) and $a=2$ under Assumption~\ref{ass_smooth}(ii).
\end{proof}

\subsection{Proof of Theorem \ref{thm_main}}

\begin{proof}
By Assumption~\ref{ass_design},
the matrix
$
    \sum_{g=1}^G n_g \bm{X}_g \bm{X}_g'
$
is nonsingular with probability approaching one, and 
$
        (
            \frac{1}{N}\sum_{g=1}^G n_g \bm{X}_g \bm{X}_g'
        )^{-1}
    =
        O_p(1)
$.
Hence, the estimator admits the representation
\[
\hat{\bm{\beta}}
=
\left(
\sum_{g=1}^G n_g \bm{X}_g \bm{X}_g'
\right)^{-1}
\sum_{g=1}^G n_g \bm{X}_g \hat m_g,
\]
on an event whose probability tends to one.
In what follows, we work on this event.

Under Assumption~\ref{ass_linear},
$
        m_g 
    = 
        E(Y\mid \bm{X}\in g)
    =
        \bm{X}_g'\bm{\beta }_0
$
for each $g=1,\dots,G$.
Therefore,
\[
        \sum_{g=1}^G n_g \bm{X}_g m_g
    =
        \sum_{g=1}^G n_g \bm{X}_g \bm{X}_g' \bm{\beta }_0,
\]
and thus
\[
        \hat{\bm{\beta}}-\bm{\beta }_0
    =
        \left(
            \frac{1}{N}
            \sum_{g=1}^G n_g \bm{X}_g \bm{X}_g'
        \right)^{-1}
        \frac{1}{N}
        \sum_{g=1}^G n_g \bm{X}_g (\hat m_g-m_g).
\]

We have
\[
        \left\|
            \frac{1}{N}\sum_{g=1}^G n_g \bm{X}_g (\hat m_g-m_g)
        \right\|
    \le
        \left(
            \max_{1\le g\le G} |\hat m_g-m_g|
        \right)
        \frac{1}{N}\sum_{g=1}^G n_g \|\bm{X}_g\|.
\]
Because $G$ is fixed and $\|\bm{X}_g\|<\infty$ for each $g=1,\dots,G$ by Assumption~\ref{ass_design},
$
    \max_{1\le g\le G}\|\bm{X}_g\| < \infty.
$
Hence
\[
        \frac{1}{N}\sum_{g=1}^G n_g \|\bm{X}_g\|
    \le
        \max_{1\le g\le G}\|\bm{X}_g\|
        \frac{1}{N}\sum_{g=1}^G n_g
    =
        \max_{1\le g\le G}\|\bm{X}_g\|
    =
        O(1).
\]
Therefore,
\[
       \frac{1}{N}\sum_{g=1}^G n_g \bm{X}_g \bigl( \hat m_g - m_g \bigr)
    =
        O_p\!\left(
            \max_{1\le g\le G} |\hat m_g - m_g|
        \right)
    =
        O_p\!\left(
            \Delta(\mathcal T_{K_N})^a+r_N
        \right),
\]
where the last equality holds by applying Lemma~\ref{lem_mean}.

Combining the preceding displays gives
\[
        \hat{\bm{\beta}}-\bm{\beta }_0
    =
        O_p\!\left(
            \Delta(\mathcal T_{K_N})^a+r_N
        \right),
\]
where $a=1$ under Assumption~\ref{ass_smooth}(i) and $a=2$ under Assumption~\ref{ass_smooth}(ii).
\end{proof}

\subsection{Proof of Lemma \ref{lem_uniform_quantile}}
\begin{proof}
Define
\[
    \delta_{n,g}
    =
    \sup_{y\in\mathbb R}
    \left|
        F_{n,g}(y)-F_g(y)
    \right|.
\]
By the Dvoretzky--Kiefer--Wolfowitz inequality  (\citealp{dkw56,massart90}),
$
    \delta_{n,g}
    =
    O_p\!\left(n_g^{-1/2}\right).
$

For the purpose of this proof, extend \(Q_g\) continuously to
\([0,1]\) by setting
\[
    Q_g(0)=\underline y_g,
    \qquad
    Q_g(1)=\overline y_g.
\]
Since \(F_g\) is strictly increasing and continuous on its support,
\(Q_g=F_g^{-1}\) on \([0,1]\). Moreover, for any
\(\underline y_g\le x<y\le\overline y_g\),
\[
    F_g(y)-F_g(x)
    =
    \int_x^y f_g(s)\,ds
    \ge
    c_g(y-x).
\]
It follows that \(Q_g\) is Lipschitz continuous on \([0,1]\) with
Lipschitz constant at most \(c_g^{-1}\). That is,
\[
    \left|
        Q_g(u)-Q_g(v)
    \right|
    \le
    c_g^{-1}|u-v|,
    \qquad
    u,v\in[0,1].
\]

Fix any \(\tau\in(0,1)\), and define
\[
    \tau^-_{n,g}(\tau)
    =
    (\tau-\delta_{n,g})\vee0,
    \qquad
    \tau^+_{n,g}(\tau)
    =
    (\tau+\delta_{n,g})\wedge1.
\]
We first show that
\[
    Q_g\!\left(\tau^-_{n,g}(\tau)\right)
    \le
    \widehat Q_g(\tau)
    \le
    Q_g\!\left(\tau^+_{n,g}(\tau)\right).
\]

For the upper inequality, suppose first that \(\tau+\delta_{n,g}<1\). 
Then, by the definition of $\delta_{n,g}$,
\[
\begin{aligned}
    F_{n,g}
    \left(
        Q_g(\tau+\delta_{n,g})
    \right)
    &\ge
    F_g
    \left(
        Q_g(\tau+\delta_{n,g})
    \right)
    -
    \delta_{n,g} 
    =
    \tau.
\end{aligned}
\]
Thus, by the definition of the empirical quantile,
\[
    	\widehat Q_g(\tau)
    =
    	    \inf
	    \left\{
	        y\in\mathbb R:
        		F_{n,g}(y)\ge\tau
	    \right\}
    \le
	    Q_g(\tau+\delta_{n,g}).
\]
If instead \(\tau+\delta_{n,g}\ge1\), then
$
    \widehat Q_g(\tau)
    \le
    \overline y_g
    =
    Q_g(1).
$
Thus, the upper inequality,
$
    \widehat Q_g(\tau)
    \le
    Q_g\!\left(
       \tau^+_{n,g}(\tau)
    \right),
$
holds.

For the lower inequality, suppose first that \(\tau-\delta_{n,g}>0\). 
For every
$
    	y
    <
    	Q_g(\tau-\delta_{n,g}),
$
strict monotonicity of \(F_g\) implies
$
    	F_g(y)
    <
    	\tau-\delta_{n,g}.
$
Consequently,
\[
    F_{n,g}(y)
    \le
    F_g(y)+\delta_{n,g}
    <
    \tau.
\]
Therefore, no \(y<Q_g(\tau-\delta_{n,g})\) can satisfy
\(F_{n,g}(y)\ge\tau\), and hence
\[
    \widehat Q_g(\tau)
    \ge
    Q_g(\tau-\delta_{n,g}).
\]
If instead \(\tau-\delta_{n,g}\le0\), then trivially
$
    	\widehat Q_g(\tau)
    \ge
   	 \underline y_g
    =
    	Q_g(0).
$
Thus, the lower inequality,
$
    \widehat Q_g(\tau)
    \ge
    Q_g\!\left(
        \tau^-_{n,g}(\tau)
    \right),
$
holds.


Next, using these inequalities, we obtain the following:
\begin{align*}
	\text{~~if~~} 
		\widehat Q_g(\tau)\ge Q_g(\tau),
	\text{~~then~~} 
		|\widehat Q_g(\tau)-Q_g(\tau)|
	&= 
		\widehat Q_g(\tau)-Q_g(\tau)
	\le
		Q_g( \tau^+_{n,g}(\tau) )-Q_g(\tau)\\
	\text{~~if~~} 
		\widehat Q_g(\tau)< Q_g(\tau),
	\text{~~then~~} 
		|\widehat Q_g(\tau)-Q_g(\tau)|
	&=
		Q_g(\tau)-\widehat Q_g(\tau)
	\le
		Q_g(\tau)
		-
		Q_g(\tau^-_{n,g}(\tau)).		
\end{align*}
Then, using the Lipschitz continuity of \(Q_g\) shown above, we obtain
\[
\begin{aligned}
    	\left|
        		\widehat Q_g(\tau)-Q_g(\tau)
    	\right|
    &\le
    	\max
    	\Bigl\{
        		Q_g(\tau)
        		-
        		Q_g\!\left(
           		\tau^-_{n,g}(\tau)
        		\right),
        	~
       		 Q_g\!\left(
          		 \tau^+_{n,g}(\tau)
       		 \right)
        		-
        		Q_g(\tau)
    	\Bigr\}
    \\
    &\le
    	c_g^{-1}\delta_{n,g}.
\end{aligned}
\]
Since the right hand side is independent of $\tau$, taking supremum over  the disclosed grid
\(T_{K_N}=\{\tau_1,\ldots,\tau_{K_N}\}\) yields,
\[
    	\sup_{1\le k\le K_N}
    	\left|
       	 	\widehat Q_g(\tau_k)-Q_g(\tau_k)
   	 \right|
    \le
    	c_g^{-1}\delta_{n,g}.
\]
Since
$
    \delta_{n,g}
    =
    O_p\!\left(n_g^{-1/2}\right),
$
it follows that
\[
    \sup_{1\le k\le K_N}
    \left|
        \widehat Q_g(\tau_k)-Q_g(\tau_k)
    \right|
    =
    O_p\!\left(n_g^{-1/2}\right).
\]
This proves the result.
\end{proof}

\subsection{Proof of Lemma \ref{lem_expansion}}
\begin{proof}
For each group $g$, define
\[
        m_{g,N}^*
    =
        \int_0^1 Q_g^*(\tau)\,d\tau,
\]
where $Q_g^*(\tau)$ denotes the piecewise linear interpolant constructed from the population quantiles
$
    \{Q_g(\tau_k):k=1,\dots,K_N\},
$
with the same extrapolation rule as in the definition of $\widetilde Q_g(\tau)$.

Recall the definitions
$
		m_g 
	= 
		\int_0^1 Q_g(\tau)\, d\tau
$
and
$
		\hat{m}_g 
	= 
		\int_0^1 \tilde{Q}_g(\tau) \, d\tau
$.
Consider the decomposition
\[
    \hat m_g-m_g
    =
    (\hat m_g-m_{g,N}^*)
    +
    (m_{g,N}^*-m_g),
\]
where term $m_{g,N}^* - m_g$ corresponds to the discretization error arising from approximating the quantile function on a finite grid, while $\hat m_g - m_{g,N}^*$ is the reconstructed mean error
 that reflects the quantile estimation error induced by sampling variability in the estimated quantiles.

\medskip
\noindent
\textbf{Step 1: discretization error.}
Define
$
		b_{g,N}
	=
	    	m_{g,N}^*-m_g   
$
Under Assumption~\ref{ass_smooth}(ii), the proof of Lemma~\ref{lem_mean} shows that
$
	    b_{g,N}
 	=
	    O\!\left(\Delta(\mathcal T_{K_N})^2\right)	    
$.

\medskip
\noindent
\textbf{Step 2: exact linear representation of the reconstructed mean error.}
By linearity of the interpolants, the difference $\widetilde Q_g(\tau)-Q_g^*(\tau)$ is piecewise linear in the nodal errors $\widehat Q_g(\tau_k)-Q_g(\tau_k)$ as shown in \eqref{eq_linearityInterpolants}. 
Integrating over each interval $[\tau_k,\tau_{k+1}]$ assigns weight $(\tau_{k+1}-\tau_k)/2$ to each endpoint error, while the two extrapolation regions contribute weights $\tau_1$ and $1-\tau_{K_N}$ at the boundaries. Collecting terms yields
\[
    \hat m_g-m_{g,N}^*
    =
    \sum_{k=1}^{K_N}\alpha_{k,N}
    \bigl(
        \widehat Q_g(\tau_k)-Q_g(\tau_k)
    \bigr),
\]
where the weights are given by
\begin{align*}
		\alpha_{1,N}
	&=
		\tau_1+\frac{\tau_2-\tau_1}{2},\\
		\alpha_{k,N}
	&=
		\frac{\tau_k-\tau_{k-1}}{2}
		+
		\frac{\tau_{k+1}-\tau_k}{2},
		\quad k=2,\dots,K_N-1,\\
		\alpha_{K_N,N}
	&=
		\frac{\tau_{K_N}-\tau_{K_N-1}}{2}
		+
		(1-\tau_{K_N}).
\end{align*}
The weights satisfy
$ \alpha_{k,N}\ge 0$
and
$  
		\sum_{k=1}^{K_N}\alpha_{k,N}
	=
		1
$.

By Lemma~\ref{lem_bahadur_grid},
\[
    		\widehat Q_g(\tau_k)-Q_g(\tau_k)
    	=
    		\frac{\tau_k-F_{n,g}(Q_g(\tau_k))}{f_g(Q_g(\tau_k))}
    		+
    		R_{g,N}(\tau_k),
\]
uniformly over $k=1,\dots,K_N$, where
$
    		\sup_{1\le k\le K_N}|R_{g,N}(\tau_k)|
    	=
    		o_p(n_g^{-1/2}).
$
Hence
\[
    		\hat m_g-m_{g,N}^*
    	=
   		 \frac{1}{n_g}\sum_{i\in\mathcal I_g}\psi_{g,N}(Y_i)
    		+
    		o_p(n_g^{-1/2}),
\]
where
\[
    		\psi_{g,N}(y)
    	=
    		-\sum_{k=1}^{K_N}\alpha_{k,N}
    		\frac{
        			\mathbf 1\{y\le Q_g(\tau_k)\}-\tau_k
    		}{
        			f_g(Q_g(\tau_k))
   	 	}.
\]

\medskip
\noindent
\textbf{Step 3: expansion for }$\hat{\bm{\beta}}$.
Combining Steps 1 and 2,
\[
    		\hat m_g-m_g
    	=
    		b_{g,N}
    		+
    		\frac{1}{n_g}\sum_{i\in \mathcal{I}_g}\psi_{g,N}(Y_i)
    		+
    		o_p(n_g^{-1/2}).
\]

From the proof of Theorem~\ref{thm_main},
\[
    		\hat{\bm{\beta}}-\bm{\beta}_0
   	 =
    		A_N
    		\frac{1}{N}\sum_{g=1}^G n_g\bm X_g
    		(\hat m_g-m_g).
\]
where
$    
		A_N
	=
    		(
        			\frac{1}{N}\sum_{g=1}^G n_g\bm X_g\bm X_g'
    		)^{-1}.
$
Substituting the expansion above yields
\begin{align*}
    		\hat{\bm{\beta}}-\bm{\beta}_0
    	&=
    		A_N
    		\frac{1}{N}\sum_{g=1}^G n_g\bm X_g b_{g,N}
    	+
    		A_N
    		\frac{1}{N}\sum_{g=1}^G \sum_{i\in \mathcal{I}_g}\bm X_g\psi_{g,N}(Y_i)
    	+
    		o_p(N^{-1/2}).
\end{align*}
The final remainder term is of order $o_p(N^{-1/2})$ because the group-specific remainders are $o_p(n_g^{-1/2})$, Assumption~\ref{ass_design} implies $n_g/N \to \pi_g \in (0,1)$ for each fixed $g$, and $G$ is fixed. Since $\|\bm X_g\|<\infty$ and
$
		A_N
	=
		O_{p}(1)
$,
the aggregation of the group-level remainder terms remains $o_p(N^{-1/2})$.

Define
$
    		S_N
    	=
    		\frac{1}{N}\sum_{g=1}^G n_g\bm X_g b_{g,N}
    		+
    		\frac{1}{N}\sum_{g=1}^G \sum_{i\in \mathcal{I}_g}\bm X_g\psi_{g,N}(Y_i).
$
Then
\[
    		\hat{\bm{\beta}}-\bm{\beta}_0
    	=
    		A_N S_N
    		+
    		o_p(N^{-1/2}).
\]

We show that
$
    S_N = O_p(N^{-1/2}).
$
First, since 
$b_{g,N}=O(\Delta(\mathcal T_{K_N})^2)$, 
$\|\bm X_g\|<\infty$, 
$n_g/N \xrightarrow{p} \pi_g\in(0,1)$ for each fixed $g$, and 
$G$ is fixed, 
we have
\[
 		\frac{1}{N}\sum_{g=1}^G n_g\bm X_g b_{g,N}
    	=
    		O_p\!\left(\Delta(\mathcal T_{K_N})^2\right).
\]

Second, 
by the definition of the conditional distribution function $F_g$,
\[
		E\!\left[
   			 \mathbf{1}\{Y_i\le Q_g(\tau_k)\}
    			\,\middle|\,
    			\bm X_i\in g
		\right]
	=
		P\!\left(
			    Y_i\le Q_g(\tau_k)
			    \,\middle|\,
			    \bm X_i\in g
		\right)
	=
		F_g(Q_g(\tau_k))
	=
		\tau_k.
\]
Hence,
$
    E[\psi_{g,N}(Y_i)\mid\bm X_i\in g]=0.
$
Moreover,
\[
		|\psi_{g,N}(y)|
	\le
		\sum_{k=1}^{K_N}\alpha_{k,N}\frac{1}{f_g(Q_g(\tau_k))}
	\le 
		c_g^{-1},
\]
and therefore
\[
		E[
		    \psi_{g,N}(Y_i)^2
		    \mid
		    \bm X_i\in g
		]
	\le
		c_g^{-2}
	<
		\infty.
\]
Consequently, under the conditional distribution of \(Y_i\)
given \(\bm X_i\in g\),
\[
		\Var\left(
		    \frac{1}{n_g}
		    \sum_{i\in \mathcal I_g}
		    \psi_{g,N}(Y_i)
		    \,\middle|\,
		    \mathcal I_g
		\right)
	=
		O(n_g^{-1}),
\]
and hence, by Chebyshev’s inequality,
\[
		\frac{1}{n_g}
		\sum_{i\in \mathcal I_g}
		\psi_{g,N}(Y_i)
	=
		O_p(n_g^{-1/2}).
\]
It follows for each fixed $g$ that
\[
    		\frac{1}{N}\sum_{i\in\mathcal I_g}\bm X_g\psi_{g,N}(Y_i)
    	=
    		\frac{n_g}{N}\bm X_g
    		\left(
        			\frac{1}{n_g}\sum_{i\in\mathcal I_g}\psi_{g,N}(Y_i)
    		\right)
   	 =
   		 O_p(N^{-1/2}).
\]
Since $G$ is fixed,
\[
    		\frac{1}{N}\sum_{g=1}^G\sum_{i\in\mathcal I_g}\bm X_g\psi_{g,N}(Y_i)
    	=
    		O_p(N^{-1/2}).
\]
Thus, we obtain
$
    		S_N 
	= 
		O_p(N^{-1/2} + \Delta(\mathcal T_{K_N})^2)
	=
		O_p(N^{-1/2}).
$

By Assumption~\ref{ass_design},
$
    A_N
    =
    \bm\Sigma_X^{-1}
    +
    o_p(1).
$
Along with 
$
    S_N = O_p(N^{-1/2}),
$
we obtain,
\[
    		(A_N-\bm\Sigma_X^{-1})S_N
    	=
    		o_p(1)\,O_p(N^{-1/2})
    	=
    		o_p(N^{-1/2}),
\]
and thus
\begin{align*}
    		\hat{\bm{\beta}}-\bm{\beta}_0
    	&=
    		\bm\Sigma_X^{-1}S_N
    		+
    		o_p(N^{-1/2})
		\\
	&=
		\bm\Sigma_X^{-1}
    		\frac{1}{N}\sum_{g=1}^G n_g\bm X_g b_{g,N}
    		+
		\bm\Sigma_X^{-1}
    		\frac{1}{N}\sum_{g=1}^G \sum_{i\in \mathcal{I}_g}\bm X_g\psi_{g,N}(Y_i)
    		+
    		o_p(N^{-1/2}).
\end{align*}

Now write
\[
    		\bm\Sigma_X^{-1}
    		\frac{1}{N}\sum_{g=1}^G n_g\bm X_g b_{g,N}
    	=
    		\bm\Sigma_X^{-1}
    		\sum_{g=1}^G \pi_g \bm X_g b_{g,N}
    		+
    		\bm\Sigma_X^{-1}
    		\sum_{g=1}^G
    		\left(
        			\frac{n_g}{N}-\pi_g
    		\right)\bm X_g b_{g,N}.
\]
Since
$
    \frac{n_g}{N}-\pi_g=o_p(1)
$
for each fixed $g$ by Assumption~\ref{ass_design}, and 
$b_{g,N}=O(\Delta(\mathcal T_{K_N})^2)$, 
$\|\bm X_g\|<\infty$, and $G$ is fixed, 
the second term on the right-hand side is
$
    o_p(\Delta(\mathcal T_{K_N})^2).
$
Under the conditions of Proposition~\ref{prop_rootN},
$
    \Delta(\mathcal T_{K_N})^2 = O(N^{-1/2}),
$
and therefore this term is
$
    o_p(N^{-1/2}).
$
Therefore, defining
\[
    		\bm B_N
    	=
   		 \bm\Sigma_X^{-1}
    		\sum_{g=1}^G \pi_g \bm X_g b_{g,N},
\]
we obtain
\[
    		\bm\Sigma_X^{-1}
    		\frac{1}{N}\sum_{g=1}^G n_g\bm X_g b_{g,N}
    	=
   		 \bm B_N
    		+
    		o_p(N^{-1/2}).
\]
Moreover, $\bm B_N = O(\Delta(\mathcal T_{K_N})^2)$.

Finally, define
\[
    		\bm\Psi_{g,N}(y)
    	=
    		\bm\Sigma_X^{-1}\bm X_g\psi_{g,N}(y).
\]
Then
\[
    		\hat{\bm{\beta}}-\bm{\beta}_0
    	=
    		\bm B_N
    		+
    		\frac{1}{N}\sum_{g=1}^G
		\sum_{i\in\mathcal I_g}
		\bm\Psi_{g,N}(Y_i)
    		+
    		o_p(N^{-1/2}),
\]
where
$
		E[\bm{\Psi}_{g, N}(Y_i) \mid \bm X_i\in g]
	=
		0
$
since
$
		E\!\left[
   			 \mathbf{1}\{Y_i\le Q_g(\tau_k)\}
    			\,\middle|\,
    			\bm X_i\in g
		\right]
	=
		\tau_k
$ as shown above. 
Moreover, 
$
		|\psi_{g,N}(y)|
	\le
		c_g^{-1}
$
and Assumption \ref{ass_design}
lead to the uniform boundedness of $\bm\Psi_{g,N}(Y_i)$.
\end{proof}

\subsection{Proof of Theorem \ref{thm_bias_clt}}
\begin{proof}
By Lemma~\ref{lem_expansion},
\[
    		\hat{\bm{\beta}}-\bm{\beta}_0-\bm B_N
    	=
    		\frac{1}{N}
   		 \sum_{g=1}^G
    		\sum_{i\in\mathcal I_g}
    		\bm\Psi_{g,N}(Y_i)
    	+
    		o_p(N^{-1/2}),
\]
where
$
    		\bm\Psi_{g,N}(y)
    	=
    		\bm\Sigma_X^{-1}\bm X_g\psi_{g,N}(y)
$
with 
$
    		\psi_{g,N}(y)
    	=
    		-\sum_{k=1}^{K_N}\alpha_{k,N}
    		\frac{
        			\mathbf 1\{y\le Q_g(\tau_k)\}-\tau_k
    		}{
        			f_g(Q_g(\tau_k))
   	 	}.
$
Multiplying both sides by $\sqrt N$ gives
\[
    		\sqrt N
    		\bigl(
        			\hat{\bm{\beta}}-\bm{\beta}_0-\bm B_N
   		 \bigr)
    	=
    		\frac{1}{\sqrt N}
   		 \sum_{g=1}^G
    		\sum_{i\in\mathcal I_g}
    		\bm\Psi_{g,N}(Y_i)
    	+
   		 o_p(1).
\]

Define
$
    		\bm Z_{i,N}
    	=
    		\sum_{g=1}^G
    		\mathbf 1\{\bm X_i\in g\}
    		\bm\Psi_{g,N}(Y_i).
$
Then
\[
    		\sum_{g=1}^G
    		\sum_{i\in\mathcal I_g}
    		\bm\Psi_{g,N}(Y_i)
    	=
    		\sum_{i=1}^N
   		 \bm Z_{i,N}.
\]

\medskip
\noindent
\textbf{Step 1: zero expectation.}
For each $g$ and $k$,
$
    		E\!\left[
			\mathbf 1\{Y_i\le Q_g(\tau_k)\}
        			\mid
       			 \bm X_i\in g
    		\right]
    	=
    		F_g(Q_g(\tau_k))
    	=
    		\tau_k.
$
Hence,
$
   		 E[
        			\psi_{g,N}(Y_i)
        			\mid
       			 \bm X_i\in g
  		]
    	=
    		0,
$
and therefore
$
    		E[
        			\bm\Psi_{g,N}(Y_i)
        			\mid
      			\bm X_i\in g
   		 ]
   	 =
    		\bm 0.
$
It follows that
\[
    		E(\bm Z_{i,N})
   	 =
    		\sum_{g=1}^G
    		P(\bm X_i\in g)
    		E[
        			\bm\Psi_{g,N}(Y_i)
        			\mid
        			\bm X_i\in g
    		]
   	 =
    		\bm 0.
\]

\medskip
\noindent
\textbf{Step 2: convergence of variance.}
For each fixed $y$, define
\[
    		h_{g,y}(\tau)
    	=
    		\frac{
       			 \mathbf 1\{y\le Q_g(\tau)\}-\tau
   		 }{
       			 f_g(Q_g(\tau))
    		}
\]
and
\[
    		\psi_g(y)
    	=
   		 -
    		\int_0^1
    		\frac{
       		 \mathbf 1\{y\le Q_g(\tau)\}-\tau
   		 }{
        		f_g(Q_g(\tau))
   		 }
   	 d\tau,
\]
where for \(y\in[\underline y_g,\overline y_g]\), a change of variables
\(x=Q_g(\tau)\) yields
\begin{align*}
		\psi_g(y)
	&=
		-\int_{\underline y_g}^{\overline y_g}
		\left\{
			1(y\le x)-F_g(x)
		\right\}
		\,dx 
	=
		-\left\{
			(\overline y_g-y)
			-
			(\overline y_g-m_g)
		\right\} 
	=
		y-m_g.
\end{align*}
Then
\[
    		\psi_g(y)
    	=
    		-
    		\int_0^1 h_{g,y}(\tau)\,d\tau
\]
and
\[
    		\psi_{g,N}(y)
    	=
    		-
    		\sum_{k=1}^{K_N}\alpha_{k,N}h_{g,y}(\tau_k).
\]

The weights $\alpha_{k,N}$ are the quadrature weights induced by the piecewise linear reconstruction and satisfy
$
    \alpha_{k,N}\ge0,
$
and
$
    \sum_{k=1}^{K_N}\alpha_{k,N}=1.
$
Moreover, $h_{g,y}$ is bounded because $f_g$ is bounded away from zero by condition (iii) of Lemma~\ref{lem_uniform_quantile}, and the numerator 
$\mathbf 1\{y\le Q_g(\tau)\}-\tau$ is bounded.
The only possible discontinuity arises from the indicator function 
$\mathbf 1\{y\le Q_g(\tau)\}$, which occurs at a value of $\tau$ satisfying 
$Q_g(\tau)=y$.
Hence $h_{g,y}$ has at most one point of discontinuity, and is therefore Riemann integrable.

Since $\Delta(\mathcal T_{K_N})\to0$, it follows that
$
    \sum_{k=1}^{K_N}\alpha_{k,N}h_{g,y}(\tau_k)
    \to
    \int_0^1 h_{g,y}(\tau)\,d\tau
$,
and hence
\[
    		\psi_{g,N}(y)
   	 \to
    		\psi_g(y)
\]
for each fixed $y$, and
\[
    		\bm\Psi_{g,N}(y)
    	=
     		 \bm\Sigma_X^{-1}\bm X_g\psi_{g,N}(y)
    		\to
    		\bm\Sigma_X^{-1}\bm X_g\psi_g(y)
    	=
    		\bm\Psi_g(y).
\]

Moreover, since
$
   		 |\psi_{g,N}(y)|
    	\le
    		c_g^{-1}
$, 
$
    		|\psi_g(y)|
    	\le
    		c_g^{-1}
$,
$\|\bm X_g\|<\infty$,
$ \| \bm\Sigma_X^{-1}\| < \infty$,
 and $G$ is fixed, there exists a finite constant $C<\infty$
such that
$
    		\|\bm\Psi_{g,N}(y)\|
    	\le 
		C,
$
and
$
    		\|\bm\Psi_g(y)\|
    	\le 
		C
$
for all $g$, $N$, and $y$.
Therefore, by the dominated convergence theorem, 
\[
    		E\!\left[
        		\|
           		 \bm\Psi_{g,N}(Y_i)
            		-
            		\bm\Psi_g(Y_i)
        		\|^2
       		 	\mid
       			 \bm X_i\in g
    		\right]
    	\to
    		0.
\]

Because the groups form a partition,
$
    \mathbf 1\{\bm X_i\in g\}
    \mathbf 1\{\bm X_i\in h\}
    =
    0
$
for
$g\ne h$.
Therefore,
\begin{align*}
		\|\bm Z_{i,N}-\bm Z_i\|^2
	&=
		\left\|
			    \sum_{g=1}^G
			    \mathbf 1\{\bm X_i\in g\}
			    \{
			        \bm\Psi_{g,N}(Y_i)-\bm\Psi_g(Y_i)
			    \}
		\right\|^2 \\
	&=
		\sum_{g=1}^G
		\mathbf 1\{\bm X_i\in g\}
		\|
		    \bm\Psi_{g,N}(Y_i)-\bm\Psi_g(Y_i)
		\|^2.
\end{align*}
Since $G$ is fixed, we obtain
\begin{align*}
    		E\!\left(
       	 	\|
            		\bm Z_{i,N}
            		-
            		\bm Z_i
        		\|^2
   		 \right)
	&=
		\sum_{g=1}^G
		\pi_g
		E\left[
		    \|
		        \bm\Psi_{g,N}(Y_i)-\bm\Psi_g(Y_i)
		    \|^2
		    \mid
		    \bm X_i\in g
		\right] 
	\longrightarrow 
		0,
\end{align*}
where
$
    		\bm Z_i
    	=
    		\sum_{g=1}^G
    		\mathbf 1\{\bm X_i\in g\}
   		 \bm\Psi_g(Y_i).
$
This $L^2$ convergence and $E(\bm Z_{i,N})=\bm 0$ for all $N$
lead to $E(\bm Z_i)=\bm 0$.
Moreover, the same $L^2$ convergence implies
$
    		E(\bm Z_{i,N}\bm Z_{i,N}')
    	\to
    		E(\bm Z_i\bm Z_i').
$
Therefore,
\[
    		\Var(\bm Z_{i,N})
    	=
    		E(\bm Z_{i,N}\bm Z_{i,N}')
    	\to
    		E(\bm Z_i\bm Z_i')
    	=
    		\Var(\bm Z_i)
	=
		\bm V_\beta.
\]

\medskip
\noindent
\textbf{Step 3: central limit theorem.}
We now verify the Lindeberg condition.
Since $\bm Z_{i,N}$ is uniformly bounded, for any $\eta>0$,
\[
   		 E\!\left(
       			 \|\bm Z_{i,N}\|^2
        			\mathbf 1
       			 \{
            			\|\bm Z_{i,N}\|>\eta\sqrt N
        			\}
    		\right)
    	\to
    		0.
\]
Thus, the Lindeberg condition holds.

Since $(Y_i,\bm X_i)$ are i.i.d., the triangular array
$\{\bm Z_{i,N}:i=1,\dots,N\}$ consists of independent and identically distributed
mean-zero random vectors for each $N$
and 
$
    		\Var (
        			\frac{1}{\sqrt N}\sum_{i=1}^N \bm Z_{i,N}
   		 )
    	=
    		\Var(\bm Z_{i,N})
    	\to
   		\bm V_\beta
$.
Therefore, by the multivariate Lindeberg--Feller central limit theorem (see for example Proposition 2.27 of \citealp{vandervaart98}),
\[
    		\frac{1}{\sqrt N}
    		\sum_{i=1}^N
    		\bm Z_{i,N}
    \overset{d}{\longrightarrow}
    		\mathcal N(\bm 0, \bm V_\beta).
\]
Combining this result with
\[
    		\sqrt N
   		 \bigl(
       			 \hat{\bm{\beta}}-\bm{\beta}_0-\bm B_N
    		\bigr)
    	=
    		\frac{1}{\sqrt N}
    		\sum_{i=1}^N
    		\bm Z_{i,N}
    	+
    		o_p(1),
\]
and applying Slutsky's theorem yields
\[
    		\sqrt N
    		\bigl(
       			 \hat{\bm{\beta}}-\bm{\beta}_0-\bm B_N
    		\bigr)
    \overset{d}{\longrightarrow}
    		\mathcal N(\bm 0, \bm V_\beta).
\]
This proves the theorem.
\end{proof}

\subsection{Proof of Corollary \ref{cor_centered_clt}}
\begin{proof}
The convergence in distribution follows from Theorem~\ref{thm_bias_clt} together with Slutsky's theorem.

For an evenly spaced grid,
$
    \Delta(\mathcal T_{K_N})
    =
    \frac{1}{K_N+1}.
$
Thus
$
    \Delta(\mathcal T_{K_N})^2=o(N^{-1/2})
$
is equivalent to
$
    K_N^{-2}=o(N^{-1/2}),
$
or
$
    N^{1/4}=o(K_N).
$
\end{proof}

\subsection{Proof of Theorem~\ref{thm_nonidentification}}
\begin{proof}
Since the disclosure grid is finite, there exists a nonempty open interval \(I\subset(0,1)\) containing no disclosed quantile. 
Choose a nonzero twice continuously differentiable function \(b:[0,1]\to\mathbb R\) whose support is contained in \(I\) and such that
$
	 \int_0^1 b(\tau)\,d\tau \neq 0
$
and
$
    b(\tau_k)=0
$
for
$
    k=1,\ldots,K.
$

Fix any vector $\bm\beta_0^{(1)}$ and any constant $s>0$.
For each group $g$, define
\[
    Q_g^{(1)}(\tau)
    =
    \bm X_g'\bm\beta_0^{(1)}
    +
    s\left(\tau-\frac12\right),
    \qquad
    \tau\in[0,1].
\]
Then $Q_g^{(1)}$ is strictly increasing and twice continuously differentiable with the corresponding conditional mean $m_g^{(1)}$ satisfying
\[
		m_g^{(1)}
	=
	    \int_0^1
	    Q_g^{(1)}(\tau)\,d\tau
    	=
	    \bm X_g'\bm\beta_0^{(1)}.
\]

Take any nonzero vector $\bm d_0$ and let
$
    \bm d
    =
    \varepsilon\bm d_0,
$
where $\varepsilon>0$ will be chosen sufficiently small.
Since $\bm\Sigma_X$ is positive definite,
$
    \bm d_0'\bm\Sigma_X\bm d_0
    =
    \sum_{g=1}^G
    \pi_g
    \left(
        \bm X_g'\bm d_0
    \right)^2
    >
    0.
$
Hence
$
    \bm X_g'\bm d_0\neq0
$
for at least one group $g$.
Define
$
    \bm\beta_0^{(2)}
    =
    \bm\beta_0^{(1)}
    +
    \bm d
$
and, for each group $g$,
$
    \lambda_g
    =
    \frac{\bm X_g'\bm d}{\int_0^1 b(\tau)\,d\tau}.
$
Now define
\[
    Q_g^{(2)}(\tau)
    =
    Q_g^{(1)}(\tau)
    +
    \lambda_g b(\tau).
\]

Because $G$ is fixed,
$\|b'\|_\infty = \sup_{\tau \in[0,1]} |b'(\tau)|<\infty$, and 
$\int_0^1 b(\tau)\,d\tau \neq0$, 
we can choose
$\varepsilon>0$ sufficiently small that
$
    \max_{1\le g\le G}
    |\lambda_g|
    \|b'\|_\infty
    <
    \frac{s}{2}.
$
It then follows that
$
    \frac{d}{d\tau}
    Q_g^{(2)}(\tau)
    =
    s+\lambda_g b'(\tau)
    \ge
    \frac{s}{2}
    >
    0
$
for every $g$ and every $\tau\in[0,1]$.
Thus $Q_g^{(2)}$ is also strictly increasing.
Moreover,
\[
    \sup_{\tau\in(0,1)}
    \left|
        \frac{d^2}{d\tau^2}
        Q_g^{(2)}(\tau)
    \right|
    =
    |\lambda_g|
    \|b''\|_\infty
    <
    \infty.
\]
Hence both $Q_g^{(1)}$ and $Q_g^{(2)}$ satisfy
Assumption~\ref{ass_smooth}(ii).

Because
$b(\tau_k)=0$ for every disclosed quantile index,
\[
    Q_g^{(2)}(\tau_k)
    =
    Q_g^{(1)}(\tau_k),
    \qquad
    g=1,\ldots,G,
    \quad
    k=1,\ldots,K.
\]
Thus the two collections of conditional distributions generate exactly the same disclosed population quantiles.

On the other hand,
\begin{align*}
	m_g^{(2)}
	=
    \int_0^1 Q_g^{(2)}(\tau)\,d\tau
    &=
    \int_0^1 Q_g^{(1)}(\tau)\,d\tau
    +
    \lambda_g
    \int_0^1 b(\tau)\,d\tau
    \\
    &=
    \bm X_g'\bm\beta_0^{(1)}
    +
    \bm X_g'\bm d
    \\
    &=
    \bm X_g'\bm\beta_0^{(2)}.
\end{align*}
Therefore,
\[
    m_g^{(\ell)}
    =
    \bm X_g'\bm\beta_0^{(\ell)},
    \qquad
    \ell=1,2,
\]
where
$
    \bm\beta_0^{(1)}
    \neq
    \bm\beta_0^{(2)}
$
since $\bm d\neq\bm 0$.

Finally, retain the same marginal distribution of $\bm X$ in the two constructions. For $\ell=1,2$ and $\bm x\in\mathcal G_g$, define the conditional distribution of $Y$ given $\bm X=\bm x$ to be the distribution whose quantile function is $Q_g^{(\ell)}$. Because the groups form a partition, this assignment is well defined and induces a valid joint distribution of $(Y,\bm X)$ for each $\ell=1,2$.

Moreover, the two joint distributions are distinct. As shown above, there exists at least one group $g$ such that $\bm X_g'\bm d_0\neq0$, and hence $\lambda_g\neq0$. Since $b$ is not zero, it follows that
$
    Q_g^{(2)}
    \neq
    Q_g^{(1)}
$
for that group.

The two joint distributions have the same marginal distribution of $\bm X$ and generate the same disclosed conditional quantiles, but they imply
$
    \bm\beta_0^{(1)}
    \neq
    \bm\beta_0^{(2)}.
$
Hence, $\bm\beta_0$ is not point identified.
\end{proof}

\subsection{Proof of Proposition~\ref{prop_var}}

Let $E_g(\cdot)$ and $\Var_g (\cdot)$ denote expectation and variance under the conditional distribution of $Y_i$ given $\bm X_i \in g$.

\begin{proof}[Proof of Proposition~\ref{prop_var}]
Let
\[
		\widehat{\bm\Sigma}_X
	=
		\frac{1}{N}
		\sum_{g=1}^G n_g \bm X_g\bm X_g'.
\]
Since $\bm\Sigma_X$ is positive definite by Assumption~\ref{ass_design},
and $\widehat{\bm\Sigma}_X \xrightarrow{p} \bm\Sigma_X$,
it follows that $\widehat{\bm\Sigma}_X$ is nonsingular with probability approaching one,
and hence $\widehat{\bm\Sigma}_X^{-1} \xrightarrow{p} \bm\Sigma_X^{-1}$.

Rewrite
\begin{align*}
		N\widehat{\bm V}_\beta
	&=
		\widehat{\bm\Sigma}_X^{-1}
		\left[
			\sum_{g=1}^G
			\frac{n_g}{N}
			\bm X_g\bm X_g'
			\left\{
				n_g\widehat{\Var}(\hat m_g)
			\right\}
		\right]
		\widehat{\bm\Sigma}_X^{-1}.
\end{align*}

By Lemma~\ref{lem_group_var},
$
		n_g\widehat{\Var}(\hat m_g)
	\xrightarrow{p}
		\Var_g(\psi_g(Y_i)).
$
Hence, by Assumption~\ref{ass_design},
\[
		\sum_{g=1}^G
		\frac{n_g}{N}
		\bm X_g\bm X_g'
		\left\{
			n_g\widehat{\Var}(\hat m_g)
		\right\}
	\xrightarrow{p}
		\sum_{g=1}^G
		\pi_g
		\bm X_g\bm X_g'
		\Var_g(\psi_g(Y_i)).
\]

Finally, because the groups form a partition,
$
    \mathbf 1\{\bm X_i\in g\}
    \mathbf 1\{\bm X_i\in h\}
    =0
$
for
$g\ne h.$
Hence all cross-group terms vanish. Since
\(E_g[\psi_g(Y_i)]=0\),
\begin{align*}
		\bm V_\beta
	&=
		\Var\!\left(
			\sum_{g=1}^G
			\mathbf 1\{\bm X_i\in \mathcal G_g\}
			\bm\Sigma_X^{-1}\bm X_g\psi_g(Y_i)
		\right)
		\\
	&=
		E\left[
			\sum_{g=1}^G
			\mathbf 1\{\bm X_i\in \mathcal G_g\}
			\bm\Sigma_X^{-1}\bm X_g\bm X_g'
			\bm\Sigma_X^{-1}
			\psi_g(Y_i)^2
		\right]
		\\
	&=
		\bm\Sigma_X^{-1}
		\left[
			\sum_{g=1}^G
			\pi_g
			\bm X_g\bm X_g'
			\Var_g(\psi_g(Y_i))
		\right]
		\bm\Sigma_X^{-1},
\end{align*}
where $\psi_g(Y_i) = Y_{i} - m_{g}$ as shown in the proof of Theorem \ref{thm_bias_clt},
so that
\[
		N\widehat{\bm V}_\beta
	\xrightarrow{p}
		\bm V_\beta.
\]
\end{proof}

\subsection{Proof of Theorem~\ref{thm_mean_bounds}}

\begin{proof}
Fix a group $g$. We show 
(1) $\mathcal M_g(\mathcal T_K)
    =
    \left\{
        \int_{\mathbb R} y\,dF(y):
        F\in\mathcal F_g(\mathcal T_K)
    \right\}$,
(2) 
$\mathcal M_g(\mathcal T_K)\subseteq
[\underline m_g,\overline m_g]$,
and
(3)
the bounds are sharp.

\emph{(1)}:
We first show that
$
    \mathcal M_g(\mathcal T_K)
    =
    \{
        \int y\,dF(y):F\in\mathcal F_g(\mathcal T_K)
    \}.
$

Take any $Q\in\mathcal Q_g(\mathcal T_K)$, let
$V\sim\operatorname{Uniform}(0,1)$, and define
$Y_Q=Q(V)$ and $F_Q(y)=P(Y_Q\le y)$.
Since $Q$ is nondecreasing and has endpoint values
$\underline y_g$ and $\overline y_g$, it is measurable and
bounded between these values.
For every $\tau\in(0,1)$, monotonicity gives
\[
    \{Q(V)<Q(\tau)\}\subseteq\{V<\tau\},
    \qquad
    \{V\le\tau\}\subseteq\{Q(V)\le Q(\tau)\}.
\]
Consequently, using 
$
F_Q(y-)=P\{Q(V)<y\}
$ and
since $V\sim\operatorname{Uniform}(0,1)$,
\[
    F_Q(Q(\tau)-)\le\tau\le F_Q(Q(\tau)).
\]
Because $Q(\tau_k)=Q_g(\tau_k)$, the distribution function
$F_Q$ belongs to $\mathcal F_g(\mathcal T_K)$, and
\[
    \int y\,dF_Q(y)
    =E[Q(V)]
    =\int_0^1Q(u)\,du.
\]

Conversely, take any $F\in\mathcal F_g(\mathcal T_K)$ and define
\[
    H_F(u)=\inf\{y\in\mathbb R:F(y)\ge u\},
    \qquad 0<u<1.
\]
The support restriction implies
$\underline y_g\le H_F(u)\le\overline y_g$.
For $u<\tau_k$, the inequality
$F(Q_g(\tau_k))\ge\tau_k>u$ implies
$H_F(u)\le Q_g(\tau_k)$.
For $u>\tau_k$, every $y<Q_g(\tau_k)$ satisfies
$F(y)\le F(Q_g(\tau_k)-)\le\tau_k<u$,
so $H_F(u)\ge Q_g(\tau_k)$.
Define $Q^F:[0,1]\to\mathbb R$ by
\[
    Q^F(u)=
    \begin{cases}
        Q_g(\tau_k),
        &u=\tau_k,\quad k=0,\ldots,K+1,\\
        H_F(u),
        &u\in(0,1)\setminus\mathcal T_K.
    \end{cases}
\]
The preceding inequalities, together with the ordering of
$Q_g(\tau_0),\ldots,Q_g(\tau_{K+1})$, show that $Q^F$
is nondecreasing.
It matches all prescribed node values, and hence
$Q^F\in\mathcal Q_g(\mathcal T_K)$.
The definition of $H_F$ and right-continuity of $F$ give
\[
    P\{H_F(V)\le y\}
    =P\{V\le F(y)\}
    =F(y),
    \qquad y\in\mathbb R.
\]
Since $Q^F$ and $H_F$ differ at only finitely many points
of $(0,1)$, $Q^F(V)=H_F(V)$ almost surely.
Thus $Q^F(V)$ has distribution function $F$ and
\[
    \int_0^1Q^F(u)\,du=\int y\,dF(y).
\]
This establishes the claimed equality of the two mean sets.

\emph{(2)}:
Second, we show 
$\mathcal M_g(\mathcal T_K)\subseteq
[\underline m_g,\overline m_g]$.
For any $Q\in\mathcal Q_g(\mathcal T_K)$, monotonicity implies
\[
    Q_g(\tau_k)\le Q(\tau)\le Q_g(\tau_{k+1}),
    \qquad
    \tau\in[\tau_k,\tau_{k+1}],
    \quad k=0,\ldots,K.
\]
Integrating over these intervals and summing yields
\begin{align*}
    \underline m_g
    &=\sum_{k=0}^{K}(\tau_{k+1}-\tau_k)Q_g(\tau_k)\\
    &\le\int_0^1Q(\tau)\,d\tau\\
    &\le\sum_{k=0}^{K}(\tau_{k+1}-\tau_k)Q_g(\tau_{k+1})
    =\overline m_g.
\end{align*}
Therefore
$\mathcal M_g(\mathcal T_K)\subseteq
[\underline m_g,\overline m_g]$.

\emph{(3)}:
Finally, we show that the bounds are sharp.
Define
\[
    Q_g^L(\tau)=Q_g(\tau_k),
    \qquad
    \tau\in[\tau_k,\tau_{k+1}),
    \quad k=0,\ldots,K,
\]
with $Q_g^L(\tau_{K+1})=Q_g(\tau_{K+1})$, and
\[
    Q_g^U(\tau)=Q_g(\tau_{k+1}),
    \qquad
    \tau\in(\tau_k,\tau_{k+1}],
    \quad k=0,\ldots,K,
\]
with $Q_g^U(\tau_0)=Q_g(\tau_0)$.
Both functions are nondecreasing and satisfy
$Q_g^L(\tau_k)=Q_g^U(\tau_k)=Q_g(\tau_k)$ for
$k=0,\ldots,K+1$.
Hence $Q_g^L,Q_g^U\in\mathcal Q_g(\mathcal T_K)$, and
\[
    \int_0^1Q_g^L(\tau)\,d\tau=\underline m_g,
    \qquad
    \int_0^1Q_g^U(\tau)\,d\tau=\overline m_g.
\]

If $\underline m_g=\overline m_g$, the only value in this
interval is already attained by $Q_g^L$.
Otherwise, for any $m\in[\underline m_g,\overline m_g]$, set
\[
    \lambda=
    \frac{m-\underline m_g}{\overline m_g-\underline m_g}
    \in[0,1],
    \qquad
    Q_g^\lambda=(1-\lambda)Q_g^L+\lambda Q_g^U.
\]
The function $Q_g^\lambda$ is nondecreasing and matches every
prescribed node value, so
$Q_g^\lambda\in\mathcal Q_g(\mathcal T_K)$.
Moreover,
\[
    \int_0^1Q_g^\lambda(\tau)\,d\tau
    =(1-\lambda)\underline m_g+\lambda\overline m_g
    =m.
\]
By (1), each such function generates a distribution in
$\mathcal F_g(\mathcal T_K)$ with mean $m$.
Thus every value in $[\underline m_g,\overline m_g]$, including
both endpoints, is attainable under the quantile probability
restrictions and support bounds.
It follows that
\[
    \mathcal M_g(\mathcal T_K)
    =[\underline m_g,\overline m_g],
\]
and the bounds are sharp.
\end{proof}

\subsection{Proof of Proposition~\ref{prop_beta}}

\begin{proof}
First, let $\bm\beta\in\Theta$ be compatible with the group-mean
model, the quantile probability restrictions, and the support
bounds in Section~\ref{sec_partial_identification}.
By Theorem~\ref{thm_mean_bounds}, the corresponding group means
satisfy
\[
    \underline m_g\le\bm X_g'\bm\beta\le\overline m_g,
    \qquad g=1,\ldots,G.
\]
Hence every compatible coefficient vector belongs to
$\mathcal B(\mathcal T_K)$.

Conversely, take any $\bm\beta\in\mathcal B(\mathcal T_K)$.
For each group $g$, Theorem~\ref{thm_mean_bounds} provides a
function $Q_g^{\bm\beta}\in\mathcal Q_g(\mathcal T_K)$ such that
\[
    \int_0^1Q_g^{\bm\beta}(u)\,du=\bm X_g'\bm\beta.
\]
Let $F_g^{\bm\beta}$ denote the distribution function of
$Q_g^{\bm\beta}(V)$ for
$V\sim\operatorname{Uniform}(0,1)$.
By (1) of the proof of Theorem~\ref{thm_mean_bounds},
this distribution has support contained in
$[\underline y_g,\overline y_g]$ and satisfies
\[
    F_g^{\bm\beta}(Q_g(\tau_k)-)
    \le\tau_k
    \le F_g^{\bm\beta}(Q_g(\tau_k)),
    \qquad k=1,\ldots,K.
\]

Retain the given marginal distribution of $\bm X$ and take
$V\sim\operatorname{Uniform}(0,1)$ independently of $\bm X$.
Define
\[
    Y^{\bm\beta}
    =\sum_{g=1}^{G}
      \mathbf 1\{\bm X\in\mathcal G_g\}
      Q_g^{\bm\beta}(V).
\]
Because the groups form a partition, exactly one term is active almost surely.
Conditional on $\bm X\in\mathcal G_g$, the variable
$Y^{\bm\beta}$ has distribution function $F_g^{\bm\beta}$ and
\[
    E(Y^{\bm\beta}\mid\bm X\in\mathcal G_g)
    =\int_0^1Q_g^{\bm\beta}(u)\,du
    =\bm X_g'\bm\beta.
\]
Thus the constructed joint distribution of $(Y^{\bm\beta},\bm X)$ preserves the original
marginal distribution of $\bm X$ and satisfies the group-mean
model, the group-specific support bounds, and all quantile
probability restrictions.
Every $\bm\beta\in\mathcal B(\mathcal T_K)$ is therefore attainable.
Together with the necessity argument, this proves that
$\mathcal B(\mathcal T_K)$ is the sharp identified set under these restrictions.

Finally, $\mathcal B(\mathcal T_K)$ is the intersection of
$\Theta$ and finitely many closed half-spaces.
It is a closed subset of the compact set $\Theta$ and is
therefore compact.
It is convex because $\Theta$ and all of these half-spaces
are convex.
\end{proof}

\clearpage
\def\thesection{}
\renewcommand{\thesubsection}{B-\arabic{subsection}} 
 \def\theequation{B.\arabic{equation}}
 \def\theassumption{B.\arabic{assumption}}	
 \makeatletter
 \@addtoreset{equation}{section}
 \@addtoreset{figure}{section}
 \@addtoreset{table}{section}
 \@addtoreset{assumption}{section}
 \makeatother
 \def\thelemma{B.\arabic{lemma}}	  

\setcounter{equation}{0} 
\setcounter{subsection}{0} 
\setcounter{lemma}{0}
\setcounter{page}{1}
\section{Supplemental Material B: Technical Lemmas} \label{appendix_technical}
This supplement collects technical lemmas.
Lemmas \ref{lem_vw_local}--\ref{lem_bahadur_grid} are used to show Lemma \ref{lem_expansion}, while Lemmas \ref{lem_derivative} and \ref{lem_group_var} are for Proposition \ref{prop_var}.

Let 
$
		P_g(A) 
	= 
		P(Y_i \in A \mid \bm{X}_i \in g)
$ 
denote the conditional distribution of $Y$ given $\bm{X}_i \in g$, and let
$
		P_{n,g}(A) 
	= 
		\frac{1}{n_g}
		\sum_{i\in \mathcal{I}_g} 
		\mathbf{1}\{Y_i \in A\}
$
denote the corresponding empirical measure,
where 
$\mathcal{I}_g = \{i: \bm X_i \in g\}$ is the index set of observations in group $g$.
Let $\delta \in (0,1)$. 
Define, for a fixed group $g$,
\[
		\mathcal H_\delta
    	=
    		\{(x,y] : P_g((x,y])\le \delta\}.
\]
The class $\mathcal H_\delta$ consists of intervals whose probability mass under $P_g$ is at most $\delta$.
We derive an empirical process bound uniformly over this localized class of sets, which in turn yields a bound on the modulus of continuity of the empirical distribution function.
This result plays a key role in establishing a uniform Bahadur representation over the disclosed grid.

For any measurable function $f$, define
$
    	P_g f = \int f \, dP_g,
$
and
$
    	P_{n,g} f = \int f \, dP_{n,g}.
$
For an interval $I$, let $\mathbf{1}_I(t) = \mathbf{1}\{t \in I\}$ denote its indicator function. 
With this notation,
$
    P_{n,g}(I) = P_{n,g}\mathbf{1}_I,
$
and
$
    P_g(I) = P_g\mathbf{1}_I.
$

We restate  Theorem 2.1 of \cite{vandervaart11} in our notation.
To do this, let
\[
		\mathbb G_{n_g} f
    	=
    		\sqrt{n_g}\,(P_{n,g}-P_g)f
\]
denote the empirical process based on the observations in group $g$.
We denote by $N(\varepsilon, \mathcal G, L_2(Q))$ the $\varepsilon$-covering number of a class $\mathcal G$ with respect to the $L_2(Q)$ metric.

\phantomsection
\addcontentsline{toc}{subsection}{Lemma \ref{lem_vw_local}}
\begin{lemma}[Localized maximal inequality, \citealp{vandervaart11}]
\label{lem_vw_local}
Let $\mathcal G$ be a $P_g$-measurable class of real-valued functions with envelope $G$ satisfying $|f|\le G \le 1$ for all $f\in\mathcal G$.
Suppose that for some $\rho\in(0,1)$,
\[
    P_g f^2 \le \rho^2 P_g G^2
    \quad \text{for all } f\in\mathcal G.
\]
Then
\[
		E \sup_{f\in\mathcal G} |\mathbb G_{n_g} f|
	\lesssim
		J(\rho,\mathcal G,G,L_2)
		\|G\|_{L_2(P_g)}
		\left(
			1 + \frac{J(\rho,\mathcal G,G,L_2)}{\rho^2 \sqrt{n_g}\,\|G\|_{L_2(P_g)}}
		\right),
\]
where
\[
		J(\rho, \mathcal G, G, L_2)
	=
		\int_0^\rho
		\sup_Q
		\sqrt{
			\log N\bigl(
				\varepsilon \|G\|_{L_2(Q)},
				\mathcal G,
				L_2(Q)
			\bigr)
		}
		\, d\varepsilon
\]
denotes the corresponding uniform entropy integral, where the supremum is taken over all finitely discrete probability measures $Q$, and $\lesssim$ means smaller than up to a universal constant.
\end{lemma}

Consequently, since
$
    		(P_{n,g}-P_g)f
    	=
    		\frac{1}{\sqrt{n_g}}\,\mathbb G_{n_g} f,
$		
we obtain
\[
    		E \sup_{f\in\mathcal G}
    		|(P_{n,g}-P_g) f|
    	\lesssim
    		\frac{1}{\sqrt{n_g}}\,
    		J\!\left(
       			 \rho,\mathcal G,G,L_2
   		 \right)\,
    		\|G\|_{L_2(P_g)}
    		\left(
        			1 +
       			 \frac{
           			 J\!\left(\rho,\mathcal G,G,L_2\right)
        			}{
            			\rho^2 \sqrt{n_g}\,\|G\|_{L_2(P_g)}
       			 }
    		\right).
\]

This result is used to show the following Lemma.

\phantomsection
\addcontentsline{toc}{subsection}{Lemma \ref{lem_interval_process}}
\begin{lemma}[Localized empirical process bound for short intervals]
\label{lem_interval_process}
Assume that the conditions of Lemma~\ref{lem_uniform_quantile} are satisfied.
Then, as $\delta\downarrow0$,
\[
    		\sup_{I\in\mathcal H_\delta}
    		\left|
        			(P_{n,g}-P_g)(I)
    		\right|
    	=
    		O_{p}\!\left(
       			 \sqrt{\frac{\delta\log(1/\delta)}{n_g}}
        			+
        			\frac{\log(1/\delta)}{n_g}
    		\right)
\]
\end{lemma}
\begin{proof}
Let
$
		\mathcal F_\delta
	=
		\{ \mathbf{1}_I : I \in \mathcal H_\delta\}.
$
Then
$
    		\sup_{I\in\mathcal H_\delta}
    		\left|
        			(P_{n,g}-P_g)(I)
   		 \right|
    	=
    		\sup_{f\in\mathcal F_\delta}
   		 \left|
        			(P_{n,g}-P_g)f
   		 \right|.
$

For every $f\in\mathcal F_\delta$, we have $f=\mathbf 1_I$ for some interval
$I=(x,y]$ satisfying $P_g(I)\le\delta$. Hence
\[
    P_g f^2 = P_g f = P_g(I)\le\delta.
\]
Moreover, since $f=\mathbf{1}_I$ takes values in $\{0,1\}$, we have
$
	|f(t)| \le 1
$
for all $t$
so that the constant function
$G=1$ is an envelope of $\mathcal F_\delta$.

Since the class of interval indicators has finite VC dimension, 
Theorem 2.6.4 of \citet{vandervaart96} implies that, for some constants $A,v>0$,
\[
    		N\!\left(
			\varepsilon,
        			\mathcal F_\delta,
        			L_2(Q)
    		\right)
    	\le
   	 	\left(
        			\frac{A}{\varepsilon}
    		\right)^v
\]
for all probability measures $Q$ and all $0<\varepsilon\le1$.

Applying Lemma~\ref{lem_vw_local} to the class $\mathcal F_\delta$ with theorem parameter
$
    \rho = \sqrt{\delta}
$
and envelope function $G$ being the constant function $1$, we obtain
\[
    		E \sup_{f\in\mathcal F_\delta}
    		|(P_{n,g}-P_g) f|
    \lesssim
    		\frac{1}{\sqrt{n_g}}\,
    		J\!\left(
        			\sqrt{\delta},\mathcal F_\delta, 1, L_2
   		 \right)\,
    		\left(
        			1 +
       			 \frac{
            			J\!\left(\sqrt{\delta},\mathcal F_\delta, 1, L_2\right)
        			}{
           			 \delta \sqrt{n_g}
        			}
    		\right).
\]
Since the covering number is bounded, 
\[
    		J(\sqrt{\delta}, \mathcal F_\delta, 1, L_2)
    	=
    		\int_0^{\sqrt{\delta}}
    		\sup_Q
    		\sqrt{\log N(\varepsilon,\mathcal F_\delta,L_2(Q))} \, d\varepsilon
    	\le
    		\int_0^{\sqrt{\delta}} \sqrt{v \log(A/\varepsilon)} \, d\varepsilon.		
\]

Using the change of variable $\varepsilon=\sqrt{\delta}\,u$, we obtain
\[
		\int_0^{\sqrt{\delta}} \sqrt{v\log(A/\varepsilon)}\,d\varepsilon
	=
		\sqrt{\delta}\int_0^1
		\sqrt{v\Bigl(\log(A/\sqrt{\delta})+\log(1/u)\Bigr)}\,du.
\]
Since $\sqrt{a+b}\le \sqrt a+\sqrt b$ for $a,b\ge0$ and
$
		\int_0^1 \sqrt{\log(1/u)}\,du < \infty,
$
it follows that
\[
		\int_0^{\sqrt{\delta}} \sqrt{v\log(A/\varepsilon)}\,d\varepsilon
	\lesssim
		\sqrt{\delta}\,\sqrt{\log(A/\sqrt{\delta})}
		+
		\sqrt{\delta}
	\lesssim
		\sqrt{\delta\log(1/\delta)}
		+
		\sqrt{\delta}
\]
Moreover, if $\delta \le e^{-1}$, then
$
		\sqrt{\delta}
	\le
		\sqrt{\delta\log(1/\delta)}.
$
Hence,
\[
    		J(\sqrt{\delta}, \mathcal F_\delta, 1, L_2)
    	=
    		O\!\left(\sqrt{\delta\log(1/\delta)}\right)
\]
as $\delta\downarrow0$.

Therefore, we obtain
\[
    		E \sup_{f\in\mathcal F_\delta}
    		\left|
        			(P_{n,g}-P_g)f
    		\right|
    =
    		O\!\left(
       			 \sqrt{\frac{\delta\log(1/\delta)}{n_g}}
        			+
        			\frac{\log(1/\delta)}{n_g}
    		\right)
\]
as
$\delta\downarrow0$.
Finally, Markov's inequality yields
\[
    		\sup_{I\in\mathcal H_\delta}
    		\left|
        			(P_{n,g}-P_g)(I)
   		 \right|
   	 =
    		O_{p}\!\left(
       			 \sqrt{\frac{\delta\log(1/\delta)}{n_g}}
        			+
        			\frac{\log(1/\delta)}{n_g}
    		\right)
	\qquad
	\text{as $\delta\downarrow0$.}
\]
\end{proof}

Let $F_{n,g}$ denote the empirical distribution function based on the $n_g$ observations in group $g$.
Fix a group $g$, and define
\[
    		\omega_{g,n}(h)
    	=
    		\sup_{|x-y|\le h}
    		\bigl|
        			(F_{n,g}-F_g)(x) - (F_{n,g}-F_g)(y)
    		\bigr|.
\]
The following lemma shows the convergence rate of $\omega_{g,n}(h)$.
\phantomsection
\addcontentsline{toc}{subsection}{Lemma \ref{lem_modulus}}
\begin{lemma}[Modulus of continuity of the empirical distribution function]
\label{lem_modulus}

Assume that the conditions of Lemma~\ref{lem_uniform_quantile} are satisfied.
Then, for any sequence $h_n \downarrow 0$,
\[
    		\omega_{g,n}(h_n)
    	=
    		O_p\!\left(
        			\sqrt{\frac{h_n \log(1/h_n)}{n_g}}
			+
        			\frac{\log(1/h_n)}{n_g}
    		\right).
\]
In particular, for any fixed $C>0$,
$
    		\omega_{g,n}(C n_g^{-1/2})
    	=
    		o_p(n_g^{-1/2}).
$
\end{lemma}

\begin{proof}
For $x<y$, we have
$
    		(F_{n,g}-F_g)(y)-(F_{n,g}-F_g)(x)
    	=
    		(P_{n,g}-P_g)((x,y]).
$
Therefore,
\[
    		\omega_{g,n}(h)
    	=
    		\sup_{|x-y|\le h}
    		\left|
        			(P_{n,g}-P_g)((x,y])
    		\right|.
\]

Since the support of $F_g$ is compact and $f_g$ is continuous, there exists a finite constant
$
    		\bar f_g 
	= 
		\sup_{y\in[\underline y_g,\overline y_g]} 
		f_g(y) 
	< 
		\infty.
$
Hence, for any $x<y$,
\[
    		P_g((x,y])
    	=
    		F_g(y)-F_g(x)
    	\le
     		(y-x)\bar f_g.
\]
It follows that 
\[
    		\omega_{g,n}(h)
    	\le
    		\sup_{I\in\mathcal H_{\bar f_g h}}
    		\left|
        			(P_{n,g}-P_g)(I)
    		\right|.
\]

Apply Lemma \ref{lem_interval_process} yields
\[
    		\omega_{g,n}(h_n)
   	 =
    		O_p\!\left(
        			\sqrt{\frac{h_n \log(1/h_n)}{n_g}}
			+
        			\frac{\log(1/h_n)}{n_g}
    		\right).
\]

Finally, if $h_n = C n_g^{-1/2}$ for fixed $C>0$, then
\[
    		\omega_{g,n}(C n_g^{-1/2})
	=
    		O_p\!\left(
        			n_g^{-3/4}\sqrt{\log n_g}
			+
			\frac{\log n_{g}}{n_{g}}
   		 \right)
    	=
    		o_p(n_g^{-1/2}),
\]
which proves the result.
\end{proof}

\phantomsection
\addcontentsline{toc}{subsection}{Lemma \ref{lem_bahadur_grid}}
\begin{lemma}[Uniform Bahadur representation over the disclosed grid]
\label{lem_bahadur_grid}
Fix a group $g$. 
Assume that the conditions of Lemma~\ref{lem_uniform_quantile} are satisfied. 
Then
\[
    		\sup_{1\le k\le K_N}
    		\left|
        			\widehat Q_g(\tau_k)-Q_g(\tau_k)
        			-
        			\frac{\tau_k-F_{n,g}(Q_g(\tau_k))}{f_g(Q_g(\tau_k))}
    		\right|
    	=
    		o_p(n_g^{-1/2}).
\]
\end{lemma}
\begin{proof}
For simplicity, write
\[
    		\widehat Q_k = \widehat Q_g(\tau_k),
    	\qquad
    		Q_k = Q_g(\tau_k),
    	\qquad
    		f_k = f_g(Q_g(\tau_k)).
\]

Because $F_g$ is strictly increasing and continuous, we have
$
    F_g(Q_k)=\tau_k
$
for all  $k=1,\dots,K_N$.
By the mean value theorem, for each $k$ there exists $\xi_{gk,N}$ between $\widehat Q_k$ and $Q_k$ such that
\[
    		F_g(\widehat Q_k)-F_g(Q_k)
    	=
    		f_g(\xi_{gk,N})(\widehat Q_k-Q_k).
\]
Therefore,
\[
    		\widehat Q_k-Q_k
    	=
    		\frac{F_g(\widehat Q_k)-\tau_k}{f_k}
    		+
    		\left(
       		 	\frac{1}{f_g(\xi_{gk,N})}
        			-
        			\frac{1}{f_k}
    		\right)
    		\bigl(F_g(\widehat Q_k)-\tau_k\bigr).
\]
Substituting
$
    		F_g(\widehat Q_k)-\tau_k
    	=
    		(F_g(\widehat Q_k)-F_{n,g}(\widehat Q_k))
    		+
    		(F_{n,g}(\widehat Q_k)-\tau_k)
$
and adding and subtracting $(F_{n,g}-F_g)(Q_k)$, we obtain
\begin{align*}
    		\widehat Q_k-Q_k
    	=
    		\frac{\tau_k-F_{n,g}(Q_k)}{f_k}
    		+
    		R_{1,gk,N}
    		+
    		R_{2,gk,N}
    		+
    		R_{3,gk,N},
\end{align*}
where
\begin{align*}
    		R_{1,gk,N}
    	&=
    		\frac{F_{n,g}(\widehat Q_k)-\tau_k}{f_k},
    		\\
    		R_{2,gk,N}
    	&=
    		\frac{
        			(F_{n,g}-F_g)(Q_k) - (F_{n,g}-F_g)(\widehat Q_k)
    		}{f_k},
   		 \\
    		R_{3,gk,N}
    	&=
    		\left(
        			\frac{1}{f_g(\xi_{gk,N})}
        			-
        			\frac{1}{f_k}
    		\right)
   		 \bigl(F_g(\widehat Q_k)-\tau_k\bigr).
\end{align*}

It suffices to show that each remainder term is $o_p(n_g^{-1/2})$ uniformly over $1\le k\le K_N$.

\medskip
\noindent
\textbf{Step 1: the term $R_{1,gk,N}$.}
By definition of the sample quantile,
$
		\tau_k 
	\le 
		F_{n,g}(\widehat Q_k).
$
Since 
$
		F_{n,g}(y)
	=
		\frac{1}{n_g}
		\sum_{i\in\mathcal I_g}
		\mathbf{1}\{Y_i\le y\}
$ 
is a step function with jumps of size $1/n_g$, we have
\[
    		0 
	\le 
		F_{n,g}(\widehat Q_k)-\tau_k \le \frac{1}{n_g}
    \qquad\text{for all } 
    		k=1,\dots,K_N.
\]
Hence
\[    		
		\sup_{1\le k\le K_N}
    		\bigl|
        			F_{n,g}(\widehat Q_k)-\tau_k
    		\bigr|
    	=
    		O(n_g^{-1}).
\]
By the lower bound $f_k \ge c_g$ from the assumption of Lemma~\ref{lem_uniform_quantile},
\[
   	 	\sup_{1\le k\le K_N}
    		|R_{1,gk,N}|
    	\le
    		c_g^{-1}
    		\sup_{1\le k\le K_N}
    		|F_{n,g}(\widehat Q_k)-\tau_k|
    	=
    		O(n_g^{-1})
    	=
    		o(n_g^{-1/2}).
\]

\medskip
\noindent
\textbf{Step 2: the term $R_{2,gk,N}$.}
By Lemma~\ref{lem_uniform_quantile},
$
    		\sup_{1\le k\le K_N}
    		|\widehat Q_k-Q_k|
    	=
    		O_p(n_g^{-1/2}).
$
Hence, for any fixed $M>0$,
\[
   	 	P\!\left(
        			\sup_{1\le k\le K_N}
        			|\widehat Q_k-Q_k|
        			\le
        			M n_g^{-1/2}
    		\right)
    	\to 
		1
\]
as $n_g\to\infty$.
Since $f_k \ge c_g$,
\[
    		|R_{2,gk,N}|
   	 \le
    		c_g^{-1}
    		\omega_{g,n}(|\widehat Q_k-Q_k|),
\]
Therefore,
\[
    \sup_{1\le k\le K_N}|R_{2,gk,N}|
    \le
    c_g^{-1}
    \omega_{g,n}\!\left(
        \sup_{1\le k\le K_N}
        |\widehat Q_k-Q_k|
    \right).
\]
Combining this with Lemma~\ref{lem_modulus} and
$
    		\sup_{1\le k\le K_N}
    		|\widehat Q_k-Q_k|
    	=
    		O_p(n_g^{-1/2}),
$ by Lemma~\ref{lem_uniform_quantile},
we obtain
\[
    		\sup_{1\le k\le K_N}|R_{2,gk,N}|
    	=
   		 o_p(n_g^{-1/2}).
\]

\medskip
\noindent
\textbf{Step 3: the term $R_{3,gk,N}$.}
Since $f_g$ is continuously differentiable and bounded away from zero on the compact support $[\underline y_g,\overline y_g]$, the function $1/f_g$ is uniformly continuous on $[\underline y_g,\overline y_g]$.
Since 
$
		|\xi_{gk,N}-Q_k|
	\le 
		|\widehat Q_k-Q_k|,
$
Lemma~\ref{lem_uniform_quantile} implies
$
		\sup_k |\xi_{gk,N}-Q_k|
	\le
		\sup_k |\widehat Q_k-Q_k|
	=
		o_p(1).
$
Thus, the uniform continuity implies
\[
    		\sup_{1\le k\le K_N}
    		\left|
        			\frac{1}{f_g(\xi_{gk,N})}
			-
        			\frac{1}{f_k}
   		 \right|
   	 =
   		 o_p(1).
\]
Moreover, because $f_g$ is bounded on $[\underline y_g,\overline y_g]$ and
$
    		\sup_{1\le k\le K_N}
    		|\widehat Q_k-Q_k|
    	=
    		O_p(n_g^{-1/2}),
$
the mean value theorem yields
\[
    \sup_{1\le k\le K_N}
    |F_g(\widehat Q_k)-\tau_k|
    =
    O_p(n_g^{-1/2}).
\]
Hence
\[
    \sup_{1\le k\le K_N}|R_{3,gk,N}|
    =
    o_p(1)\,O_p(n_g^{-1/2})
    =
    o_p(n_g^{-1/2}).
\]
\end{proof}

\phantomsection
\addcontentsline{toc}{subsection}{Lemma \ref{lem_derivative}}
\begin{lemma}[Uniform consistency of the derivative estimator]
\label{lem_derivative}
Suppose that Assumption \ref{ass_smooth}(ii) and the conditions of Lemma \ref{lem_uniform_quantile} hold.
If
$
	\Delta(\mathcal T_{K_N}) \to 0
$
and
$
	n_{g}^{-1/2} = o_{p}(\underline\Delta(\mathcal T_{K_N}))
$,
then
\[
		\max_{1\le k\le K_N}
		\left|
			\widehat{Q_g'}(\tau_k)-Q_g'(\tau_k)
		\right|
	\xrightarrow{p}
		0.
\]
\end{lemma}

\begin{proof}
For interior points $k=2,\dots,K_N-1$,
write
\begin{align*}
		\widehat{Q_g'}(\tau_k) - Q_g'(\tau_k)
	&=
		\left[
			\frac{
				Q_g(\tau_{k+1}) - Q_g(\tau_{k-1})
			}{
				\tau_{k+1} - \tau_{k-1}
			}
			- Q_g'(\tau_k)
		\right]
	+
		\frac{
			(\widehat Q_g - Q_g)(\tau_{k+1}) - (\widehat Q_g - Q_g)(\tau_{k-1})
		}{
			\tau_{k+1} - \tau_{k-1}
		}.
\end{align*}
By Assumption~\ref{ass_smooth}(ii), the first term is
$
	O(\Delta(\mathcal T_{K_N})).
$
For the second term,
\[
		\left|
			\frac{
				(\widehat Q_g - Q_g)(\tau_{k+1}) - (\widehat Q_g - Q_g)(\tau_{k-1})
			}{
				\tau_{k+1} - \tau_{k-1}
			}
		\right|
	\le
		\frac{
			2
			\sup_{\tau \in \mathcal T_{K_N}}
			|\widehat Q_g(\tau)-Q_g(\tau)|
		}{
			\underline\Delta(\mathcal T_{K_N})
		}.
\]
Since 
$
		\sup_{\tau}|\widehat Q_g(\tau)-Q_g(\tau)| 
	= 
		O_p(n_g^{-1/2}),
$
by Lemma~\ref{lem_uniform_quantile}, the second term is
\[
		O_p\!\left(\frac{n_g^{-1/2}}{\underline\Delta(\mathcal T_{K_N})}\right)
	=
		o_{p}(1).
\]
Thus, result follows for the interior points.

For boundary points $k=1$ and $k=K_N$, the derivative estimator is defined using one-sided differences:
\[
\widehat{Q_g'}(\tau_1)
=
\frac{\widehat Q_g(\tau_2)-\widehat Q_g(\tau_1)}{\tau_2-\tau_1},
\qquad
\widehat{Q_g'}(\tau_{K_N})
=
\frac{\widehat Q_g(\tau_{K_N})-\widehat Q_g(\tau_{K_N-1})}{\tau_{K_N}-\tau_{K_N-1}}.
\]
These cases can be handled similarly.
\end{proof}

Let $E_g(\cdot)$, $\Var_g (\cdot)$, and $\Cov_g(\cdot,\cdot)$ denote expectation, variance, and covariance under the conditional distribution of $Y_i$ given $\bm X_i \in g$.

\phantomsection
\addcontentsline{toc}{subsection}{Lemma \ref{lem_group_var}}
\begin{lemma}[Consistency of group-level variance estimator]
\label{lem_group_var}
Suppose that the conditions of Proposition~\ref{prop_rootN} hold under Assumption~\ref{ass_smooth}(ii).
If
$
	N^{-1/2} = o(\underline\Delta(\mathcal T_{K_N}))
$,
then,
\[
		n_g \widehat{\Var}(\hat m_g)
	\xrightarrow{p}
		 \Var(Y_i-m_g\mid\bm X_i\in g).
\]
\end{lemma}

\begin{proof}
By definition,
\[
		n_g\widehat{\Var}(\hat m_g)
	=
		\sum_{k=1}^{K_N}\sum_{l=1}^{K_N}
		\alpha_{k,N}\alpha_{l,N}
		\frac{
			\min(\tau_k,\tau_l)-\tau_k\tau_l
		}{
			\hat f_g(\widehat Q_g(\tau_k))\,\hat f_g(\widehat Q_g(\tau_l))
		}.
\]

By Assumption \ref{ass_design} , 
\(n_g/N\overset p\to\pi_g>0\), so the spacing condition implies \(n_g^{-1/2}=o_p(\underline\Delta(\mathcal T_{K_N}))\).
Then by Lemma~\ref{lem_derivative} and the identity
$
		f_g(Q_g(\tau)) 
	= 
		\frac{1}{Q_g'(\tau)},
$
we have
\[
		\max_k
		\left|
			\frac{1}{\hat f_g(\widehat Q_g(\tau_k))}
			-
			\frac{1}{f_g(Q_g(\tau_k))}
		\right|
	\xrightarrow{p}
		0.
\]

Recall that the reconstruction weights satisfy
$
    \alpha_{k,N} \ge 0,
$
and
$
    \sum_{k=1}^{K_N}\alpha_{k,N}=1.
$
Hence,
$
    	\sum_{k=1}^{K_N}\sum_{l=1}^{K_N}
    	\alpha_{k,N}\alpha_{l,N}
    =
    	1.
$
Therefore, 
\begin{align*}
		n_g\widehat{\Var}(\hat m_g)
	&=
		\sum_{k=1}^{K_N}\sum_{l=1}^{K_N}
		\alpha_{k,N}\alpha_{l,N}
		\frac{
			\min(\tau_k,\tau_l)-\tau_k\tau_l
		}{
			f_g(Q_g(\tau_k))\,f_g(Q_g(\tau_l))
		}
	+ 
		o_p(1).
\end{align*}

Under the conditions of (i) and (iii) in Lemma~\ref{lem_uniform_quantile}, the function
\[
    		H_g(\tau,s)
    	=
    		\frac{
        			\min(\tau,s)-\tau s
    		}{
        			f_g(Q_g(\tau))f_g(Q_g(s))
    		}
\]
is bounded and continuous on $[0,1]^2$. Since
$\Delta(\mathcal T_{K_N})\to0$ and the weights
$\{\alpha_{k,N}\}$ are the quadrature weights, we have
\[
    		\sum_{k=1}^{K_N}\sum_{l=1}^{K_N}
    		\alpha_{k,N}\alpha_{l,N}
    		H_g(\tau_k,\tau_l)
    	\to
    		\int_0^1\int_0^1 H_g(\tau,s)\,d\tau ds.
\]

On the other hand,
\[
		\psi_g(y)
	=
		-
		\int_0^1
		\frac{
			\mathbf 1\{y\le Q_g(\tau)\}-\tau
		}{
			f_g(Q_g(\tau))
		}
		d\tau.
\]

Since $E_g[\psi_g(Y_i)]=0$,
\begin{align*}
		\Var_g (\psi_g(Y_i))
	&=
		\int_0^1\int_0^1
		\frac{
			\Cov_g(\mathbf 1\{Y_i\le Q_g(\tau)\},\mathbf 1\{Y_i\le Q_g(s)\})
		}{
			f_g(Q_g(\tau))f_g(Q_g(s))
		}
		d\tau ds
		\\
	&=
		\int_0^1\int_0^1
		\frac{
			\min(\tau,s)-\tau s
		}{
			f_g(Q_g(\tau))f_g(Q_g(s))
		}
		d\tau ds.
\end{align*}

Thus,
\[
		n_g\widehat{\Var}(\hat m_g)
	\xrightarrow{p}
		\Var_g(\psi_g(Y_i)).
\]
Since \(\psi_g(Y_i)=Y_i-m_g\), as established in the proof of
Theorem~\ref{thm_bias_clt},
\[
    \Var_g\{\psi_g(Y_i)\}
    =
    \Var(Y_i-m_g\mid\bm X_i\in g).
\]
\end{proof}

\clearpage
\def\thesection{}
\renewcommand{\thesubsection}{C-\arabic{subsection}} 
 \def\theequation{C.\arabic{equation}}
 \def\theassumption{C.\arabic{assumption}}	
 \makeatletter
 \@addtoreset{equation}{section}
 \@addtoreset{figure}{section}
 \@addtoreset{table}{section}
 \@addtoreset{assumption}{section}
 \makeatother
 \def\thelemma{C.\arabic{lemma}}	  

\setcounter{equation}{0} 
\setcounter{subsection}{0} 
\setcounter{lemma}{0}
\setcounter{page}{1}
\section{Supplemental Material C: A Standard Uniform Bahadur Representation on Compact Interior Sets}
For comparison with the boundary-robust representation in Lemma \ref{lem_bahadur_grid}, this section records a standard uniform Bahadur representation over a fixed compact interior subset of the unit interval.

Lemma \ref{lem_uniform_bahadur_compact} below concerns a fixed interior index set. 
The expanding-grid representation used in the main analysis is established separately in Lemma \ref{lem_bahadur_grid} under the conditions of Lemma \ref{lem_uniform_quantile}.

To show a sufficient condition for a standard uniform Bahadur representation, we introduce some notation.
Let  $\varepsilon \in (0,1/2)$
and
$
    		I_\varepsilon
    	=
    		[\varepsilon,1-\varepsilon].
$
For a compact set 
$
    		K_\varepsilon
    	=
    		[
       			 Q_g(\varepsilon/2),
        			Q_g(1-\varepsilon/2)
    		]
$,
 let
\[
    		\ell^\infty(K_\varepsilon)
    	=
    		\Bigl\{
        			f:K_\varepsilon \to \mathbb{R}
       		 \;\big|\;
        		\sup_{y\in K_\varepsilon} |f(y)| < \infty
    \Bigr\}
\]
denote the space of bounded functions on $K_\varepsilon$, equipped with the sup-norm
\[
    		\|f\|_\infty
    	=
    		\sup_{y\in K_\varepsilon} |f(y)|.
\]
Let
$
    		C(K_\varepsilon)
   	 \subset 
    		\ell^\infty(K_\varepsilon)
$
denote the space of continuous functions on $K_\varepsilon$.
We write $\rightsquigarrow$ to denote weak convergence.

\begin{lemma}[Uniform Bahadur representation on compact interior sets]
\label{lem_uniform_bahadur_compact}
Fix a group $g$ and let $\varepsilon \in (0,1/2)$.
Suppose that conditions (i)--(iii) of
Lemma~\ref{lem_uniform_quantile} are satisfied, and that
$
		\sqrt{n_g}(F_{n,g}-F_g) 
	\rightsquigarrow 
		\mathbb G_g
$
in $\ell^\infty(K_\varepsilon)$, where $\mathbb G_g$ has continuous sample paths almost surely.
Then
\[
    \sup_{\tau\in[\varepsilon,1-\varepsilon]}
    \left|
        \widehat Q_g(\tau)-Q_g(\tau)
        -
        \frac{\tau-F_{n,g}(Q_g(\tau))}{f_g(Q_g(\tau))}
    \right|
    =
    o_p(n_g^{-1/2}).
\]
\end{lemma}
\begin{proof}
Consider the quantile map \(\Phi\) defined on a neighborhood of \(F_g\) in
\(\ell^\infty(K_\varepsilon)\) by
\[
    		\Phi(H)(\tau)
    	=
    		\inf
    		\{
        		y \in K_\varepsilon : H(y) \ge \tau
    		\},
    \qquad
    		\tau \in I_\varepsilon.
\]

for nondecreasing \(H\).
Since $F_g$ is strictly increasing on a neighborhood of $K_\varepsilon$, we have
\[
    		\Phi(F_g)(\tau)
    	=
    		Q_g(\tau),
    \qquad
    		\tau \in I_\varepsilon.
\]

By conditions (i)--(iii) of Lemma~\ref{lem_uniform_quantile}, 
$f_g$ is continuous and bounded away from zero on $K_\varepsilon$.
Therefore, the inverse map $\Phi$ is Hadamard differentiable at $F_g$
tangentially to $C(K_\varepsilon)$, with derivative
\[
    \Phi'_{F_g}(h)(\tau)
    =
    -
    \frac{
        h(Q_g(\tau))
    }{
        f_g(Q_g(\tau))
    },
    \qquad
    \tau \in I_\varepsilon,
\]
for any $h \in C(K_\varepsilon)$.

Since $\Phi:\ell^\infty(K_\varepsilon)\to \ell^\infty(I_\varepsilon)$ is
Hadamard differentiable at $F_g$ tangentially to $C(K_\varepsilon)$, and
$
		\sqrt{n_g}(F_{n,g}-F_g) 
	\rightsquigarrow 
		\mathbb G_g
$ 
in $\ell^\infty(K_\varepsilon)$,
where $\mathbb G_g$ has continuous sample paths almost surely (so that the limit
takes values in $C(K_\varepsilon)$), 
the functional delta method (see, for example Theorem 20.8 of \citealp{vandervaart98}) yields
\[
    \left\|
        \sqrt{n_g}
        \bigl(
            \Phi(F_{n,g})-\Phi(F_g)
        \bigr)
        -
        \Phi'_{F_g}
        \bigl(
            \sqrt{n_g}(F_{n,g}-F_g)
        \bigr)
    \right\|_\infty
    =
    o_p(1).
\]

Hence,
\[
    \sup_{\tau\in I_\varepsilon}
    \left|
        \sqrt{n_g}
        \bigl(
            \Phi(F_{n,g})(\tau)-Q_g(\tau)
        \bigr)
        +
        \frac{
            \sqrt{n_g}
            (F_{n,g}(Q_g(\tau))-\tau)
        }{
            f_g(Q_g(\tau))
        }
    \right|
    =
    o_p(1).
\]
Equivalently,
\[
    \sup_{\tau\in I_\varepsilon}
    \left|
        \Phi(F_{n,g})(\tau)-Q_g(\tau)
        -
        \frac{
            \tau-F_{n,g}(Q_g(\tau))
        }{
            f_g(Q_g(\tau))
        }
    \right|
    =
    o_p(n_g^{-1/2}).
\]

It remains to show that $\Phi(F_{n,g})(\tau)=\widehat Q_g(\tau)$ uniformly over $\tau\in I_\varepsilon$
with probability tending to one. 
By definitions of $\Phi(F_{n,g})$ and $\widehat Q_g$, it is equivalent to show
$\widehat Q_g(\tau)\in K_\varepsilon$
for all
$ \tau\in I_\varepsilon$ with probability tending to one.

Since
$
		\sqrt{n_g}(F_{n,g}-F_g) 
	\rightsquigarrow 
		\mathbb G_g
$
in $\ell^\infty(K_\varepsilon)$,
the sequence \(\sqrt{n_g}(F_{n,g}-F_g)\) is tight in \(\ell^\infty(K_\varepsilon)\). Hence,
$
		\sup_{y\in K_\varepsilon}|F_{n,g}(y)-F_g(y)|
	=
		o_p(1).
$
Then, for any fixed $\eta \in (0,\varepsilon/2)$,
\[
   		 \sup_{y\in K_\varepsilon}
    		|F_{n,g}(y)-F_g(y)|
    	\le 
		\eta
\]
with probability tending to one.
On this event,
\[
    		F_{n,g}(Q_g(\varepsilon/2))
   	 \le
   		 \varepsilon/2+\eta
   	 <
   		 \varepsilon
\]
and
\[
    		F_{n,g}(Q_g(1-\varepsilon/2))
    	\ge
    		1-\varepsilon/2-\eta
    	>
    		1-\varepsilon.
\]
Therefore, for all $\tau\in I_\varepsilon$, the sample quantile $\widehat Q_g(\tau)$ belongs to
$K_\varepsilon$, and hence
$
    		\widehat Q_g(\tau)
    	=
   		 \Phi(F_{n,g})(\tau)
$    		
for all 
$\tau \in I_\varepsilon$
with probability tending to one.
\end{proof}

\clearpage
\def\thesection{}
\renewcommand{\thesubsection}{D-\arabic{subsection}} 
 \def\theequation{D.\arabic{equation}}
 \def\theassumption{D.\arabic{assumption}}	
 \def\thelemma{D.\arabic{lemma}}
\renewcommand{\thefigure}{D.\arabic{figure}}
\renewcommand{\thetable}{D.\arabic{table}}

 \makeatletter
 \@addtoreset{equation}{section}
 \@addtoreset{figure}{section}
 \@addtoreset{table}{section}
 \@addtoreset{assumption}{section}
 \makeatother
 \def\thelemma{D.\arabic{lemma}}	  

\setcounter{equation}{0} 
\setcounter{subsection}{0} 
\setcounter{lemma}{0}
\setcounter{page}{1}
\section{Supplemental Material D:  Simulation}

For a given number of disclosed quantiles $K$, let
$
    p_k=\frac{k}{K+1},
$
for
$
    k=1,\ldots,K.
$
The evenly spaced grid is defined by
$
    \tau_k=p_k.
$
The mildly and severely right-tail-coarse grids are defined by
$
    \tau_k=p_k^{3/2}
$
and
$
    \tau_k=p_k^3,
$
respectively.

\subsection{Evenly Spaced Grids: Numerical Coverage and Region Geometry}
\label{app_sim_even_details}

Table~\ref{tab:coverage_even} gives numerical coverage probabilities for the
Gaussian point-estimation experiment in Section~\ref{sec_simulation},
complementing Figure~\ref{fig_simulation}.
Coverage approaches its nominal 95\% level as $K$ increases.
Because the Gaussian distribution and the evenly spaced grid are symmetric,
population mean-reconstruction error cancels for every finite $K$;
these results therefore primarily reflect sample-quantile and feasible
variance-estimation error.

\input{QOR_Replication/simulation/results_smooth_qlm/table_coverage_even.tex}

\begin{figure}[htbp]
\centering
\includegraphics[width=\linewidth]{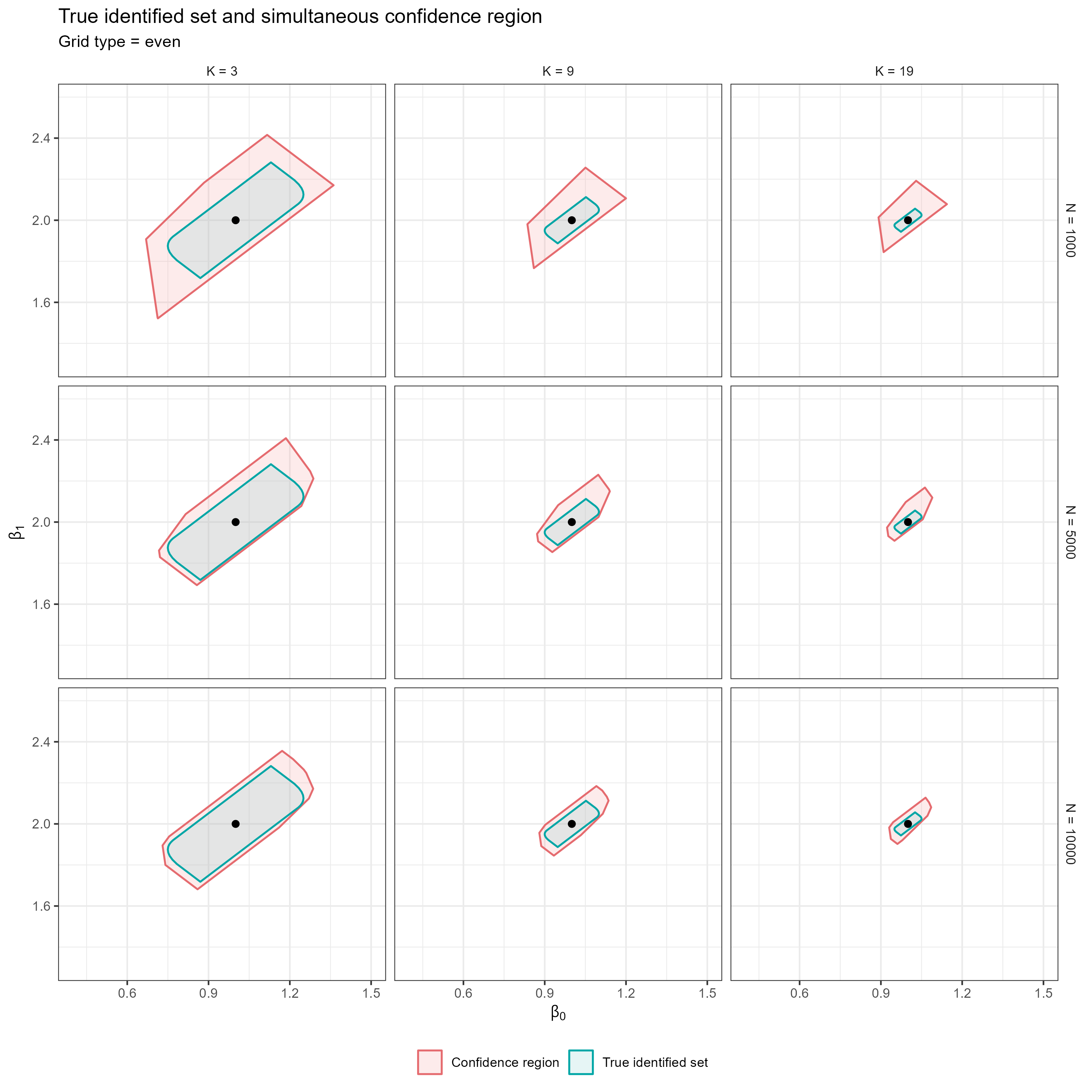}
\caption{\label{fig_simulation_partial}
True identified sets and simultaneous confidence regions under evenly spaced disclosed quantiles.
}
\end{figure}

For the bounded-support uniform-error experiment,
Figure~\ref{fig_simulation_partial} illustrates a representative replication
of the true identified set and the simultaneous confidence region.
For fixed $K$, the confidence region contracts toward the identified set
as $N$ increases.
For fixed $N$, both become smaller as more quantiles are disclosed,
reflecting reduced identification uncertainty.

\subsection{All Sample Sizes}
\begin{figure}[htbp]
\centering
\includegraphics[width=0.9\textwidth]{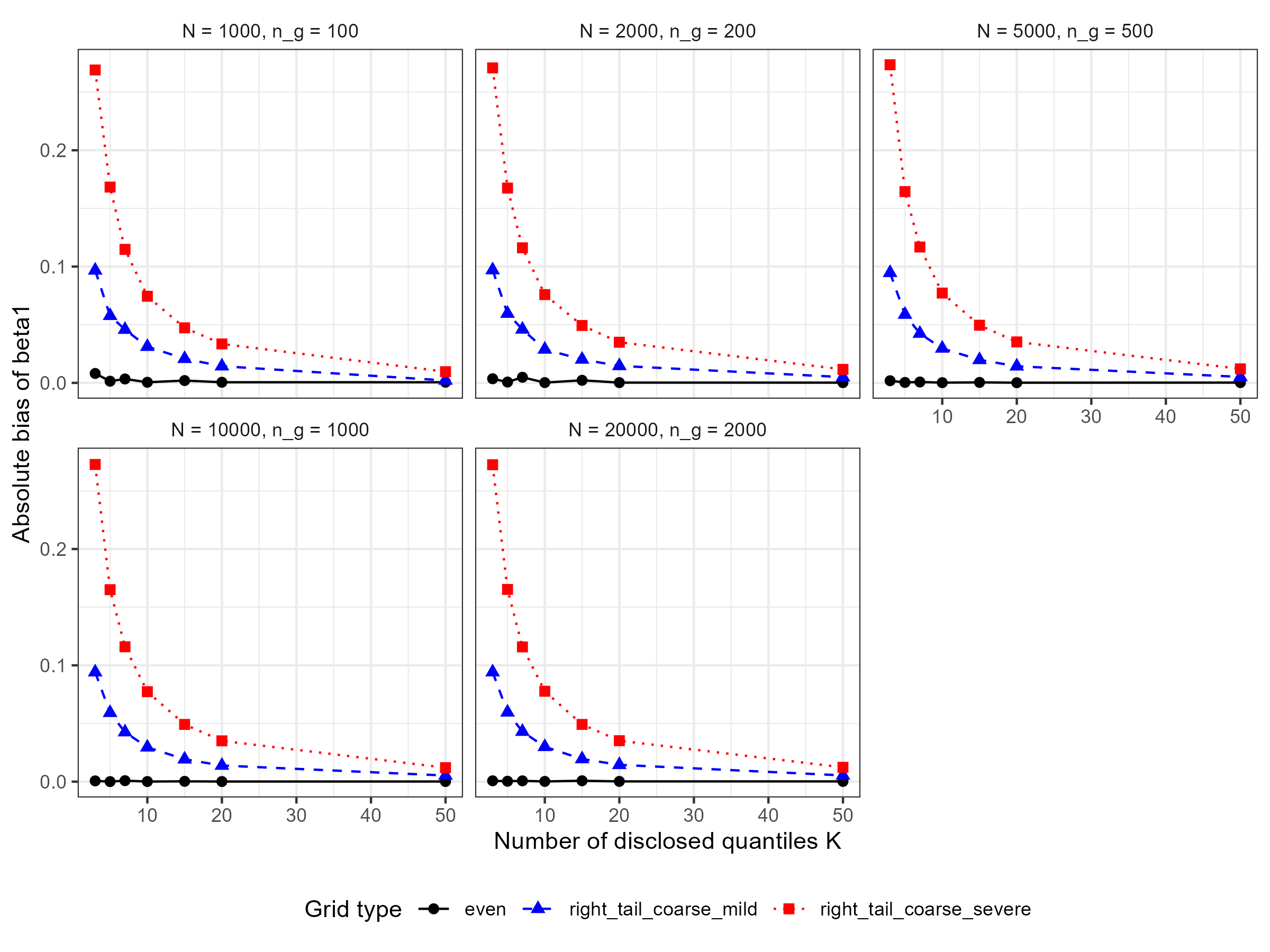}
\caption{Absolute bias across all sample sizes.}
\label{fig_appendix_bias}
\end{figure}

\begin{figure}[htbp]
\centering
\includegraphics[width=0.9\textwidth]{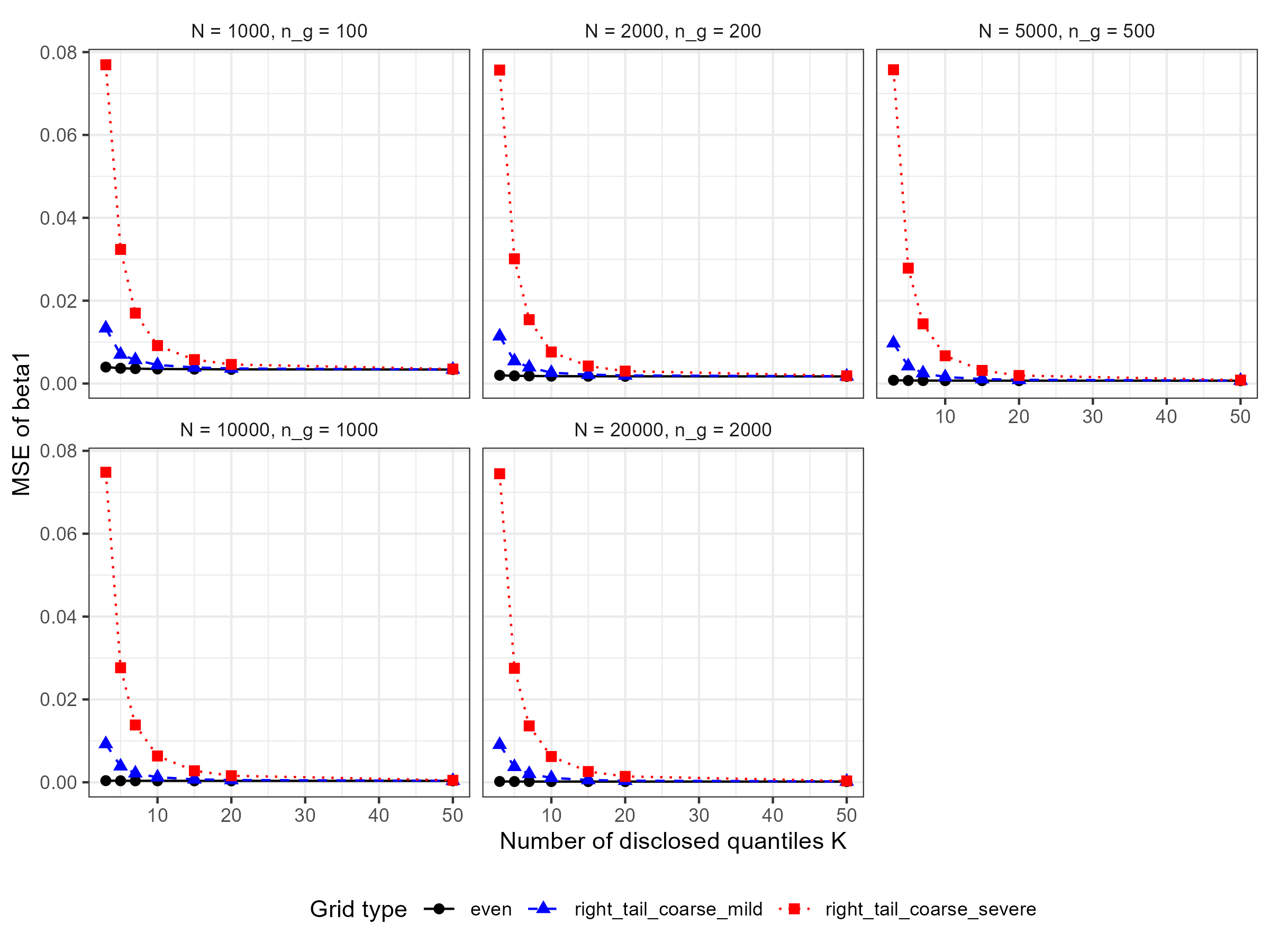}
\caption{Mean squared error across all sample sizes.}
\label{fig_appendix_mse}
\end{figure}

\begin{figure}[htbp]
\centering
\includegraphics[width=0.9\textwidth]{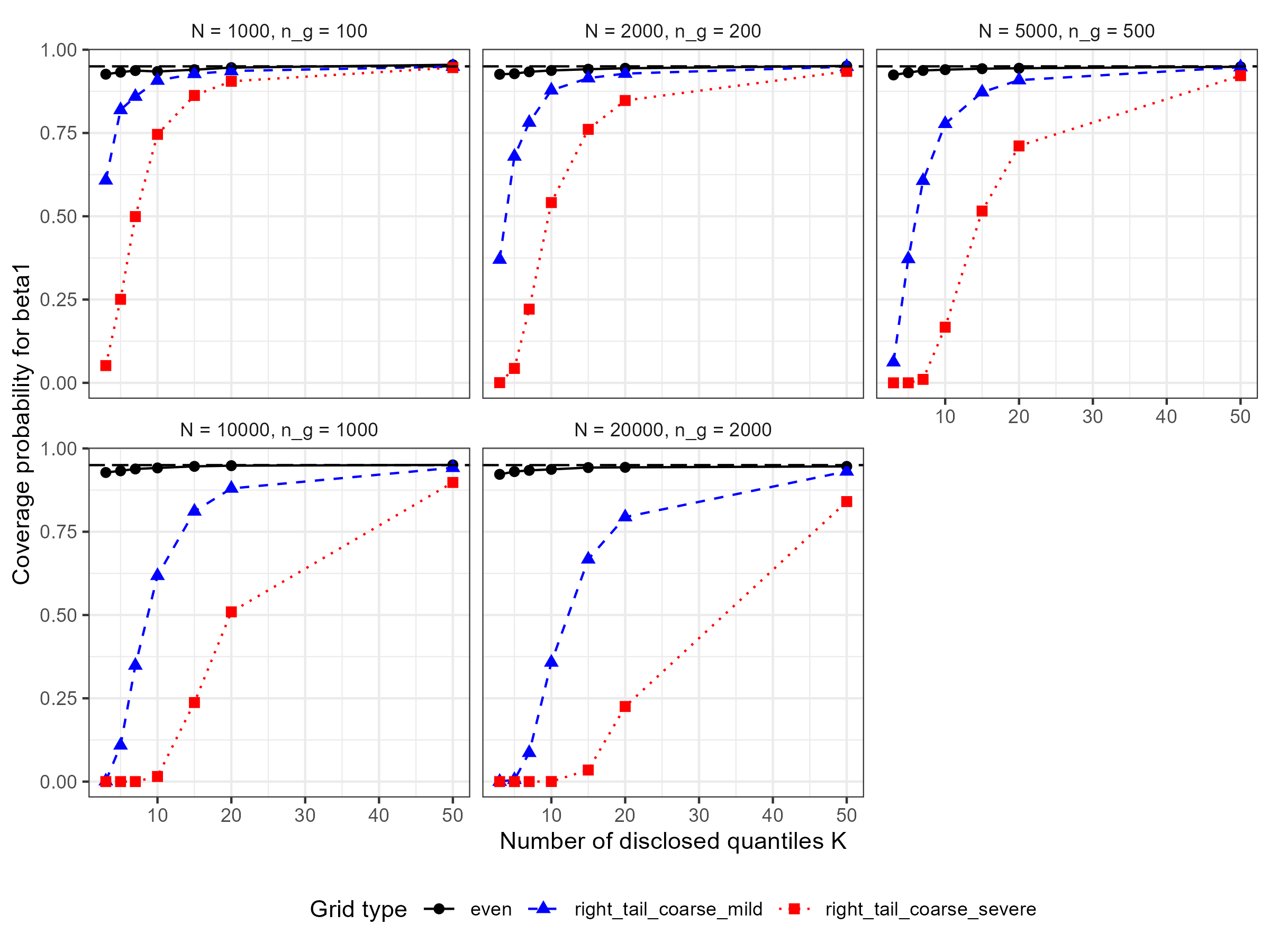}
\caption{Coverage probability across all sample sizes.}
\label{fig_appendix_coverage}
\end{figure}

Figures \ref{fig_appendix_bias}--\ref{fig_appendix_coverage} report the results for all sample sizes. 
The qualitative patterns are unchanged across $N$, and the interpretation is the same as in the main text.
In particular, the relative performance of different grid designs is stable across sample sizes.

\clearpage
\subsection{The Maximal Spacing \texorpdfstring{$\Delta(\mathcal T_K)$}{Delta(T K)}}

\begin{figure}[htbp]
\centering
\includegraphics[width=0.9\textwidth]{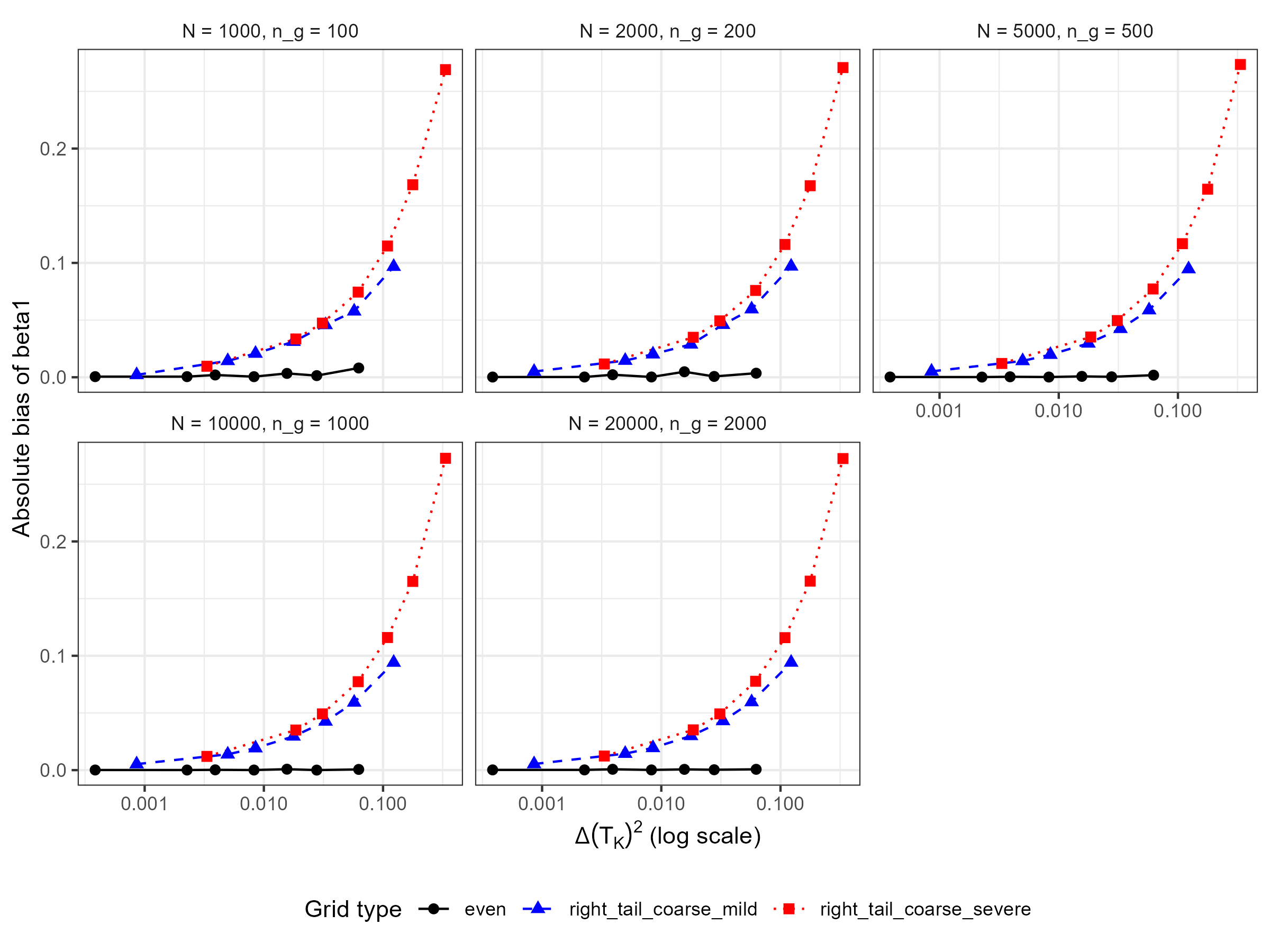}
\caption{Relationship between bias and $\Delta(\mathcal T_K)^2$.}
\label{fig_appendix_delta}
\end{figure}

Figure~\ref{fig_appendix_delta} reports the absolute bias of $\hat\beta_1$ as a function of the squared maximal spacing $\Delta(\mathcal T_K)$, with the horizontal axis displayed on a logarithmic scale.
The results are qualitatively identical to those obtained when plotting against $K$, confirming that the discretization measure $\Delta(\mathcal T_K)$ is the key determinant of the estimation error.

The horizontal range differs across grid designs because $\Delta(\mathcal T_K)$ depends not only on the number of quantiles but also on their placement. 
In particular, uneven grids generate larger values of $\Delta(\mathcal T_K)$, leading to a wider support in the figure.

\subsection{Simulation under the Maintained Compact-Support Conditions}
This subsection examines the finite-sample performance of the proposed point estimator under a data-generating process that satisfies Assumption~\ref{ass_smooth}(ii) and the primitive conditions of Lemma~\ref{lem_uniform_quantile}.
The simulation design is identical to the Gaussian design in the main text except for the distribution of the error term.

Let
\[
    Z_i^*
    \sim
    \mathcal N(0,1)
    \mid
    a \leq Z_i^* \leq b,
    \qquad
    a=-2,\quad b=3.
\]
Let \(\mu_T\) and \(s_T^2\) denote the mean and variance of this truncated normal distribution:
\[
    \mu_T
    =
    \frac{\phi(a)-\phi(b)}{\Phi(b)-\Phi(a)},
\]
and
\[
    s_T^2
    =
    1+
    \frac{a\phi(a)-b\phi(b)}{\Phi(b)-\Phi(a)}
    -\mu_T^2,
\]
where $\phi$ and $\Phi$ denote the standard normal density and distribution functions.
We generate the error term as
\[
    u_i
    =
    \sigma_g
    \frac{Z_i^*-\mu_T}{s_T},
\]
where
$\sigma_g=\exp(0.5X_g).$
This normalization implies
$
    E(u_i\mid X_i\in g)=0,
$
and
$
    \Var(u_i\mid X_i\in g)=\sigma_g^2.
$
The conditional mean and variance therefore coincide with those in the Gaussian design, so that the difference between the two experiments is the shape and support of the conditional error distribution.

The corresponding group-specific conditional quantile function is
\[
    Q_g(\tau)
    =
    \beta_0+\beta_1X_g
    +
    \frac{\sigma_g}{s_T}
    \left[
        \Phi^{-1}
        \left\{
            \Phi(a)+\tau [\Phi(b)-\Phi(a)]
        \right\}
        -\mu_T
    \right],
    \qquad \tau\in(0,1).
\]
Because the argument of $\Phi^{-1}$ remains between $\Phi(a)$ and $\Phi(b)$, the quantile function is twice continuously differentiable with a bounded second derivative on $(0,1)$.
The design therefore satisfies Assumption~\ref{ass_smooth}(ii).

Moreover, for every group, the conditional distribution has compact interval support.
Its density is continuously differentiable and strictly positive over its support, including at the support endpoints.
Since the support is compact, the density is bounded away from zero.
Thus, the primitive conditions of Lemma~\ref{lem_uniform_quantile} are also satisfied.

The truncation interval is deliberately asymmetric.
This produces a nonlinear and asymmetric conditional quantile function while avoiding the exact cancellation of discretization errors that can arise under a symmetric distribution and a symmetric quantile grid.
All remaining features of the experiment, including the sample sizes, group structure, quantile grids, values of $K$, and number of Monte Carlo replications, are the same as in the Gaussian design.

\begin{figure}[!htbp]
\centering
\includegraphics[width=\linewidth]
{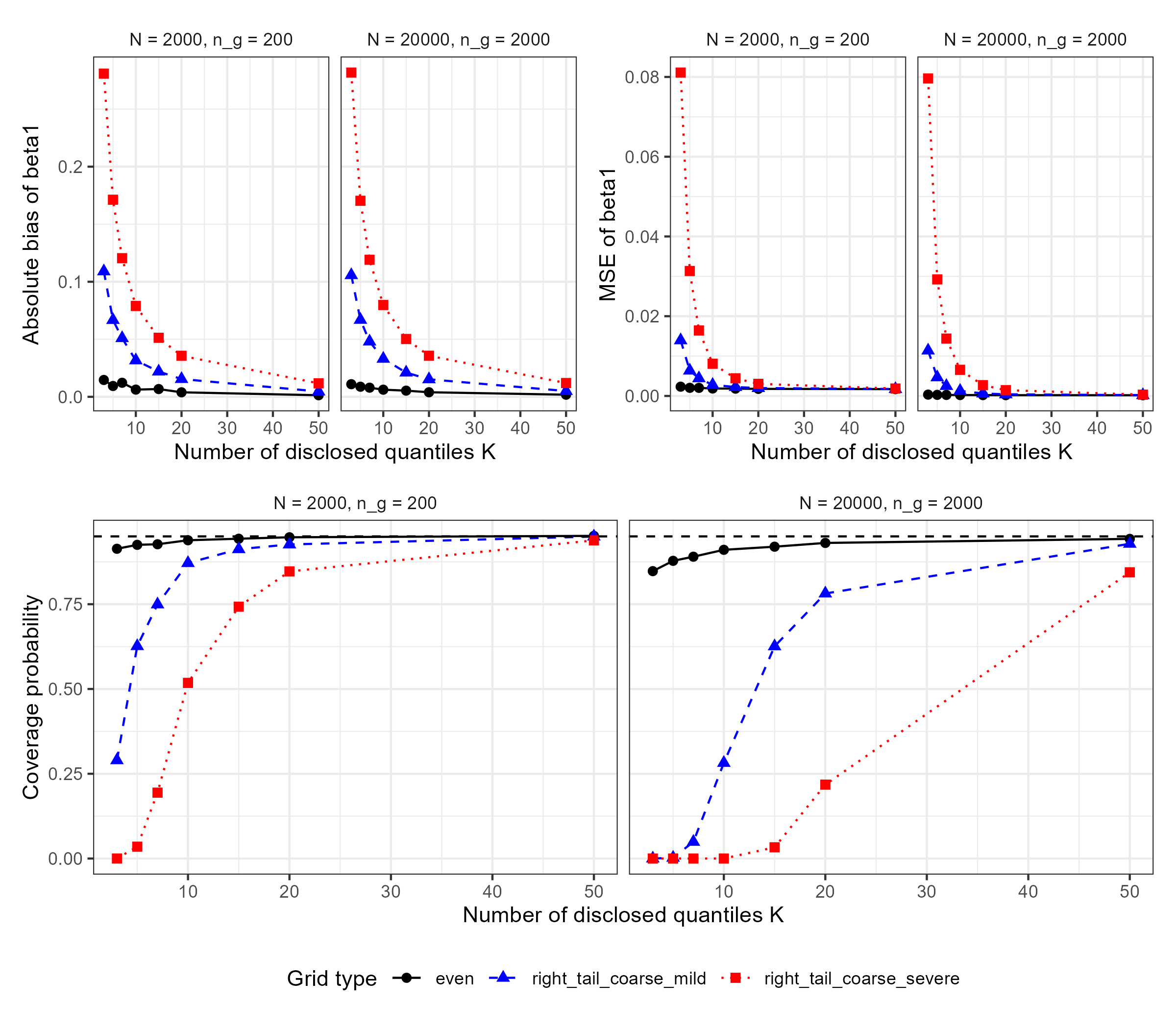}
\caption{\label{fig_compact_support}
Simulation results under the asymmetric compact-support design.
The upper panels report the absolute bias and MSE of $\hat\beta_1$,
and the lower panels report the coverage probability of nominal 95\% confidence intervals.
}
\end{figure}

Figure~\ref{fig_compact_support} reports the absolute bias and mean squared error of $\hat\beta_1$, together with the coverage probability of nominal 95\% confidence intervals.
The qualitative results are similar to those under the Gaussian design.
Bias and mean squared error decrease as the number of disclosed quantiles increases, and evenly spaced grids consistently perform best.
For a given $K$, grids with larger values of the maximal spacing produce greater discretization bias and more severe undercoverage.

\input{QOR_Replication/simulation/results_compact_truncnorm_qlm/table_coverage_even.tex}

Table~\ref{tab:coverage_even_compact} reports the  Monte Carlo coverage probabilities for evenly spaced grids. 
For \(N=2{,}000\), coverage increases from \(0.913\) at \(K=3\) to \(0.952\) at \(K=50\). 
For \(N=20{,}000\), it increases from \(0.848\) at \(K=3\) to \(0.943\) at \(K=50\). 
For a fixed coarse grid, coverage is lower at \(N=20{,}000\) than at \(N=2{,}000\). 
This occurs because sampling uncertainty and confidence interval width decrease with the sample size, whereas the discretization bias associated with a fixed grid remains non-negligible.

Coverage under coarse grids is also lower than in the Gaussian design reported in Table~\ref{tab:coverage_even}, particularly when \(N\) is large. 
This difference reflects the asymmetry of the truncated-normal design rather than compact support itself. 
Under the symmetric Gaussian design and an evenly spaced grid, approximation errors in the lower and upper portions of the quantile function largely cancel. 
This cancellation does not occur under the asymmetric truncated-normal design, resulting in greater discretization bias. 
As \(K\) increases and the grid becomes denser, the discretization bias declines, coverage approaches the nominal level, and the difference between the two designs becomes small.

Overall, the results confirm the main qualitative conclusions under a data-generating process that satisfies the maintained theoretical conditions.
They also illustrate that the finite-sample magnitude of the discretization error depends on the shape of the conditional quantile function, which is suppressed by the rate bound
$O(\Delta(\mathcal T_K)^2)$.

\subsection{Simulation under Non-smooth Quantile Function}
\begin{figure}[htbp]
\centering
\includegraphics[width=\linewidth]{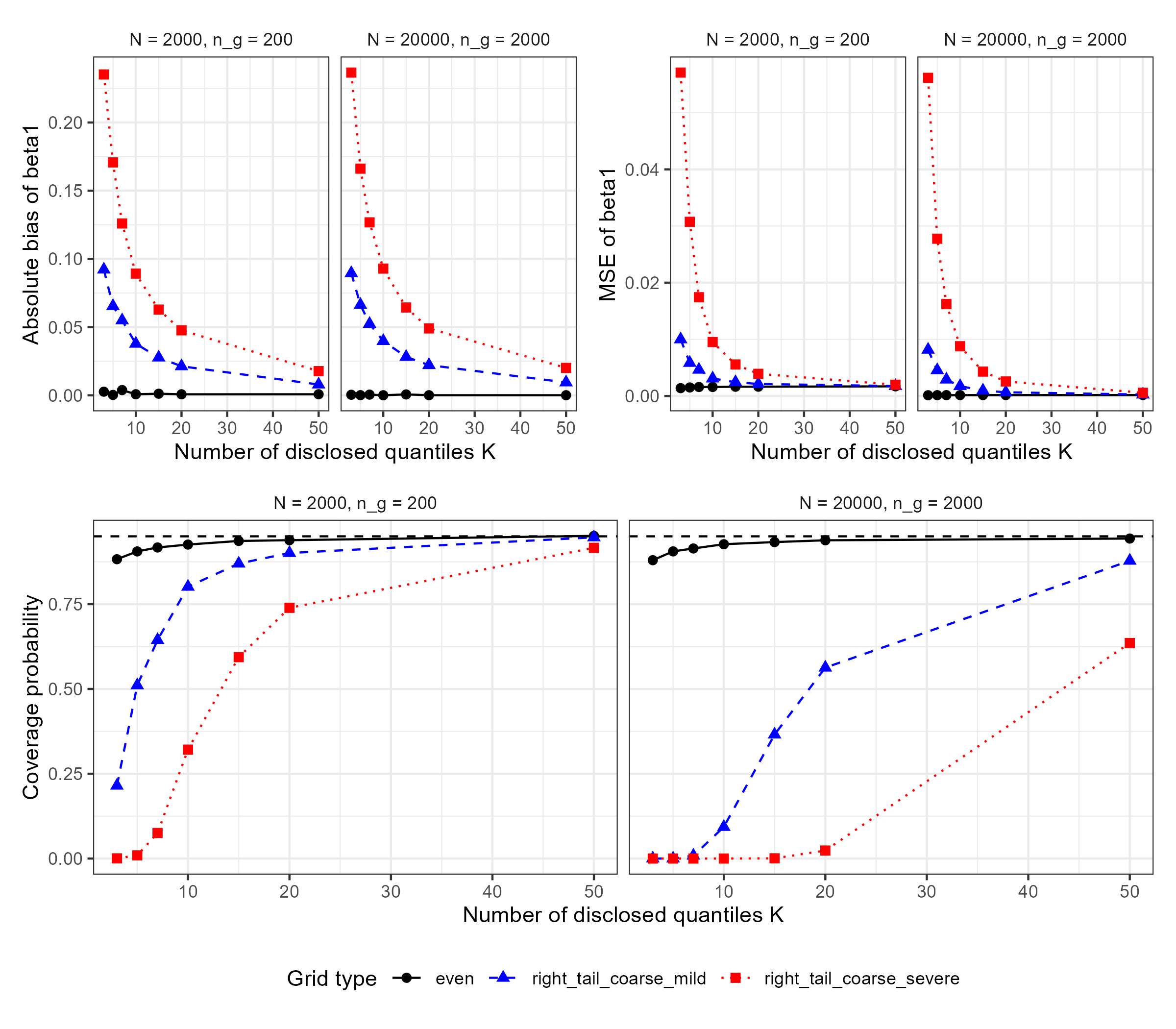}
\caption{\label{fig_nonsmooth}
Simulation results under different quantile designs.
The upper panels report the absolute bias and MSE of $\hat\beta_1$,
and the lower panel reports the coverage probability of nominal 95\% confidence intervals.
}
\end{figure}

Figure~\ref{fig_nonsmooth} and Table~\ref{tab:coverage_even_nonsmooth} report simulation results under a non-smooth design in which the error term follows a Laplace distribution. 
This specification induces a quantile function that is continuously differentiable but not twice continuously differentiable at the median, and therefore violates Assumption~\ref{ass_smooth}(ii).

Specifically, we generate the error term as 
$
		u_i = \sigma_g \varepsilon_i / \sqrt{2},
$
where $\varepsilon_i$ follows a Laplace distribution with density $f(\varepsilon) = \frac{1}{2}\exp(-|\varepsilon|)$. 
This normalization ensures that $\Var(u_i \mid X_i \in g) = \sigma_g^2$, matching the variance in the Gaussian design.

The results are qualitatively similar to those obtained under the Gaussian design in the manuscript.
In particular, evenly spaced grids continue to perform best across all metrics.

However, coverage approaches the nominal level more slowly, reflecting the slower decay of the discretization error in the non-smooth setting.

\input{QOR_Replication/simulation/results_nonsmooth/table_coverage_even.tex}

\subsection{Confidence Region under Mild and Coarse Grids}
This subsection shows partial identification results under mild and severe right-tail-coarse disclosed quantiles.

\input{QOR_Replication/simulation/results_partial_qlm/table_region_coverage_right_tail_coarse_mild.tex}

\input{QOR_Replication/simulation/results_partial_qlm/table_region_coverage_right_tail_coarse_severe.tex}

Tables~\ref{tab:partial_identification_region_coverage_right_tail_coarse_mild}
and~\ref{tab:partial_identification_region_coverage_right_tail_coarse_severe}
report the corresponding results for uneven quantile grids. 
The qualitative pattern is similar to that under evenly spaced grids: coverage improves as the sample size increases and approaches the nominal level for large \(N\).

Compared with the evenly spaced grid, however, right-tail-coarse grids tend to produce wider projection intervals, especially when \(K\) is small. 
This reflects the larger maximal spacing \(\Delta\), which increases the uncertainty induced by partial disclosure. 
The effect is most pronounced under the severely right-tail-coarse grid with \(K=3\), where both the identified set and the associated confidence region are substantially wider.

\begin{figure}[htbp]
\centering
\includegraphics[width=\linewidth]{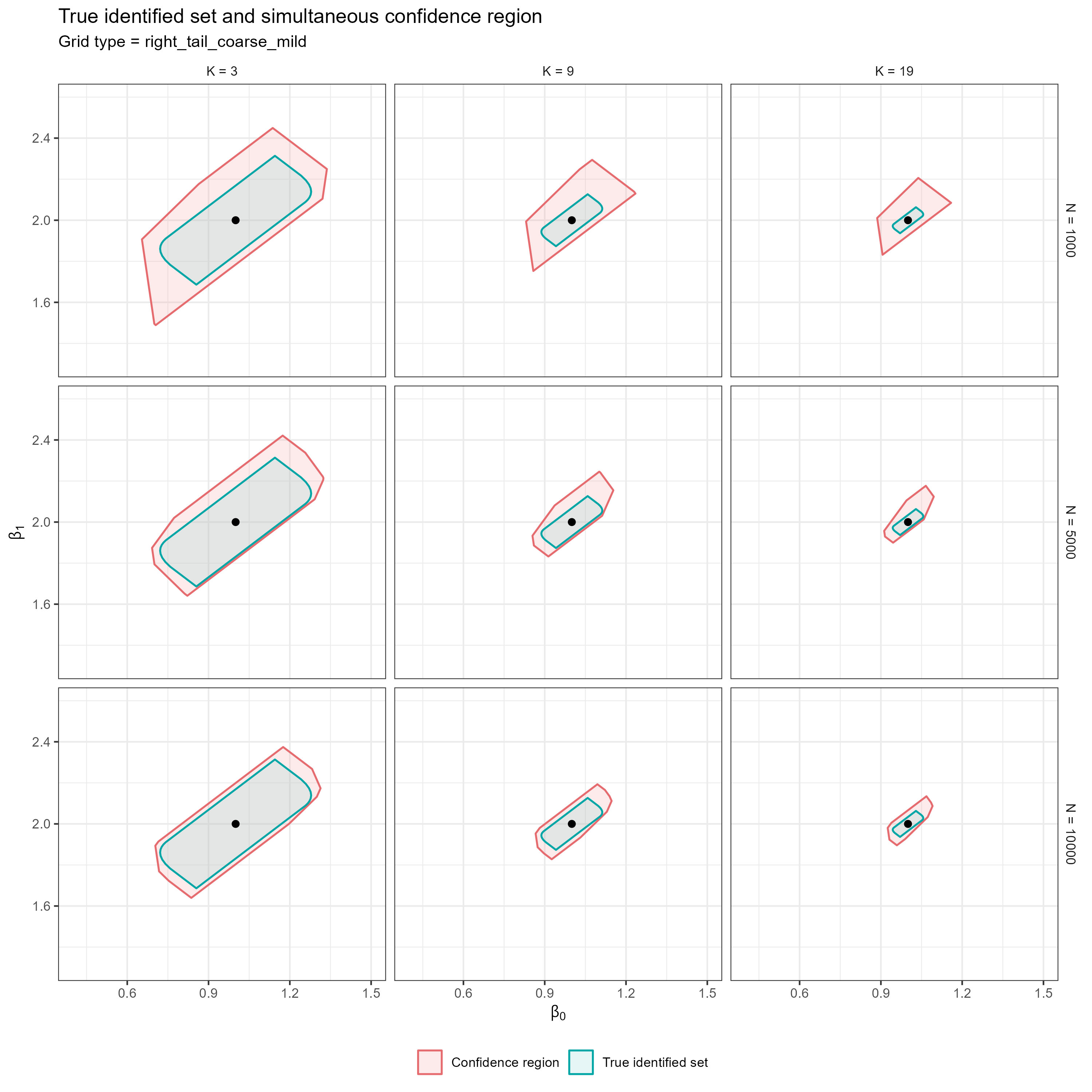}
\caption{\label{fig_simulation_partial_mild}
True identified sets and simultaneous confidence regions under mild right-tail-coarse disclosed quantiles.
}
\end{figure}

\begin{figure}[htbp]
\centering
\includegraphics[width=\linewidth]{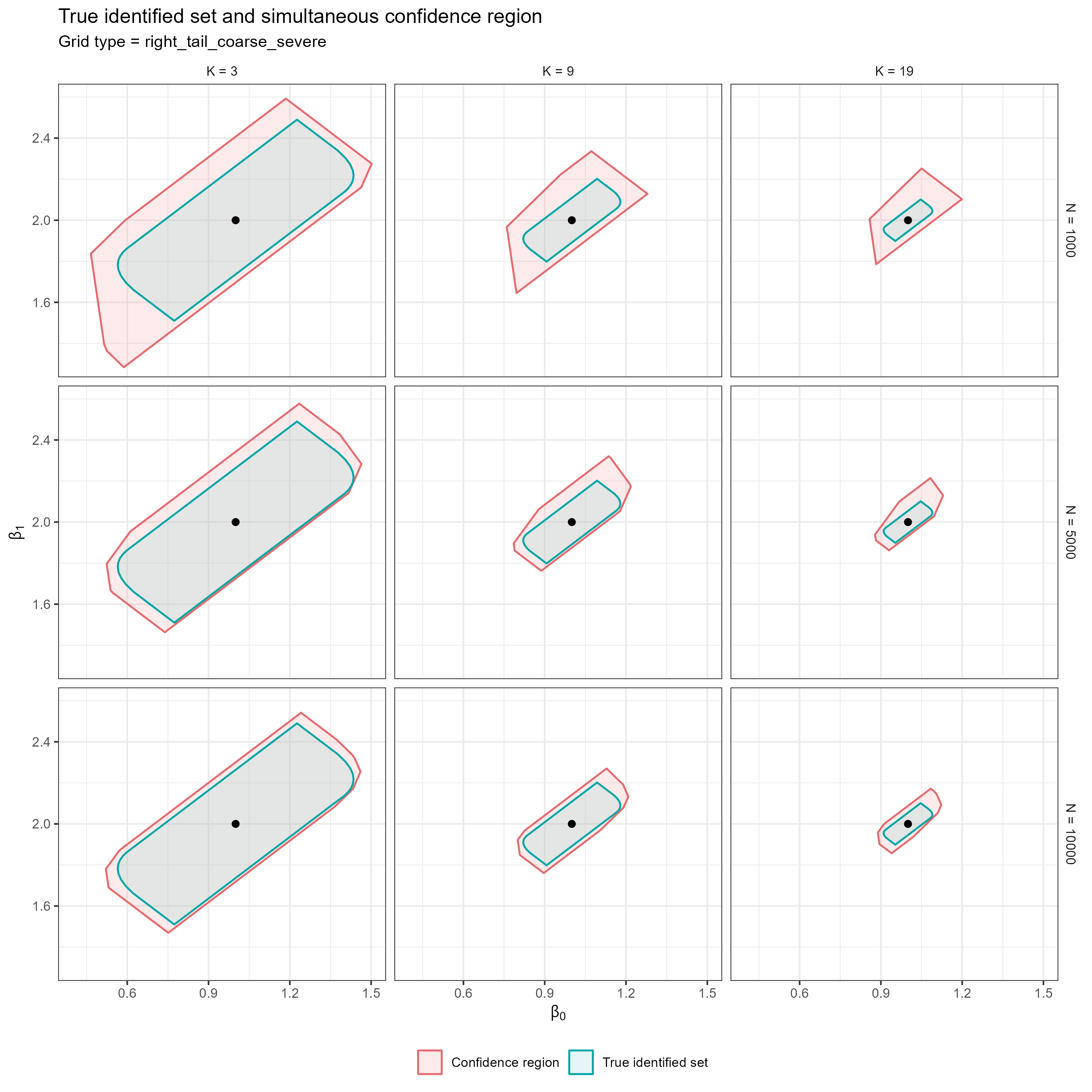}
\caption{\label{fig_simulation_partial_severe}
True identified sets and simultaneous confidence regions under severe right-tail-coarse disclosed quantiles.
}
\end{figure}

Figures~\ref{fig_simulation_partial_mild} and~\ref{fig_simulation_partial_severe}
illustrate the geometry of the identified set and the simultaneous confidence region under uneven quantile disclosure grids. 
The qualitative pattern remains similar to that under evenly spaced grids: the confidence region contracts toward the identified set as the sample size increases.

However, having relatively few disclosed quantiles in the upper tail substantially enlarges both the identified set and the confidence region, particularly when \(K\) is small.
The effect is especially pronounced under the severely right-tail-coarse design with \(K=3\), where the large maximal spacing induces substantial identification uncertainty.

The simulations indicate that evenly spaced grids perform well in the present setting and typically yield tighter identified sets than coarse uneven grids. 
At the same time, the results illustrate that the geometry of the identified set is sensitive to the quantile disclosure scheme, suggesting that the problem of optimal quantile disclosure warrants further theoretical investigation.

\clearpage
\def\thesection{}
\renewcommand{\thesubsection}{E-\arabic{subsection}} 
 \def\theequation{E.\arabic{equation}}
 \def\theassumption{E.\arabic{assumption}}	
 \def\thelemma{E.\arabic{lemma}}
 \def\thetheorem{E.\arabic{theorem}}
\def\theproposition{E.\arabic{proposition}}
\setcounter{proposition}{0}

\renewcommand{\thefigure}{E.\arabic{figure}}
\renewcommand{\thetable}{E.\arabic{table}}

 \makeatletter
 \@addtoreset{equation}{section}
 \@addtoreset{figure}{section}
 \@addtoreset{table}{section}
 \@addtoreset{assumption}{section}
 \makeatother
 \def\thelemma{E.\arabic{lemma}}	  

\setcounter{equation}{0} 
\setcounter{subsection}{0} 
\setcounter{lemma}{0}
\setcounter{theorem}{0}
\setcounter{page}{1}
\section{Supplemental Material E: Inference}
This supplement gives results supporting estimation and simultaneous inference under fixed disclosure in Section~\ref{sec_fixedDC}.

\subsection{Estimation of the Identified Set}
\label{app_set_consistency_statement}

The estimated mean bounds used in Section~\ref{subsec_inference} are
\[
    \widehat{\underline m}_g
    =\sum_{k=0}^{K}(\tau_{k+1}-\tau_k)\widehat Q_g(\tau_k),
    \qquad
    \widehat{\overline m}_g
    =\sum_{k=0}^{K}(\tau_{k+1}-\tau_k)\widehat Q_g(\tau_{k+1}),
\]
where the boundary-node values at $\tau_0=0$ and
$\tau_{K+1}=1$ are set equal to the known support bounds
$\underline y_g$ and $\overline y_g$, respectively.
The following proposition establishes Hausdorff consistency of
$\widehat{\mathcal B}(\mathcal T_K)$.

\begin{proposition}[Consistency of the Estimated Identified Set]
\label{prop_set_consistency}
Suppose that $\Theta\subset\mathbb R^p$ is compact and convex and that
$\mathcal B(\mathcal T_K)$ has nonempty interior.
Suppose further that
$
		\widehat Q_g(\tau_k)
	\xrightarrow{p}
		Q_g(\tau_k)
$
for all $g=1,\dots,G$ and $k=0,\dots,K+1$.
Then
\[
		d_H
		\bigl(
			\widehat{\mathcal B}(\mathcal T_K),
			\mathcal B(\mathcal T_K)
		\bigr)
	\xrightarrow{p}
		0,
\]
where $d_H$ denotes the Hausdorff distance.
\end{proposition}

\begin{proof}
For a point $\bm b\in\mathbb R^p$ and a set $A\subset\mathbb R^p$, let
$
        d(\bm b,A)
        =
        \inf_{\bm a\in A}
        \|\bm b-\bm a\|.
$
The Hausdorff distance is defined by
\[
        d_H(A,B)
        =
        \max
        \left\{
        \sup_{\bm a\in A} d(\bm a,B),
        \sup_{\bm b\in B} d(\bm b,A)
        \right\}.
\]

Let
\[
		\delta_n
	=
		\max_{1\le g\le G}
		\max
		\left\{
			|\widehat{\underline m}_g-\underline m_g|,
			|\widehat{\overline m}_g-\overline m_g|
		\right\}.
\]
Since $K$ and $G$ are fixed and
$
		\widehat Q_g(\tau_k)
	\xrightarrow{p}
		Q_g(\tau_k)
$
for all $g$ and $k$, we have
$
        \delta_n\xrightarrow{p}0.
$

Since $\mathcal B(\mathcal T_K)$ has nonempty interior there exist $\bm\beta^\circ\in\Theta$ and $\eta>0$
such that
\[
        		\underline m_g+\eta
        \le
        		\bm X_g'\bm\beta^\circ
        \le
        		\overline m_g-\eta,
        		\qquad
        		g=1,\dots,G.
\]
Let
$
D=\sup_{\bm\beta,\bm b\in\Theta}\|\bm\beta-\bm b\|
$
denote the diameter of $\Theta$,
which is finite by compactness.

Since $\delta_n\xrightarrow{p}0$ and $\eta>0$ is fixed,
$
P(\delta_n<\eta)\to1
$.
Then, on the event $ \{\delta_n<\eta\}$, define $\lambda_n=\delta_n/\eta$.

First take any $\bm\beta\in\mathcal B(\mathcal T_K)$ and define
\[
        \bm\beta_n
        =
        (1-\lambda_n)\bm\beta
        +
        \lambda_n\bm\beta^\circ .
\]
By convexity of $\Theta$, we have $\bm\beta_n\in\Theta$.
Moreover, for each $g$,
\[
        \bm X_g'\bm\beta_n
        =
        (1-\lambda_n)\bm X_g'\bm\beta
        +
        \lambda_n\bm X_g'\bm\beta^\circ
        \ge
        \underline m_g+\lambda_n\eta
        =
        \underline m_g+\delta_n
        \ge
        \widehat{\underline m}_g,
\]
and similarly,
\[
        \bm X_g'\bm\beta_n
        \le
        \overline m_g-\delta_n
        \le
        \widehat{\overline m}_g .
\]

Hence
$\bm\beta_n\in\widehat{\mathcal B}(\mathcal T_K)$.
Thus
\[
		d\bigl(
			\bm\beta,
			\widehat{\mathcal B}(\mathcal T_K)
		\bigr)
	=
		\inf_{\bm a \in\widehat{\mathcal B}(\mathcal T_K)}
		\|
			\bm\beta- \bm a
		\|
	\le
		\|
			\bm\beta-\bm\beta_n
		\|
	=
		\lambda_{n} \| \bm\beta- \bm \beta^{o} \|
	\le
		\frac{D}{\eta}\delta_n.
\]
which implies
\begin{equation}
		\sup_{\bm\beta\in\mathcal B(\mathcal T_K)}
		d\bigl(
			\bm\beta,
			\widehat{\mathcal B}(\mathcal T_K)
		\bigr)
	\le
		\frac{D}{\eta}\delta_n.
		\label{app_partial_1}
\end{equation}

Conversely, take any
$
\bm\beta^\dagger\in\widehat{\mathcal B}(\mathcal T_K).
$
Define
\[
        \widetilde{\bm\beta}_n
        =
        (1-\lambda_n)\bm\beta^\dagger
        +
        \lambda_n\bm\beta^\circ.
\]
By convexity of $\Theta$, we have $\widetilde{\bm\beta}_n\in\Theta$.
Since
$
\bm\beta^\dagger\in\widehat{\mathcal B}(\mathcal T_K),
$
we have
$
        \bm X_g'\bm\beta^\dagger
        \ge
        \widehat{\underline m}_g
        \ge
        \underline m_g-\delta_n
$
and
$
        \bm X_g'\bm\beta^\dagger
        \le
        \widehat{\overline m}_g
        \le
        \overline m_g+\delta_n.
$
Recall that
$
        \underline m_g+\eta
        \le
        \bm X_g'\bm\beta^\circ
        \le
        \overline m_g-\eta.
$
Then,
\[
\begin{aligned}
		\bm X_g'\widetilde{\bm\beta}_n
	&=
		(1-\lambda_n)\bm X_g'\bm\beta^\dagger
		+
		\lambda_n\bm X_g'\bm\beta^\circ\\
	&\ge
		(1-\lambda_n)(\underline m_g-\delta_n)
		+
		\lambda_n(\underline m_g+\eta)\\
	&=
		\underline m_g
		-(1-\lambda_n)\delta_n
		+\delta_{n}
	\ge
		\underline m_g 
\end{aligned}
\]
and
\[
\begin{aligned}
		\bm X_g'\widetilde{\bm\beta}_n
	&\le
		(1-\lambda_n)(\overline m_g+\delta_n)
		+
		\lambda_n(\overline m_g-\eta)\\
	&=
		\overline m_g
		+(1-\lambda_n)\delta_n
		-\delta_n
	\le
		\overline m_g,
\end{aligned}
\]
which implies
$
        \underline m_g
        \le
        \bm X_g'\widetilde{\bm\beta}_n
        \le
        \overline m_g
$
for
$
        g=1,\dots,G.
$
Hence
$
\widetilde{\bm\beta}_n\in\mathcal B(\mathcal T_K).
$
Therefore,
\[
        d\bigl(
            \bm\beta^\dagger,
            \mathcal B(\mathcal T_K)
        \bigr)
	=
		\inf_{\bm a\in\mathcal B(\mathcal T_K)}
		\|
			\bm\beta^\dagger - \bm a
		\|        
        \le
        \bigl\|
            \bm\beta^\dagger-\widetilde{\bm\beta}_n
        \bigr\|
        =
        \lambda_{n}
        \bigl\|
            \bm\beta^\dagger- \bm\beta^{o}
        \bigr\|        
        \le
        \frac{D}{\eta}\delta_n,
\]
which implies
\begin{equation}
        \sup_{\bm\beta^\dagger\in\widehat{\mathcal B}(\mathcal T_K)}
        d\bigl(
            \bm\beta^\dagger,
            \mathcal B(\mathcal T_K)
        \bigr)
        \le
        \frac{D}{\eta}\delta_n.
        \label{app_partial_2}
\end{equation}

Combining \eqref{app_partial_1} and \eqref{app_partial_2} yields
$
        d_H
        (
            \widehat{\mathcal B}(\mathcal T_K),
            \mathcal B(\mathcal T_K)
        )
        \le
        \frac{D}{\eta}\delta_n .
$
Since
$
\delta_n\xrightarrow{p}0
$,
we obtain
\[
        d_H
        \bigl(
            \widehat{\mathcal B}(\mathcal T_K),
            \mathcal B(\mathcal T_K)
        \bigr)
        \xrightarrow{p}
        0.
\]
\end{proof}

\subsection{Joint Asymptotic Normality for Fixed Quantile Grids}
\label{app_joint_asymptotic_normality}

For each group $g$, write
\[
	        \bm Q_g
        =
	        \left(
        			Q_g(\tau_1),
		        \dots,
		        Q_g(\tau_K)
      		  \right)',
        \qquad
		\widehat{\bm Q}_g
        =
	        \left(
		        \widehat Q_g(\tau_1),
		        \dots,
		        \widehat Q_g(\tau_K)
	        \right)'.
\]

We first state a finite-dimensional joint asymptotic normality result for the disclosed sample quantiles.
The asymptotic normality of a fixed sample quantile is standard; see, for example, Chapter 21 of \citealp{vandervaart98}.
The result follows from the Bahadur representation and the multivariate central limit theorem applied to the vector of indicator functions.

\begin{lemma}[Joint asymptotic normality of disclosed sample quantiles]
\label{lem_joint_quantile_normality}
Fix a group $g$.
Suppose that, for each $k=1,\dots,K$,
\[
        		\widehat Q_g(\tau_k)-Q_g(\tau_k)
        	=
        		\frac{
                		\tau_k-F_{n,g}(Q_g(\tau_k))
       		}{
                		f_g(Q_g(\tau_k))
        		}
        		+
        		o_p(n_g^{-1/2}),
\]
where
$0<f_g(Q_g(\tau_k))<\infty$.
Then
\[
		\sqrt{n_g}
		( \widehat{\bm Q}_g-\bm Q_g)
	\xrightarrow{d}
		\mathcal N(\bm 0,\bm\Sigma_g),
\]
where the $(k,\ell)$ element of $\bm\Sigma_g$ is
\[
        [\bm\Sigma_g]_{k\ell}
        =
        \frac{
                \min(\tau_k,\tau_\ell)-\tau_k\tau_\ell
        }{
                f_g(Q_g(\tau_k))f_g(Q_g(\tau_\ell))
        },
        \qquad
        k,\ell=1,\dots,K.
\]
\end{lemma}

\begin{proof}
By the Bahadur representation,
\[
        \sqrt{n_g}
        (
        \widehat{\bm Q}_g-\bm Q_g
        )
        =
        \bm D_g^{-1}
        \sqrt{n_g}
        \left(
        \bm\tau-\bm F_{n,g}
        \right)
        +
        o_p(1),
\]
where
\[
        \bm D_g
        =
        \mathrm{diag}
        \left(
        f_g(Q_g(\tau_1)),
        \dots,
        f_g(Q_g(\tau_K))
        \right),
\]
\[
        \bm\tau
        =
        (\tau_1,\dots,\tau_K)',
\]
and
\[
        \bm F_{n,g}
        =
        \left(
        F_{n,g}(Q_g(\tau_1)),
        \dots,
        F_{n,g}(Q_g(\tau_K))
        \right)'.
\]

By the multivariate central limit theorem applied to the vector of indicator functions
\[
        \left(
        \mathbf 1\{Y_i\le Q_g(\tau_1)\},
        \dots,
        \mathbf 1\{Y_i\le Q_g(\tau_K)\}
        \right)',
\]
we have
\[
        \sqrt{n_g}
        \left(
        \bm F_{n,g}-\bm\tau
        \right)
        \xrightarrow{d}
        \mathcal N(\bm 0,\bm V_g),
\]
where
\[
        [\bm V_g]_{k\ell}
        =
        \min(\tau_k,\tau_\ell)-\tau_k\tau_\ell.
\]
Therefore,
\[
        \sqrt{n_g}
        (
        \widehat{\bm Q}_g-\bm Q_g
        )
        \xrightarrow{d}
        \mathcal N
        \left(
        \bm 0,
        \bm D_g^{-1}\bm V_g\bm D_g^{-1}
        \right).
\]
Thus
$
\bm\Sigma_g=\bm D_g^{-1}\bm V_g\bm D_g^{-1}
$,
which gives the stated covariance formula.
\end{proof}

Using Lemma \ref{lem_joint_quantile_normality}, we can show the joint asymptotic normality of estimated mean bounds.

\begin{lemma}[Joint asymptotic normality of estimated mean bounds]
\label{lem_joint_bound_normality}
Suppose that the conditions of Lemma~\ref{lem_joint_quantile_normality}
hold for each $g=1,\dots,G$.
Assume further that
$
		n_g/N
	    \xrightarrow{p}
		\pi_g\in(0,1)
$
for each $g$.
Then
\[
        \sqrt N
        (
        \widehat{\bm s}-\bm s
        )
        \xrightarrow{d}
        \mathcal N(\bm 0,\bm\Omega),
\]
where
$
        \bm\Omega
        =
        \mathrm{diag}
        (
        \bm\Omega_1,\dots,\bm\Omega_G
        ),
$
and
\[
		\bm\Omega_g
	=
		\pi_g^{-1}
		\begin{pmatrix}
			(\bm a^L)'\bm\Sigma_g\bm a^L
			&
			(\bm a^L)'\bm\Sigma_g\bm a^U
			\\
			(\bm a^U)'\bm\Sigma_g\bm a^L
			&
			(\bm a^U)'\bm\Sigma_g\bm a^U
\end{pmatrix}.
\]
with
$
		\bm a^L
	=
		(
			\tau_2-\tau_1,
			\tau_3-\tau_2,
			\dots,
			\tau_{K+1}-\tau_K
		)'
$
and
$
		\bm a^U
	=
		(
			\tau_1-\tau_0,
			\tau_2-\tau_1,
			\dots,
			\tau_K-\tau_{K-1}
		)'
$.
\end{lemma}
\begin{proof}
Recall that we write $\bm X_i\in g$ as shorthand for $\bm X_i\in\mathcal G_g$.
Define
\[
		    \bm J_g(Y_i)
	    =
		    \begin{pmatrix}
		        1\{Y_i\leq Q_g(\tau_1)\}\\
		        \vdots\\
		        1\{Y_i\leq Q_g(\tau_K)\}
		    \end{pmatrix},
	   \qquad
    		\bm\tau
	    =
		    (\tau_1,\ldots,\tau_K)',
\]
and let
$
    \bm D_g
    =
    \operatorname{diag}
    \left(
        f_g(Q_g(\tau_1)),\ldots,
        f_g(Q_g(\tau_K))
    \right).
$

By the Bahadur representation in Lemma~\ref{lem_joint_quantile_normality},
\[
		    \widehat{\bm Q}_g-\bm Q_g
	    =
		    \bm D_g^{-1}
		    \left(
		        \bm\tau-\bm F_{n,g}
		    \right)
		    +
		    o_p(n_g^{-1/2}),
\]
where
$
		    \bm F_{n,g}
	    =
		    \frac{1}{n_g}
		    \sum_{i=1}^N
		    \mathbf 1\{ \bm X_i\in g\}\bm J_g(Y_i).
$
Since
$
  		  n_g
	    =
		    \sum_{i=1}^N \mathbf 1\{\bm X_i\in g\},
$
we have
\[
		    \bm\tau-\bm F_{n,g}
	    =
		    \frac{1}{n_g}
		    \sum_{i=1}^N
		    \mathbf 1\{\bm X_i\in g\}
		    \left\{
		        \bm\tau-\bm J_g(Y_i)
		    \right\}.
\]
Therefore,
\[
\begin{aligned}
		    \sqrt N
		    (\widehat{\bm Q}_g-\bm Q_g)
	    &=
		    \bm D_g^{-1}
		    \frac{N}{n_g}
		    \frac{1}{\sqrt N}
		    \sum_{i=1}^N
		    \mathbf 1\{\bm X_i\in g\}
		    \left\{
		        \bm\tau-\bm J_g(Y_i)
		    \right\}
	    +
		    o_p(1)
	    \\
	    &=
		    \frac{1}{\sqrt N}
		    \sum_{i=1}^N
		    \bm\phi_g(Y_i, \bm X_i)
	    +
		    o_p(1),
\end{aligned}
\]
where
\[
    \bm\phi_g(Y_i, \bm X_i)
    =
    \frac{\mathbf 1\{\bm X_i\in g\}}{\pi_g}
    \bm D_g^{-1}
    \left\{
        \bm\tau-\bm J_g(Y_i)
    \right\},
\]
and the second equality follows from
$n_g/N \xrightarrow{p} \pi_g$.

Since
$
    E\!\left[
        \bm J_g(Y_i)
        \mid
        \bm X_i\in\mathcal G_g
    \right]
    =
    \bm\tau,
$
we have
$
	    E[  \bm\phi_g(Y_i,\bm X_i)]
    =
	    \bm 0
$
and
\[
\begin{aligned}
	    E[
	        \bm\phi_g(Y_i,\bm X_i)
	        \bm\phi_g(Y_i,\bm X_i)'
	    ]
    &=
	    \pi_g^{-1}
	    \bm D_g^{-1}
	    \Var
	    \left(
	        \bm J_g(Y_i)
	        \mid
	        \bm X_i\in\mathcal G_g
	    \right)
	    \bm D_g^{-1}
    \\
    &=
	    \pi_g^{-1}\bm\Sigma_g
\end{aligned}
\]
for each $g$.
For two distinct groups $g\neq h$, the maintained partition structure implies
$
    \mathbf 1\{\bm X_i\in g\}
    \mathbf 1\{\bm X_i\in h\}
    =
    0
$
for every $i$.
Consequently, for $g\neq h$,
$
    E[
        \bm\phi_g(Y_i, \bm X_i)
        \bm\phi_h(Y_i, \bm X_i)'
    ]
    =
    \bm 0.
$
Thus, stacking the group-specific influence functions gives
\[
	    \sqrt N
	    \begin{pmatrix}
	        \widehat{\bm Q}_1-\bm Q_1\\
	        \vdots\\
	        \widehat{\bm Q}_G-\bm Q_G
	    \end{pmatrix}
	    \overset{d}{\longrightarrow}
	    N
	    \left(
	        \bm 0,
	        \operatorname{diag}
	        \left(
	            \pi_1^{-1}\bm\Sigma_1,\ldots,
	            \pi_G^{-1}\bm\Sigma_G
	        \right)
   	 \right)
\]
by the multivariate central limit theorem.

For each group $g$, since the support bounds
$\underline y_g$ and $\overline y_g$ are known constants and
therefore do not contribute to sampling variation,
\[
    \begin{pmatrix}
        \widehat{\underline m}_g-\underline m_g\\
        \widehat{\overline m}_g-\overline m_g
    \end{pmatrix}
    =
    \begin{pmatrix}
        (\bm a^L)'\\
        (\bm a^U)'
    \end{pmatrix}
    (\widehat{\bm Q}_g-\bm Q_g).
\]
Hence, stacking across groups and applying this linear transformation
to the preceding joint limit yields
\[
    \sqrt N(\widehat{\bm s}-\bm s)
    \overset{d}{\longrightarrow}
    N(\bm 0,\bm\Omega).
\]
\end{proof}

The block-diagonal covariance does not require finite-sample independence of the group-specific quantile estimators. 
It follows from the mutually exclusive group indicators in the full-sample influence functions, which imply zero cross-group covariances.

\subsection{Asymptotic Validity of the Confidence Region}
\label{app_CR}

Let $q_{1-\alpha}$ denote the $1-\alpha$ quantile of
\[
        \max_{1\le j\le 2G}
        \left|
        Z_j/\omega_j
        \right|,
        \qquad
        \bm Z\sim\mathcal N(\bm 0,\bm\Omega),
\]
where $\omega_j^2$ denotes the $j$th diagonal element of $\bm\Omega.$
$
\hat q_{1-\alpha}
$
is the feasible analog defined in Section~\ref{subsec_inference}.

\begin{theorem}[Coverage of the simultaneous confidence region]
\label{thm_confidence_region_coverage}
Suppose that the conditions of Lemma~\ref{lem_joint_bound_normality} hold.
Assume further that
$
		\widehat{\bm\Omega}
	\xrightarrow{p}
		\bm\Omega
$
and that the diagonal elements of $\bm\Omega$ are positive.
Assume that the distribution of
$
        \max_{1\le j\le 2G}
        \left|
        Z_j/\omega_j
        \right|
$
is continuous at $q_{1-\alpha}$.
Then
\[
        \liminf_{N\to\infty}
        P\left(
        \mathcal B(\mathcal T_K)
        \subseteq
        \widehat{\mathcal C}_{1-\alpha}
        \right)
        \ge
        1-\alpha .
\]
\end{theorem}

\begin{proof}
Write
\[
        \bm s
        =
        (s_1,\dots,s_{2G})'
        =
        (
        \underline m_1,
        \overline m_1,
        \dots,
        \underline m_G,
        \overline m_G
        )',
\]
and
\[
        \widehat{\bm s}
        =
        (\widehat s_1,\dots,\widehat s_{2G})'
        =
        (
        \widehat{\underline m}_1,
        \widehat{\overline m}_1,
        \dots,
        \widehat{\underline m}_G,
        \widehat{\overline m}_G
        )'.
\]
Define
\[
        T_N
        =
        \max_{1\le j\le 2G}
        \left|
        \frac{
        \sqrt N(\widehat s_j-s_j)
        }{
        \widehat\omega_j
        }
        \right|.
\]

By Lemma~\ref{lem_joint_bound_normality},
\[
        \sqrt N
        (
        \widehat{\bm s}-\bm s
        )
        \xrightarrow{d}
        \mathcal N(\bm 0,\bm\Omega).
\]

Together with
$
		\widehat{\bm\Omega}
	\xrightarrow{p}
		\bm\Omega
$
and the positivity of the diagonal elements of $\bm\Omega$,
the continuous mapping theorem and Slutsky's theorem imply that
\[
        T_N
        \xrightarrow{d}
        T
        =
        \max_{1\le j\le 2G}
        \left|
        Z_j/\omega_j
        \right|,
        \qquad
        \bm Z\sim\mathcal N(\bm 0,\bm\Omega).
\]

Since the distribution of $T$
is continuous at $q_{1-\alpha}$,
\[
        	P(T_N\le q_{1-\alpha})
        \to
        	1-\alpha .
\]

Moreover, since
$
		\widehat{\bm\Omega}
	\xrightarrow{p}
		\bm\Omega
$,
the Gaussian distributions
$
		\mathcal N(\bm0,\widehat{\bm\Omega})
$
converge weakly to
$
		\mathcal N(\bm0,\bm\Omega).
$
By continuity of the limiting distribution at $q_{1-\alpha}$,
the corresponding quantiles satisfy
\[
		\hat q_{1-\alpha}
	\xrightarrow{p}
		q_{1-\alpha}.
\]
Hence, by the convergence $T_N\xrightarrow{d}T$
and
$
		\hat q_{1-\alpha}
	\xrightarrow{p}
		q_{1-\alpha},
$
together with the continuity of the distribution of $T$ at $q_{1-\alpha}$,
\[
		P(T_N\le \hat q_{1-\alpha})
	\to
		1-\alpha.
\]

On the event
$
\{T_N\le \hat q_{1-\alpha}\},
$
for each
$
g=1,\dots,G
$
we have
\[
        \widehat{\underline m}_g
        -
        \hat q_{1-\alpha}
        \frac{\widehat\omega_{2g-1}}{\sqrt N}
        \le
        \underline m_g
\]
and
\[
        \overline m_g
        \le
        \widehat{\overline m}_g
        +
        \hat q_{1-\alpha}
        \frac{\widehat\omega_{2g}}{\sqrt N}.
\]

Now take any
$
\bm\beta\in\mathcal B(\mathcal T_K).
$
By definition,
$
        \underline m_g
        \le
        \bm X_g'\bm\beta
        \le
        \overline m_g,
$
for all $g=1,\dots,G.$
Combining the preceding inequalities gives
\[
        \widehat{\underline m}_g
        -
        \hat q_{1-\alpha}
        \frac{\widehat\omega_{2g-1}}{\sqrt N}
        \le
        \bm X_g'\bm\beta
        \le
        \widehat{\overline m}_g
        +
        \hat q_{1-\alpha}
        \frac{\widehat\omega_{2g}}{\sqrt N}
\]
for all
$g=1,\dots,G$.
Therefore
$
\bm\beta\in\widehat{\mathcal C}_{1-\alpha}.
$

Since this holds for every
$
		\bm\beta
	\in
		\mathcal B(\mathcal T_K)
$,
we have
\[
        \mathcal B(\mathcal T_K)
        \subseteq
        \widehat{\mathcal C}_{1-\alpha}
\]
on the event $\{T_N\le \hat q_{1-\alpha}\}$.
Hence
\[
        P\left(
        \mathcal B(\mathcal T_K)
        \subseteq
        \widehat{\mathcal C}_{1-\alpha}
        \right)
        \ge
        P(T_N\le \hat q_{1-\alpha})
        \to
        1-\alpha.
\]
Therefore,
\[
        \liminf_{N\to\infty}
        P\left(
        \mathcal B(\mathcal T_K)
        \subseteq
        \widehat{\mathcal C}_{1-\alpha}
        \right)
        \ge
        1-\alpha .
\]
\end{proof}

The consistency condition 
$
		\widehat{\bm\Omega}
	\xrightarrow{p}
		\bm\Omega
$
in Theorem~\ref{thm_confidence_region_coverage} can be satisfied when the data provider additionally releases consistent estimates of the asymptotic covariance matrices $\bm\Sigma_g$ of the disclosed sample quantiles, or of the local densities $f_g(Q_g(\tau_k))$.
These estimates may be computed from the underlying microdata without releasing those data.
Together with $n_g/N \xrightarrow{p} \pi_g$, they yield a consistent estimator of $\bm\Omega$ through the covariance formula above.

The plug-in estimator $\widehat{\bm\Omega}$ constructed using the finite-difference estimator $\hat f_g(\widehat Q_g(\tau_k))$ is also consistent in special cases where the finite-difference formulas,
evaluated at the population quantiles, yield the exact derivatives $Q_g'(\tau_k)$ at all disclosed quantile indices.
This occurs, for example, when each population quantile function $Q_g(\tau)$ is linear in $\tau$ on all intervals used for differencing.
In general, however, this estimator is a plug-in approximation whose consistency is not guaranteed under fixed quantile disclosure.

Another possibility is to estimate $\bm\Omega$ using additional sample quantiles disclosed in shrinking neighborhoods of the fixed quantile indices $\tau_1,\ldots,\tau_K$.
For example, suppose that $\widehat Q_g(\tau_k-h_N)$ and $\widehat Q_g(\tau_k+h_N)$ are available for each $g$ and $k$. 
With a suitably chosen sequence $h_N\to0$ and under appropriate regularity conditions, these additional quantiles may allow consistent estimation of $Q_g'(\tau_k)$ using local finite differences.
This would provide a way to construct a consistent estimator of $\bm\Omega$ without requiring the disclosure grid to become dense over the entire unit interval.

\clearpage
\def\thesection{}
\renewcommand{\thesubsection}{F-\arabic{subsection}} 
 \def\theequation{F.\arabic{equation}}
 \def\theassumption{F.\arabic{assumption}}	
 \def\thelemma{F.\arabic{lemma}}
\renewcommand{\thefigure}{F.\arabic{figure}}
\renewcommand{\thetable}{F.\arabic{table}}

 \makeatletter
 \@addtoreset{equation}{section}
 \@addtoreset{figure}{section}
 \@addtoreset{table}{section}
 \@addtoreset{assumption}{section}
 \makeatother
 \def\thelemma{F.\arabic{lemma}}	  

\setcounter{equation}{0} 
\setcounter{subsection}{0} 
\setcounter{lemma}{0}
\setcounter{page}{1}
\section{Supplemental Material F: Details and Additional Results for the Children in Need Application}\label{app_emp}

\subsection{Data Construction and Year Matching}

To construct the measure of political control, we combine the DfE
assessment data with annual council-composition records obtained from
electionresults.uk and based on Open Council Data.\footnote{See
\url{https://electionresults.uk/councils/data}. Version information is
provided in the replication package.}
The DfE outcome data cover reporting years 2017--2025, with each reporting year running from 1 April of the preceding calendar year through 31 March of the reporting year.
A council-composition snapshot for a given calendar year is matched to the DfE reporting year ending in the following calendar year.
Thus, for example, the 2024 council-composition snapshot is matched to the DfE 2025 reporting year, which covers 1 April 2024 through 31 March 2025.
The council-composition measure is therefore an annual snapshot rather than a measure of political composition throughout the entire DfE reporting year.
The resulting matched data include years without a scheduled local election as well as election years.

\subsection{Disclosed Assessment-duration Quantiles}\label{app_CNA_quantile}

\begin{figure}[!htbp]
\centering
\includegraphics[width=\textwidth]{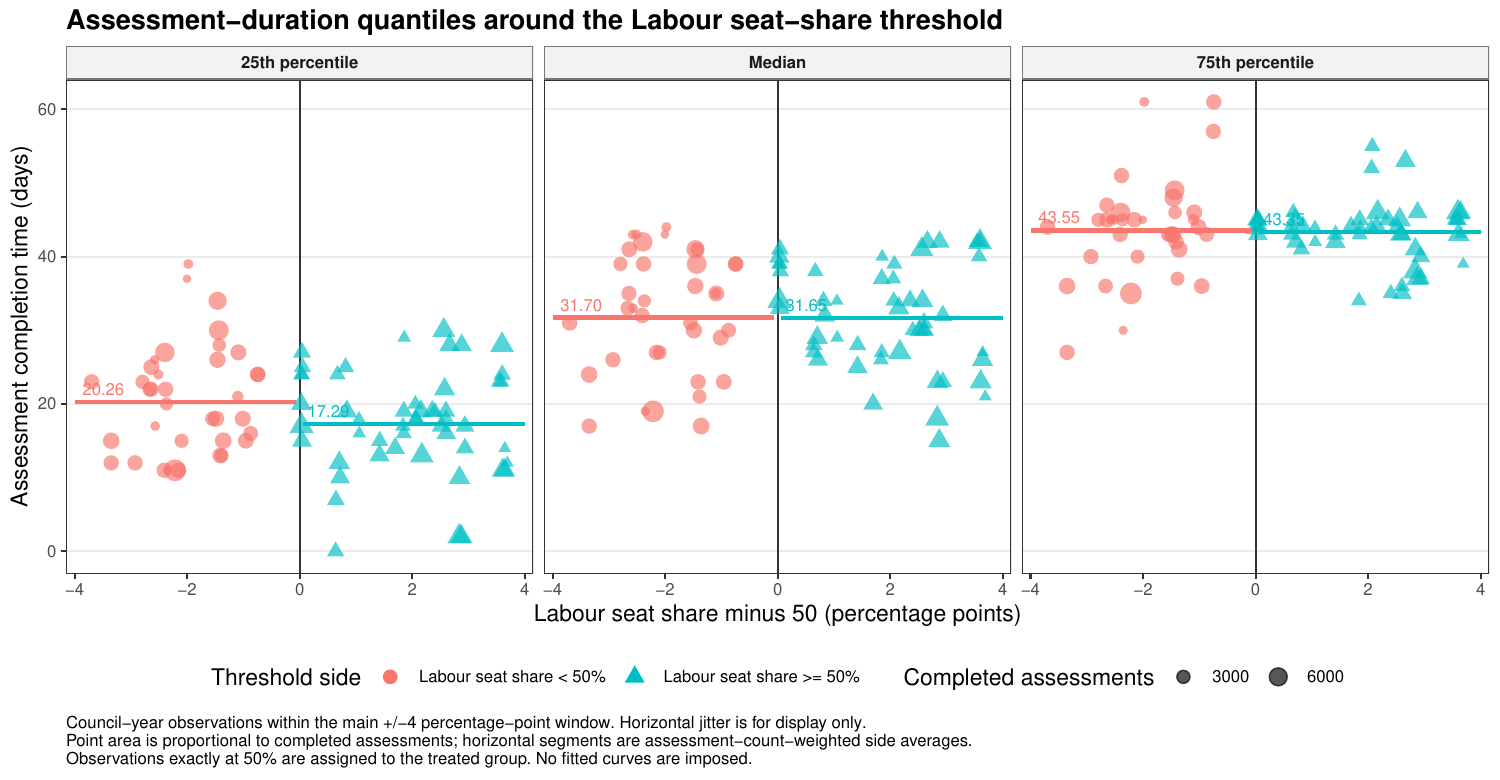}
\caption{Disclosed assessment-duration quantiles around the Labour seat-share threshold.}
\label{fig:cin_quantiles_appendix}
\end{figure}

The panels in Figure~\ref{fig:cin_quantiles_appendix}  report the 25th percentile, median, and 75th percentile for matched council-by-DfE-reporting-year observations within the primary
\(h=4\) percentage-point window. 
Point area is proportional to the number of completed assessments. 
Horizontal segments show assessment-count-weighted averages of the reported quantiles on each side of the threshold.
A small horizontal jitter is used only to reveal overlapping council-year observations.

The 25th percentile panel shows a decline in the weighted side average from 20.26 days below the threshold to 17.29 days at or above it. 
In contrast, the corresponding differences are small for the median, at 31.70 versus 31.65 days, and for 75th percentile, at 43.55 versus 43.35 days. 
Under the reconstruction weights, the 25th percentile difference contributes \(-0.743\) days to the total \(-0.830\)-day difference. 
The negative reconstructed-mean estimate therefore appears to be associated primarily with the lower disclosed quartile rather than a general downward shift in all three disclosed quantiles.

The panels do not estimate individual-level quantile treatment effects. The differences in the reported quantiles across the two sides of the threshold are comparisons of council-year-level summary statistics and should not be interpreted as discontinuities in the quantile function of individual assessment durations.

\subsection{Treatment-definition sensitivity}

\input{QOR_Replication/empirical_illustration/UK_election/output/paper_tables/table_strict_majority_sensitivity.tex}

The primary specification follows \cite{alonso20} by assigning observations with a Labour seat share of exactly 50 percent to the treated group. Because exactly half of the seats does not constitute a strict majority in the institutional sense, Table~\ref{tab:qonlyreg_strict_majority} instead defines treatment as
\[
D_{g}^{\mathrm{strict}}
=
\mathbf 1\{R_{g}>0\}.
\]
There are seven observations exactly at the threshold in each of the reported fixed-window samples.

Under the strict definition, the main-window sample is divided evenly between 44 treated and 44 control observations. The point estimate becomes \(-1.504\) days, compared with \(-0.830\) under the primary definition. 
The estimates remain negative in the \(h=3\) and \(h=5\) windows. 
The negative sign of the point estimate is therefore not attributable to assigning the exactly-50-percent observations to treatment, although its magnitude is sensitive to this assignment.
The projections of the simultaneous confidence regions continue to contain zero under every upper support  bound.

\subsection{Leave-one-council-out sensitivity}

\input{QOR_Replication/empirical_illustration/UK_election/output/paper_tables/table_leave_one_council_out.tex}

Because the primary window contains repeated observations from only 36 councils and the estimation criterion weights observations by completed assessment counts, councils with many assessments can have substantial influence. Table~\ref{tab:qonlyreg_leave_one_council_out} therefore re-estimates the primary specification after removing all observations from one council at a time. The table reports the baseline and the five exclusions that produce the largest absolute changes; complete results for all 36 councils are provided in the replication files.

The leave-one-council-out estimates range from \(-1.405\) to \(0.107\) days. They remain negative in 35 of the 36 exclusions. The exception is Bristol: removing its five council-year observations, representing 22,290 completed assessments, changes the estimate from \(-0.830\) to \(0.107\) days. Removing Kirklees or Croydon also moves the estimate close to zero, to \(-0.024\) and \(-0.045\) days, respectively.

Consequently, the negative sign is preserved in 35 of the 36 single-council exclusions. The magnitude is sensitive to several influential councils, whereas the sign changes only when Bristol is
excluded.

\subsection{Reporting-year fixed effects}

The primary analysis pools council-year observations from multiple reporting years, but the baseline specification does not include reporting-year fixed effects to absorb common shocks specific to each reporting year.
The baseline specification includes only an intercept and the threshold indicator, so that \(\delta_h\) equals the assessment-count-weighted difference in reconstructed means across the threshold.
Assessment duration and the distribution of observations across the two sides of the threshold may nevertheless change over time. 
Let $t_g$ denote the reporting year of group $g$.
As a sensitivity analysis, we augment the group-mean model
with reporting-year indicators:
\[
    m_g
    =
    \alpha+\delta_hD_g
    +\sum_{s=2018}^{2025}
      \lambda_s\mathbf 1\{t_g=s\},
    \qquad g\in\mathcal G_h,
\]
where reporting year 2017 is the omitted category. The fixed windows, assessment-count weights, quantile reconstruction, and treatment definition are otherwise unchanged. Because the augmented model remains linear in the coefficient vector, both the point estimator and the simultaneous confidence-region procedure apply without modification.

\input{QOR_Replication/empirical_illustration/UK_election/output/paper_tables/table_year_fixed_effects_sensitivity.tex}

Table~\ref{tab:qonlyreg_year_fixed_effects} reports the results. Adding reporting-year fixed effects makes the point estimates more negative in all three windows. The estimates are \(-2.434\), \(-1.698\), and \(-3.361\) days in the \(h=3\), \(h=4\), and \(h=5\) windows, respectively, compared with \(-1.638\), \(-0.830\), and \(-2.045\) days in the specifications without year fixed effects. In the primary \(h=4\) window, the standard error is \(0.067\).

The change in magnitude is consistent with differences in the temporal composition of the two threshold sides. 
The reconstructed mean assessment duration generally increases over the sample period, while observations at or above the Labour threshold are relatively more prevalent in later reporting years.
Without year fixed effects, this temporal composition raises the weighted mean on the treated side and attenuates the negative point estimate. The year indicators instead compare treated and control observations after allowing for common reporting-year shifts.

The projections of the simultaneous confidence region remain wide. For the primary window, they are 
$[-25.09,22.05]$, $[-31.34,28.30]$, and $[-37.59,34.55]$ under upper support bounds of \(150\), \(175\), and \(200\) days, respectively. Every projected region contains zero. Thus, accounting for common reporting-year variation strengthens the negative point estimate but does not alter the fixed-disclosure inference: the simultaneous confidence region does not exclude a zero effect.

Reporting-year fixed effects absorb common temporal changes in assessment duration, including changes in national conditions or reporting practices. They do not account for dependence between repeated observations from the same council. The latter remains a separate limitation of the empirical illustration.

\clearpage
\def\thesection{}
\renewcommand{\thesubsection}{G-\arabic{subsection}} 
 \def\theequation{G.\arabic{equation}}
 \def\theassumption{G.\arabic{assumption}}	
 \def\thelemma{G.\arabic{lemma}}
\renewcommand{\thefigure}{G.\arabic{figure}}
\renewcommand{\thetable}{G.\arabic{table}}

 \makeatletter
 \@addtoreset{equation}{section}
 \@addtoreset{figure}{section}
 \@addtoreset{table}{section}
 \@addtoreset{assumption}{section}
 \makeatother
 \def\thelemma{G.\arabic{lemma}}	  

\setcounter{equation}{0} 
\setcounter{subsection}{0} 
\setcounter{lemma}{0}
\setcounter{page}{1}
\section{Supplemental Material G: Relation to an Individual-Level Linear Regression Model}
\label{app_individual_level_model}

The linear restriction imposed on the group-level means can arise from a
conventional individual-level linear conditional mean model as a special
case. To see this, suppose that individual-level outcomes and covariates
satisfy
\[
    E(Y \mid \bm X)
    =
    \bm X'\bm\beta_0.
\]
For a given group \(g\), the law of iterated expectations then gives
\begin{align*}
    m_g
    &=
    E(Y \mid \bm X \in g)
    =
    E\left[
        E(Y \mid \bm X)
        \mid
        \bm X \in g
    \right]
    =
    E(\bm X \mid \bm X \in g)'\bm\beta_0.
\end{align*}
Therefore, if the group-level covariate vector consists of the group means, that is,
$
    \bm X_g
    =
    E(\bm X \mid \bm X \in g),
$
the individual-level linear conditional mean model implies
\[
    m_g
    =
    \bm X_g'\bm\beta_0,
\]
which is the group-level restriction considered in the main text.

A particularly simple case arises when the covariates entering the
regression are constant within each group. If
$
    \bm X = \bm X_g
$
for all individuals in group $g$,
then
$
    E(\bm X \mid \bm X \in g)
    =
    \bm X_g,
$
and hence the same group-level restriction follows immediately from the
individual-level model.

The converse, however, does not generally hold. The restriction
\[
    E(Y \mid \bm X \in g)
    =
    \bm X_g'\bm\beta_0,
    \qquad
    g=1,\ldots,G,
\]
only restricts the means of the outcome across the specified groups and
does not require the individual-level conditional mean
\(E(Y\mid\bm X)\) to be linear in \(\bm X\). For example, suppose
\[
    E(Y\mid\bm X)
    =
    \bm X'\bm\beta_0 + h(\bm X),
\]
where
\[
    E\{h(\bm X)\mid \bm X\in g\}
    =
    0,
    \qquad
    g=1,\ldots,G.
\]
Then
\[
    E(Y\mid\bm X\in g)
    =
    E(\bm X\mid\bm X\in g)'\bm\beta_0
\]
for every group, even though the individual-level conditional mean need
not be linear.

Thus, the group-level specification used in this paper does not require
an underlying individual-level linear regression model. Rather, the
group-level mean restriction is imposed directly and encompasses the
model obtained by aggregating a conventional individual-level linear
conditional mean model as a special case.

\clearpage
\def\thesection{}
\renewcommand{\thesubsection}{H-\arabic{subsection}} 
 \def\theequation{H.\arabic{equation}}
 \def\theassumption{H.\arabic{assumption}}	
 \def\thelemma{H.\arabic{lemma}}
  \def\theproposition{H.\arabic{proposition}}
\renewcommand{\thefigure}{H.\arabic{figure}}
\renewcommand{\thetable}{H.\arabic{table}}

 \makeatletter
 \@addtoreset{equation}{section}
 \@addtoreset{figure}{section}
 \@addtoreset{table}{section}
 \@addtoreset{assumption}{section}
 \makeatother
 \def\thelemma{H.\arabic{lemma}}	  

\setcounter{equation}{0} 
\setcounter{subsection}{0} 
\setcounter{lemma}{0}
\setcounter{proposition}{0}
\setcounter{page}{1}
\section{Supplemental Material H: Alternative Density-Free Variance Estimation}

The feasible variance estimator in Section~\ref{subsec_variance} estimates
$f_g(Q_g(\tau_k))$ using finite differences of adjacent
disclosed quantiles. This approach requires a minimum-spacing
condition to control the sampling error in the resulting
derivative estimator. In this subsection, we present an
alternative variance estimator that avoids density estimation.

Recall from Theorem~\ref{thm_bias_clt} and derivation in Section~\ref{subsec_variance} that
\[
    \bm V_\beta
    =
    \bm\Sigma_X^{-1}
    \left[
        \sum_{g=1}^G
        \pi_g
        \bm X_g\bm X_g'
        \Var
        \left(
            Y_i-m_g
            \mid
            \bm X_i\in g
        \right)
    \right]
    \bm\Sigma_X^{-1}.
\]
Thus, feasible estimation of $\bm V_\beta$ can alternatively
be based on reconstruction of the group-specific conditional
variance.

For each group $g$, define the conditional second moment
\[
		    s_{2g}
	    =
		    E
		    \left(
		        Y_i^2
		        \mid
		        \bm X_i\in g
		    \right)
	    =
		    \int_0^1 Q_g(\tau)^2\,d\tau.
\]
Using the same reconstructed quantile function
$\widetilde Q_g(\tau)$ as in the point estimator, define
\[
    \hat s_{2g}
    =
    \int_0^1
    \widetilde Q_g(\tau)^2\,d\tau.
\]
Writing
$
    q_{gk}
    =
    \widehat Q_g(\tau_k),
$
for
$
    k=1,\ldots,K_N,
$
the estimator has the closed-form expression
\begin{align*}
		    \hat s_{2g}
	    =&
		    \tau_1q_{g1}^2
		    +
		    \sum_{k=1}^{K_N-1}
		    \frac{\tau_{k+1}-\tau_k}{3}
		    \left(
		        q_{gk}^2
		        +
		        q_{gk}q_{g,k+1}
		        +
		        q_{g,k+1}^2
		    \right)
		    +
		    (1-\tau_{K_N})q_{g,K_N}^2.
\end{align*}

We then define the reconstructed conditional variance by
\begin{equation}
		    \hat\sigma_{g,\mathrm{alt}}^2
	    =
		    \hat s_{2g}
		    -
		    \hat m_g^2
	=
		    \int_0^1
		    \left\{
		        \widetilde Q_g(\tau)-\hat m_g
		    \right\}^2
		    d\tau.	
		        \label{eq_alt_conditional_variance_integral}
\end{equation}
The representation in
\eqref{eq_alt_conditional_variance_integral} makes clear that
$
    \hat\sigma_{g,\mathrm{alt}}^2
    \geq 0
$
for every sample.

Equivalently, let
$
    d_{gk}
    =
    q_{gk}-\hat m_g.
$
Then the estimator in \eqref{eq_alt_conditional_variance_integral} can also be written as
\begin{align*}
    		\hat\sigma_{g,\mathrm{alt}}^2
	    =&
		    \tau_1d_{g1}^2
		+
		    \sum_{k=1}^{K_N-1}
		    \frac{\tau_{k+1}-\tau_k}{3}
		    \left(
		        d_{gk}^2
		        +
		        d_{gk}d_{g,k+1}
		        +
		        d_{g,k+1}^2
		    \right)
		+
		    (1-\tau_{K_N})d_{g,K_N}^2.
\end{align*}

Define the alternative estimator of the sampling variance of
$\hat m_g$ by
\begin{equation}
		    \widehat{\Var}_{\mathrm{alt}}(\hat m_g)
	    =
		    \frac{
		        \hat\sigma_{g,\mathrm{alt}}^2
		    }{
		        n_g
		    }.
    \label{eq_alt_mean_variance}
\end{equation}
In analogy with the variance estimator in Section~\ref{subsec_variance}, define
the alternative plug-in estimator of the variance of
$\hat{\bm\beta}$ by
\begin{align}
		    \widehat{\bm V}_{\beta,\mathrm{alt}}
	    =&
		    \left(
		        \sum_{g=1}^G
		        n_g\bm X_g\bm X_g'
		    \right)^{-1}
		    \left(
		        \sum_{g=1}^G
		        n_g^2
		        \bm X_g\bm X_g'
		        \widehat{\Var}_{\mathrm{alt}}(\hat m_g)
		    \right)
		    \left(
		        \sum_{g=1}^G
		        n_g\bm X_g\bm X_g'
		    \right)^{-1}.
    \label{eq_alt_beta_variance}
\end{align}

The following proposition establishes consistency of the
alternative variance estimator.

\begin{proposition}[Consistency of the alternative variance estimator]
\label{prop_alt_variance}
Under the conditions of Proposition~\ref{prop_rootN} with Assumption~\ref{ass_smooth}(ii),
for each $g=1,\ldots,G$,
\[
    n_g
    \widehat{\Var}_{\mathrm{alt}}(\hat m_g)
    =
    \hat\sigma_{g,\mathrm{alt}}^2
    \xrightarrow{p}
    \Var
    \left(
        Y_i-m_g
        \mid
        \bm X_i\in g
    \right).
\]
Consequently,
\[
    N
    \widehat{\bm V}_{\beta,\mathrm{alt}}
    \xrightarrow{p}
    \bm V_\beta.
\]
\end{proposition}

Unlike the finite-difference variance estimator in Proposition~\ref{prop_var}, consistency of
$\widehat{\bm V}_{\beta,\mathrm{alt}}$
does not require the minimum-spacing condition
$
    N^{-1/2}
    =
    o
    \left(
        \underline{\Delta}(T_{K_N})
    \right).
$

\begin{proof}
Fix a group $g$, and let $Q_g^*(\tau)$ denote the piecewise linear interpolant constructed from the population quantiles, as defined in the proof of Lemma~\ref{lem_mean}. 

By the compact-support condition, there exists a finite constant
$C_g$ such that
\[
		    |Q_g(\tau)|,
	    \quad
		    |Q_g^*(\tau)|,
	    \quad
		    |\widetilde Q_g(\tau)|
	    \leq 
	    	C_g
\]
for every $\tau\in[0,1]$, with probability one.
Using
$
		    |a^2-b^2|
	    =
		    |a-b||a+b|,
$
we obtain
\begin{align*}
		    \left|
		        \hat s_{2g}-s_{2g}
		    \right|
	    &\leq
		    \int_0^1
    		\left|
		        \widetilde Q_g(\tau)^2
		        -
		        Q_g^*(\tau)^2
		    \right|
		    d\tau
	+
		    \int_0^1
		    \left|
		        Q_g^*(\tau)^2
		        -
		        Q_g(\tau)^2
		    \right|	
		    d\tau
    \\
    &\leq
		    2C_g
		    \int_0^1
		    \left|
		        \widetilde Q_g(\tau)
		        -
		        Q_g^*(\tau)
		    \right|
		    d\tau
	+
   		 2C_g
		    \int_0^1
  		  \left|
		        Q_g^*(\tau)
		        -
		        Q_g(\tau)
		    \right|
    		d\tau.
\end{align*}
By the arguments used in the proof of Lemma~\ref{lem_mean},
\[
    \int_0^1
    \left|
        \widetilde Q_g(\tau)
        -
        Q_g^*(\tau)
    \right|
    d\tau
    =
    O_p(r_{g,N}),
\]
and
\[
    \int_0^1
    \left|
        Q_g^*(\tau)
        -
        Q_g(\tau)
    \right|
    d\tau
    =
    O
    \left(
        \Delta(\mathcal T_{K_N})^a
    \right),
\]
where $a=2$ under Assumption~\ref{ass_smooth}(ii). Therefore,
\[
    \hat s_{2g}-s_{2g}
    =
    O_p
    \left(
        r_{g,N}
        +
        \Delta(\mathcal T_{K_N})^a
    \right).
\]

Under the conditions of Lemma~\ref{lem_uniform_quantile}, Assumption~\ref{ass_quantile} holds with
$
  	  r_{g,N}
    	=
	    n_g^{-1/2}.
$
Since $n_g\to\infty$ and
$\Delta(\mathcal T_{K_N})\to0$, it follows that
$
    \hat s_{2g}
    \xrightarrow{p}
    s_{2g}.
$
Similarly, Lemma~\ref{lem_mean} implies
$
    \hat m_g
    \xrightarrow{p}
    m_g.
$
Therefore,
\begin{align*}
		    \hat\sigma_{g,\mathrm{alt}}^2
	    &=
		    \hat s_{2g}
		    -
		    \hat m_g^2
	\xrightarrow{p}
		    s_{2g}
		    -
		    m_g^2
	=
		    \Var
    		\left(
		        Y_i-m_g
		        \mid
		        \bm X_i\in g
		    \right).
\end{align*}
By the definition in
\eqref{eq_alt_mean_variance}, this also gives
\[
    n_g
    \widehat{\Var}_{\mathrm{alt}}(\hat m_g)
    \xrightarrow{p}
    \Var
    \left(
        Y_i-m_g
        \mid
        \bm X_i\in g
    \right).
\]

Next, define
$
    \widehat{\bm\Sigma}_X
    =
    \frac{1}{N}
    \sum_{g=1}^G
    n_g\bm X_g\bm X_g'.
$ as in the proof of Proposition~\ref{prop_var}.
By Assumption~\ref{ass_design},
$
    \widehat{\bm\Sigma}_X
    \xrightarrow{p}
    \bm\Sigma_X,
$
and
$
    \widehat{\bm\Sigma}_X^{-1}
    \xrightarrow{p}
    \bm\Sigma_X^{-1}.
$
Using
\eqref{eq_alt_mean_variance} and
\eqref{eq_alt_beta_variance}, we can write
\begin{align*}
    		N
		    \widehat{\bm V}_{\beta,\mathrm{alt}}
	    =
		    \widehat{\bm\Sigma}_X^{-1}
		    \left(
		        \sum_{g=1}^G
		        \frac{n_g}{N}
		        \bm X_g\bm X_g'
		        \hat\sigma_{g,\mathrm{alt}}^2
		    \right)
		    \widehat{\bm\Sigma}_X^{-1}.
\end{align*}
Because $G$ is fixed, $n_g/N\xrightarrow{p}\pi_g$, and
$
		    \hat\sigma_{g,\mathrm{alt}}^2
		    \xrightarrow{p}
		    \Var
		    \left(
		        Y_i-m_g
     		   \mid
		        \bm X_i\in g
		    \right)
$
for every $g$, Slutsky's theorem yields
\begin{align*}
    N
    \widehat{\bm V}_{\beta,\mathrm{alt}}
    \xrightarrow{p}
    \bm\Sigma_X^{-1}
    \left[
        \sum_{g=1}^G
        \pi_g
        \bm X_g\bm X_g'
        \Var
        \left(
            Y_i-m_g
            \mid
            \bm X_i\in g
        \right)
    \right]
    \bm\Sigma_X^{-1}.
\end{align*}
\end{proof}

The estimator in \eqref{eq_alt_beta_variance} uses only the disclosed sample quantiles and does not require estimation of the density or the derivative of the quantile function.
Consequently, unlike the finite-difference estimator in Section~\ref{subsec_variance}, its consistency does not require a lower bound on the spacing between adjacent disclosed quantiles. 
The conditional variance is instead reconstructed from the disclosed quantile grid and therefore inherits a finite-grid approximation error when the grid is coarse. 
This approximation error vanishes as $\Delta(\mathcal T_{K_N})\to0$.

The simulations, empirical applications, and the accompanying \texttt{qonlyreg} package use the finite-difference variance estimator described in Section~\ref{subsec_variance}. 
The alternative estimator developed here is included to demonstrate that the minimum-spacing condition in Proposition~\ref{prop_var} is specific to that finite-difference implementation and is not intrinsic to
variance estimation under increasingly rich quantile disclosure.

\clearpage
\def\thesection{}
\renewcommand{\thesubsection}{I-\arabic{subsection}}
\def\theequation{I.\arabic{equation}}
\def\theassumption{I.\arabic{assumption}}
\def\thelemma{I.\arabic{lemma}}
\renewcommand{\thefigure}{I.\arabic{figure}}
\renewcommand{\thetable}{I.\arabic{table}}

\makeatletter
\@addtoreset{equation}{section}
\@addtoreset{figure}{section}
\@addtoreset{table}{section}
\@addtoreset{assumption}{section}
\makeatother

\setcounter{equation}{0}
\setcounter{subsection}{0}
\setcounter{lemma}{0}
\setcounter{table}{0}
\setcounter{figure}{0}
\setcounter{page}{1}

\section{Supplemental Material I:
Implementation Details for the Healthcare-Associated Infection
Meta-Analysis Application}
\label{app_hai}
This supplement provides implementation details for the healthcare-associated infection (HAI) meta-analysis application in Section~\ref{subsec_hai_expenditure}.

A detailed description is useful because the application adapts the baseline framework in several ways that are not developed explicitly in the main theoretical presentation.
First, the application features mixed disclosure: some comparisons report HAI- and non-HAI-specific sample means and standard deviations, whereas others report only selected quantiles.
Second, the parameter of interest is a common within-comparison mean difference, obtained after profiling out study--year-specific intercepts from the HAI- and non-HAI-specific means.
Third, the available information on sampling uncertainty also differs across comparisons: for the three annual comparisons from \citet{li17_hai}, the standard errors of the paired mean differences are directly reported, whereas for the other comparisons the required sampling covariance matrices must be constructed from the disclosed marginal summaries.

We therefore describe in detail 
the profiling step in Section \ref{app_hai_profile}, 
the treatment of directly reported and reconstructed means and the construction of the
sampling variance in Section \ref{app_hai_variance}, 
the implementation of fixed-disclosure bounds in Section \ref{supp_HAI_IR},
and simultaneous inference for the within-comparison mean differences in Section \ref{app_hai_region}.
Finally, sensitivity to the upper support bound is examined in Section~\ref{app_hai_support}.

Throughout this supplement, \(g=1,\ldots,G\) indexes the study--year comparisons, with \(G=7\), and \(a\in\{H,C\}\) indexes HAI status.
Each \((g,a)\) cell corresponds to one group in the general framework.

\subsection{Study--Year-Specific Intercepts and Profiled Estimator}
\label{app_hai_profile}
For comparison \(g\), consider the specification
\[
    \mu_g^a
    =
    \alpha_g+\beta_0D_g^a,
    \qquad
    a\in\{H,C\},
\]
where
$
    \mu_g^H=E(Y_g^H),
$
$
    \mu_g^C=E(Y_g^C),
$
$
    D_g^H=1,
$
$
    D_g^C=0.
$
Thus,
\[
    \mu_g^H=\alpha_g+\beta_0,
    \qquad
    \mu_g^C=\alpha_g,
\]
and hence
\[
    d_g
    =
    \mu_g^H-\mu_g^C
    =
    \beta_0.
\]

Let
\(\widehat\mu_g^H\)
and
\(\widehat\mu_g^C\)
denote the reconstructed or directly reported HAI- and
non-HAI-specific means.
The sample-size-weighted minimum-distance criterion is
\[
    \sum_{g=1}^{G}
    \biggl[
        n_g^H
        \left(
            \widehat\mu_g^H-\alpha_g-\beta
        \right)^2
        +
        n_g^C
        \left(
            \widehat\mu_g^C-\alpha_g
        \right)^2
    \biggr].
\]

For any fixed value of \(\beta\), minimizing this criterion with respect
to \(\alpha_g\) gives
\[
    \widehat\alpha_g(\beta)
    =
    \frac{
        n_g^H(\widehat\mu_g^H-\beta)
        +
        n_g^C\widehat\mu_g^C
    }{
        n_g^H+n_g^C
    }.
\]
Substituting this expression into the criterion yields
\[
    \sum_{g=1}^{G}
    \frac{
        n_g^Hn_g^C
    }{
        n_g^H+n_g^C
    }
    \left(
        \widehat d_g-\beta
    \right)^2,
    \qquad
    \widehat d_g
    =
    \widehat\mu_g^H-\widehat\mu_g^C.
\]

Accordingly, the profiled estimator is numerically equivalent to a weighted regression of the within-comparison mean differences on an
intercept, with weights
$
    \frac{
        n_g^Hn_g^C
    }{
        n_g^H+n_g^C
    }.
$
The point estimator can therefore be written as
\[
    \hat\beta
    =
    \sum_{g=1}^{G}
    w_g\widehat d_g,
    \qquad
    w_g
    =
    \frac{
        n_g^Hn_g^C/(n_g^H+n_g^C)
    }{
        \sum_{h=1}^{G}
        n_h^Hn_h^C/(n_h^H+n_h^C)
    }.
\]

\subsection{Mean Reconstruction and Sampling Variance}
\label{app_hai_variance}

Three comparisons disclose the first quartile, median, and third quartile separately for HAI and non-HAI patients.
Let \(Q_g^a(\tau)\) denote the \(\tau\)th quantile of \(Y_g^a\), and let \(\widehat Q_g^a(\tau)\) denote the corresponding disclosed
sample quantile.
For these comparisons, the reconstruction rule in Section~\ref{sec_model_estimation} gives
\[
    \widehat\mu_g^a
    =
    \frac{3}{8}\widehat Q_g^a(0.25)
    +
    \frac{1}{4}\widehat Q_g^a(0.50)
    +
    \frac{3}{8}\widehat Q_g^a(0.75),
    \qquad
    a\in\{H,C\}.
\]
For comparisons reporting sample means, the reported HAI- and
non-HAI-specific sample means are used directly.

In all cases, the estimator of the within-comparison mean difference is
\[
    \widehat d_g
    =
    \widehat\mu_g^H-\widehat\mu_g^C.
\]
Its sampling variance satisfies
\[
    \Var(\widehat d_g)
    =
    \Var(\widehat\mu_g^H)
    +
    \Var(\widehat\mu_g^C)
    -
    2\Cov(
        \widehat\mu_g^H,
        \widehat\mu_g^C
    ).
\]
The variance estimator is constructed differently according to the
form of the reported summaries.

For the three annual comparisons from \citet{li17_hai}, the primary
paper reports the standard error of the paired mean expenditure
difference.
We use this reported standard error directly in the variance
calculation.
Because the reported standard errors are expressed in 2015 US dollars,
we convert them to CNY using the exchange rate of 6.2 CNY per US dollar
reported in that paper and set
$
    \widehat{\Var}(\widehat d_g)
$
equal to the square of the converted standard error.

For the remaining mean-reporting comparison, let \(s_g^H\) and \(s_g^C\) denote the reported sample standard deviations
for HAI and non-HAI patients, respectively.
Because the available summaries do not identify the covariance between the two sample mean estimators, we impose the working assumption that this covariance is zero and use
\begin{equation}
    \widehat{\Var}(\widehat d_g)
    =
    \frac{(s_g^H)^2}{n_g^H}
    +
    \frac{(s_g^C)^2}{n_g^C}.
    \label{supp_var_meanReport}
\end{equation}

For the quartile-reporting comparisons,  we can write
\[
    \widehat\mu_g^a
    =
    \bm a'
    \widehat{\bm Q}_g^a,
\]
where
$
    \bm a
    =
    (
        \frac{3}{8}~~
        \frac{1}{4}~~
        \frac{3}{8}
    )'
$
and
$
    \widehat{\bm Q}_g^a
    =
    (
        \widehat Q_g^a(0.25)~~
        \widehat Q_g^a(0.50)~~
        \widehat Q_g^a(0.75)
    )'.
$
Let
\(\widehat{\bm\Sigma}_g^a\) denote the estimated sampling covariance matrix of \(\widehat{\bm Q}_g^a\).
The estimated variance of the reconstructed mean is therefore
$
    \widehat{\Var}(\widehat\mu_g^a)
    =
    \bm a'
    \widehat{\bm\Sigma}_g^a
    \bm a.
$
Because the available summaries do not identify the covariance between the HAI- and non-HAI-specific quantile estimators, we impose the same working assumption
$
    \Cov(
        \widehat{\bm Q}_g^H,
        \widehat{\bm Q}_g^C
    )
    =
    \bm 0.
$
Hence, for these comparisons,
\[
    \widehat{\Var}(\widehat d_g)
    =
    \bm a'
    \widehat{\bm\Sigma}_g^H
    \bm a
    +
    \bm a'
    \widehat{\bm\Sigma}_g^C
    \bm a.
\]
The matrices
\(\widehat{\bm\Sigma}_g^H\)
and
\(\widehat{\bm\Sigma}_g^C\)
are constructed using the finite-difference covariance estimator described in Section~\ref{subsec_variance}.

Under the assumption that the comparison-specific estimators are independent across comparisons, the variance estimator for the
profiled estimator is
\[
    \widehat{\Var}(\hat\beta)
    =
    \sum_{g=1}^{G}
    w_g^2
    \widehat{\Var}(\widehat d_g).
\]
The conventional 95\% confidence interval is
\[
    \left[
        \hat\beta
        -
        1.96
        \sqrt{
            \widehat{\Var}(\hat\beta)
        },
        \,
        \hat\beta
        +
        1.96
        \sqrt{
            \widehat{\Var}(\hat\beta)
        }
    \right].
\]
This interval accounts for sampling uncertainty in the point estimator, but not for the discrepancy between the true mean and its finite-grid reconstruction in the quartile-reporting comparisons.

\subsection{Fixed-Disclosure Bounds and Identified Set}\label{supp_HAI_IR}

For comparisons reporting quartiles, we impose upper-bound assumptions
$
    Y_g^a\in[0,U],
$
for
$
    a\in\{H,C\}.
$
For \(\tau=(0.25,0.50,0.75)\), Theorem~\ref{thm_mean_bounds} implies the following sample analog lower and upper bounds for the
HAI- and non-HAI-specific population means \(\mu_g^a\):
\[
    \widehat\ell_g^a
    =
    \frac14
    \left\{
        \widehat Q_g^a(0.25)
        +
        \widehat Q_g^a(0.50)
        +
        \widehat Q_g^a(0.75)
    \right\},
\]
and
\[
    \widehat u_g^a
    =
    \widehat\ell_g^a+\frac{U}{4}.
\]
Consequently, the lower and upper bounds on the within-comparison mean difference are
\[
    \widehat{\underline d}_g
    =
    \widehat\ell_g^H-\widehat u_g^C,
    \qquad
    \widehat{\overline d}_g
    =
    \widehat u_g^H-\widehat\ell_g^C.
\]

For comparisons reporting sample means, the sample analog constraint is
a singleton,
\[
    \widehat{\underline d}_g
    =
    \widehat{\overline d}_g
    =
    \widehat\mu_g^H-\widehat\mu_g^C.
\]
The plug-in estimator of the common-effect identified set is therefore
\[
    \widehat{\mathcal B}
    =
    \left[
        \max_g\widehat{\underline d}_g,\,
        \min_g\widehat{\overline d}_g
    \right]
\]
whenever the lower endpoint does not exceed the upper endpoint, and is
empty otherwise.

\subsection{Simultaneous Confidence Region}
\label{app_hai_region}

For comparisons other than the three annual comparisons from \citet{li17_hai}, the covariance matrix of the estimated
lower and upper within-comparison difference bounds is constructed using \texttt{qlm\_contrast\_region}, a function in the \texttt{qonlyreg} package developed for this paper.
For the quartile-reporting comparisons, the covariance matrices are estimated using finite differences based
solely on the disclosed sample quantiles.
The assumed support bounds are used to construct the mean bounds but do not enter the finite-difference estimation.
With a fixed number of disclosed quantiles, these finite differences need not consistently estimate the local
quantile densities. 
They are therefore used here as a practical plug-in approximation.

For the quartile-reporting comparisons, let
$
    \bm c
    =
   	(
        \frac14~~
        \frac14~~
        \frac14
    	)'.
$
Then
\[
    \widehat{\underline d}_g
    =
    \bm c'\widehat{\bm Q}_g^H
    -
    \bm c'\widehat{\bm Q}_g^C
    -
    \frac{U}{4},
\]
and
\[
    \widehat{\overline d}_g
    =
    \bm c'\widehat{\bm Q}_g^H
    -
    \bm c'\widehat{\bm Q}_g^C
    +
    \frac{U}{4}.
\]
Define the estimated bound vector for comparison \(g\) as
$
    \widehat{\bm b}_g
    =
    (
        \widehat{\underline d}_g~~
        \widehat{\overline d}_g
    )'.
$
Under the working assumption that the covariance between the HAI- and non-HAI-specific quantile estimators is zero, the sampling covariance matrix of the estimated lower and upper bounds is 
\[
		\widehat{\Var}
		\left(
			\widehat{\bm b}_g
		\right)
	=
		\left(
		    \bm c'
		    \widehat{\bm\Sigma}_g^H
		    \bm c
		    +
		    \bm c'
		    \widehat{\bm\Sigma}_g^C
		    \bm c
		\right)
		\begin{pmatrix}
		    1 & 1\\
		    1 & 1
		\end{pmatrix}.
\]
The matrices
\(\widehat{\bm\Sigma}_g^H\) and
\(\widehat{\bm\Sigma}_g^C\)
are constructed using the same finite-difference covariance estimator used for the point-estimator variance above.

For the four mean-reporting comparisons,
$
    \widehat{\underline d}_g
    =
    \widehat{\overline d}_g
    =
    \widehat d_g,
$
so that
\[
    \widehat{\Var}
    \left(
        \widehat{\bm b}_g
    \right)
    =
    \widehat{\Var}(\widehat d_g)
    \begin{pmatrix}
        1 & 1\\
        1 & 1
    \end{pmatrix}.
\]
For the three annual comparisons from \citet{li17_hai}, \(\widehat{\Var}(\widehat d_g)\) is obtained from the reported paired-difference standard errors as described in Section \ref{app_hai_variance}.
For the remaining mean-reporting comparison, \(\widehat{\Var}(\widehat d_g)\) is estimated using \eqref{supp_var_meanReport}.

Stack the estimated lower and upper bounds as
$
    \widehat{\bm s}
    =
    (
        \widehat{\underline d}_1~~
        \widehat{\overline d}_1~
        \cdots~
        \widehat{\underline d}_G~~
        \widehat{\overline d}_G
    )'.
$
Treating sampling errors as independent across comparisons, let
\[
    \widehat{\bm V}
    =
    \widehat{\Var}(\widehat{\bm s})
\]
denote the block-diagonal sampling covariance matrix obtained by combining the comparison-specific covariance matrices described above.
Let \(\widehat V_{jk}\) denote the \((j,k)\)th element of \(\widehat{\bm V}\).

To construct the simultaneous critical value, define
$
    \widehat{\bm\Omega}
    =
    N\widehat{\bm V},
$
so that the covariance matrix is expressed on the \(\sqrt{N}\) scale as in Section~\ref{subsec_inference}.
Let
$
    \widehat\omega_j^2
    =
    N\widehat V_{jj},
$
for
$ 
    j=1,\ldots,2G,
$
denote the diagonal elements of \(\widehat{\bm\Omega}\).
Let \(\hat q_{1-\alpha}\) denote the \(1-\alpha\) quantile of
$
    \max_{1\le j\le 2G}
    |
        Z_j / \widehat\omega_j
    |,
$
where
$
    \bm Z
    =
    (Z_1,\ldots,Z_{2G})'
    \sim
    \mathcal N(\bm 0,\widehat{\bm\Omega}).
$
In the implementation, \(\hat q_{1-\alpha}\) is computed by 100,000 Gaussian draws.\footnote{
Equivalently, since the common factor \(N\) cancels after standardization, 
the critical value can be simulated as the \(1-\alpha\) quantile of
$
    \max_{1\le j\le 2G}|Z_j^*|,
$
where
$
    \bm Z^*
    \sim
    \mathcal N(\bm 0,\widehat{\bm R}),
$
and
$
    \widehat R_{jk}
    =
    {
        \widehat V_{jk}
    }/{
        \sqrt{
            \widehat V_{jj}
            \widehat V_{kk}
        }
    }
$
is the correlation matrix.}

The adjusted comparison-specific constraints are
\[
    \widehat{\underline d}_g
    -
    \hat q_{0.95} \frac{\widehat\omega_{2g-1}}{\sqrt{N}}
    \le
    \beta_0
    \le
    \widehat{\overline d}_g
    +
    \hat q_{0.95} \frac{\widehat\omega_{2g}}{\sqrt{N}},
    \qquad
    g=1,\ldots,G.
\]
Because the common-effect model contains only \(\beta_0\) after profiling the study--year-specific intercepts, the simultaneous confidence region is
\[
    \widehat{\mathcal C}_{0.95}
    =
    \left[
        \max_g
        \left\{
            \widehat{\underline d}_g
            -
            \hat q_{0.95}\frac{\widehat\omega_{2g-1}}{\sqrt{N}}
        \right\},
        \,
        \min_g
        \left\{
            \widehat{\overline d}_g
            +
            \hat q_{0.95}\frac{\widehat\omega_{2g}}{\sqrt{N}}
        \right\}
    \right].
\]

\subsection{Sensitivity to the Upper Support Bound}
\label{app_hai_support}
In the main specification, we impose the upper support bound $U=200{,}000$ CNY on expenditure.
To analyze the sensitivity, we repeat the fixed-disclosure analysis for
\[
    U
    \in
    \{
        70{,}000,\,
        80{,}000,\,
        90{,}000,\,
        100{,}000,\,
        200{,}000,\,
        500{,}000
    \},
\]
where the smallest value remains above the largest disclosed third quartile in the analysis sample.

The sensitivity analysis changes the deterministic width of the quantile-based mean bounds used for the quartile-reporting comparisons while leaving the finite-difference covariance approximation unchanged.
The reported results therefore assess the dependence of the fixed-disclosure inference on the maintained upper-support restriction.

\begin{table}[htbp]
\centering
\caption{Sensitivity to the upper support bound}
\label{tab:hai_support_sensitivity}
\begin{tabular}{@{}ccc@{}}
\toprule
Upper support \(U\) &
Estimated identified set &
95\% simultaneous confidence region \\
\midrule
70,000  & Empty set & [28,894.94, 30,332.05] \\
80,000  & Empty set & [28,894.94, 32,832.05] \\
90,000  & Empty set & [28,894.94, 35,332.05] \\
100,000 & Empty set & [28,894.94, 37,832.05] \\
200,000 & Empty set & [28,894.94, 38,163.10] \\
500,000 & Empty set & [28,894.94, 38,163.10] \\
\bottomrule
\end{tabular}
\end{table}

Table~\ref{tab:hai_support_sensitivity} examines sensitivity to the assumed upper support bound.
The smallest value considered, \(U=70{,}000\) CNY, lies only slightly above the largest disclosed third quartile in the analysis sample, \(68{,}427.10\) CNY, and therefore represents a relatively tight support restriction.
The 95 percent simultaneous confidence region remains strictly positive as the support restriction is relaxed.

The stability of the confidence region for larger values of \(U\) arises because the quartile-reporting comparisons no longer determine the endpoints of the simultaneous confidence region once the support restriction
becomes sufficiently loose.
The binding restrictions are instead supplied by comparisons reporting sample means, whose constraints do not depend on \(U\).
Thus, the main result under \(U=200{,}000\) CNY is not driven by imposing a relatively tight upper bound on medical expenditure.

\clearpage
\def\thesection{}
\renewcommand{\thesubsection}{J-\arabic{subsection}}
\def\theequation{J.\arabic{equation}}
\def\theassumption{J.\arabic{assumption}}
\def\thelemma{J.\arabic{lemma}}
\renewcommand{\thefigure}{J.\arabic{figure}}
\renewcommand{\thetable}{J.\arabic{table}}

\makeatletter
\@addtoreset{equation}{section}
\@addtoreset{figure}{section}
\@addtoreset{table}{section}
\@addtoreset{assumption}{section}
\makeatother

\setcounter{equation}{0}
\setcounter{subsection}{0}
\setcounter{lemma}{0}
\setcounter{table}{0}
\setcounter{figure}{0}
\setcounter{page}{1}

\section{Supplemental Material J: An Additional Mixed-Disclosure Meta-Analysis Illustration}
\label{supp_VitaminD}
The main text illustrates mixed disclosure using the HAI expenditure application in Section \ref{subsec_hai_expenditure}. 
This supplement provides a second illustration in which the point estimate and the fixed-disclosure confidence region are closely aligned.

We illustrate this feature using the meta-analysis dataset of \citet{nnoaham08}, which studies the association between serum vitamin D levels and tuberculosis.
The original meta-analysis dataset consists of seven observational studies comparing tuberculosis patients and controls.
One study reports only an odds ratio for vitamin D deficiency rather than mean or quantile summaries of serum vitamin D levels and is therefore excluded from the analysis.
For studies reporting sample means, the means for tuberculosis patients and controls are directly observed, whereas for studies reporting only medians and ranges, 
we use the reported minimum, median, and maximum as interpolation nodes for piecewise-linear mean reconstruction.

The parameter of interest is the mean difference in vitamin D levels between tuberculosis patients and controls:
\[
        m_g
    =
        E(Y_g),
    \quad
        Y_g
    =
        Y_g^{\mathrm{case}}
        -
        Y_g^{\mathrm{control}},
\]
where
\(Y_g^{\mathrm{case}}\)
and
\(Y_g^{\mathrm{control}}\)
denote the serum vitamin D levels for tuberculosis patients and controls, respectively, in study \(g\).
We consider the intercept-only common-effect specification
\[
    m_g
    =
    \beta_0,
    \qquad g=1,\ldots,G,
\]
so that
\(\beta_0\)
represents the common population mean difference across studies under this specification.

For studies reporting only the minimum, median, and maximum, we assign these reported order statistics to the interpolation nodes \(\tau=(0,0.5,1)\) for the purpose of piecewise-linear mean reconstruction.
For studies reporting exact sample means, the reported means are used directly.
The observed or reconstructed means are then used to estimate \(\beta_0\) using the proposed minimum-distance estimator, with weights proportional to the total sample size in each study.\footnote{
This weighting generally differs from the inverse-variance weighting commonly used in conventional meta-analysis.
}

For the partially identified analysis, we assume that the conditional outcome distributions have lower support endpoint zero and supports contained in \([0,U]\), with \(U=150\), reflecting a clinically plausible upper bound for serum vitamin D levels measured in nmol/L.
Because
$
    Y_g
    =
    Y_g^{\mathrm{case}}
    -
    Y_g^{\mathrm{control}},
$
these restrictions imply
\(Y_g\in[-U,U]\).\footnote{
Although the support restriction is required to construct partially identified intervals for studies reporting only quantile summaries, the empirical results are largely insensitive to the choice of the upper bound because the feasible confidence region is strongly anchored by the studies reporting exact means.
}

\input{QOR_Replication/empirical_illustration/vitamin_d_meta_analysis/table_vitaminD_meta.tex}

Table~\ref{tab:vitd_main} reports the results.
The point estimator implies that tuberculosis patients have lower vitamin D levels than controls,
and the corresponding asymptotic confidence interval excludes zero.

The estimated identified set is empty in this application.
This occurs because some studies report exact sample means, which generate singleton constraints that are incompatible in finite samples due to sampling variability.
Importantly, however, the feasible simultaneous confidence region remains nonempty after accounting for estimation uncertainty.
The resulting confidence region also excludes zero, indicating that the negative association between tuberculosis and vitamin D levels remains statistically significant even under coarse quantile disclosure and partial identification.

An important empirical implication of this result is that partially identified inference can remain highly informative when some studies disclose exact means in addition to coarse quantile summaries.
Although quantile-only disclosure generally weakens identification, the presence of exact mean information substantially tightens the feasible confidence region
because the proposed method imposes simultaneous feasibility constraints across studies rather than relying on an averaging estimator alone.
In the present application, the feasible confidence region is substantially tightened by the studies reporting exact sample means.

\clearpage
\def\thesection{}
\renewcommand{\thesubsection}{K-\arabic{subsection}}
\def\theequation{K.\arabic{equation}}
\def\theassumption{K.\arabic{assumption}}
\def\thelemma{K.\arabic{lemma}}
\def\thetheorem{K.\arabic{theorem}}

\makeatletter
\@addtoreset{equation}{section}
\@addtoreset{figure}{section}
\@addtoreset{table}{section}
\@addtoreset{assumption}{section}
\makeatother

\setcounter{equation}{0}
\setcounter{subsection}{0}
\setcounter{lemma}{0}
\setcounter{theorem}{0}
\setcounter{page}{1}

\section{Supplemental Material K: Non-identification under Nonvanishing Disclosure Gaps}
\label{app_nonvanishing_gap}

Theorem~\ref{thm_nonidentification} in the main text establishes non-identification under a fixed finite disclosure grid.
This supplement shows that the same basic mechanism extends to sequences of disclosure grids whose maximal spacing remains bounded away from zero.
The result allows the number and locations of disclosed quantiles to change with the sample size.
Thus, unlike Theorem~\ref{thm_nonidentification}, the alternative joint distributions in this extension may depend on $N$.
Accordingly, the result does not assert the existence of a single pair of data-generating processes that remains observationally equivalent along the entire sequence of disclosure designs.

\begin{theorem}[Non-identification under nonvanishing disclosure gaps]
\label{thm_nonidentification_nonvanishing}
Suppose that
$
		\bm\Sigma_X
	=
		\sum_{g=1}^G
		\pi_g
		\bm X_g\bm X_g'
$		
is positive definite. 
Consider a sequence of quantile disclosure grids
$
    \mathcal T_{K_N}
    =
    \{\tau_{1,N},\ldots,\tau_{K_N,N}\}
    \subset(0,1)
$
satisfying
$
    \liminf_{N\to\infty}
    \Delta(\mathcal T_{K_N})
    >
    0.
$
Then there exist two distinct vectors
$
    \bm\beta_0^{(1)}
    \neq
    \bm\beta_0^{(2)}
$
and an integer $N_0$ such that, for every $N\ge N_0$, there exist two distinct joint distributions of $(Y,\bm X)$ with the same marginal distribution of $\bm X$ such that the induced group-specific conditional means $m_{g,N}^{(\ell)}$ satisfy
$
    m_{g,N}^{(\ell)}
    =
    \bm X_g'\bm\beta_0^{(\ell)}
$
for
$
    \ell=1,2,
$
the corresponding group-specific conditional quantile functions $Q_{g,N}^{(\ell)}$ satisfy Assumption~\ref{ass_smooth}(ii), and
$
    Q_{g,N}^{(1)}(\tau_{k,N})
    =
    Q_{g,N}^{(2)}(\tau_{k,N})
$
for
$
    g=1,\ldots,G
$
and
$
    k=1,\ldots,K_N.
$
Consequently, for every sufficiently large $N$, the regression coefficient is not point identified from the quantiles disclosed under $\mathcal T_{K_N}$.
\end{theorem}
\begin{proof}
Since
$
    \liminf_{N\to\infty}
    \Delta(\mathcal T_{K_N})
    >
    0,
$
there exist constants $\delta>0$ and $N_0<\infty$ such that
$
    \Delta(\mathcal T_{K_N})
    \ge
    \delta
$
for every $N\ge N_0$.
Hence, for every $N\ge N_0$, there exists a nonempty open interval
$
    I_N\subset(0,1)
$
containing no disclosed quantile and having length at least $\delta/2$.

For each $N\ge N_0$, choose $c_N$ such that
$
    [c_N,c_N+\frac{\delta}{4}]\subset I_N,
$
and define
\[
    b_N(\tau)
    =
    \begin{cases}
    (\tau-c_N)^3(c_N+\frac{\delta}{4}-\tau)^3,
    &
    \tau\in[c_N,c_N+\frac{\delta}{4}],
    \\[0.5ex]
    0,
    &
    \text{otherwise}.
    \end{cases}
\]
Then $b_N$ is nonnegative, nonzero, and twice continuously differentiable on $[0,1]$.
Moreover,
$
    b_N(\tau_{k,N})=0
$
for $k=1,\ldots,K_N$, because its support is contained in $I_N$.
Since \(\delta\) is fixed and \(b_N\) differs across \(N\) only through the location \(c_N\) of its support, its integral and derivative bounds do not depend on \(N\).
Hence
$
    \int_0^1 b_N(\tau)\,d\tau
    >0
$
does not depend on $N$, and
$
		\sup_{N\ge N_0}
		\sup_{\tau\in[0,1]}
		|b_N'(\tau)|
	<
		\infty
$
and
$
		\sup_{N\ge N_0}
		\sup_{\tau\in[0,1]}
		|b_N''(\tau)|
	<
		\infty
$.

Fix any vector $\bm\beta_0^{(1)}$ and any constant $s>0$.
For every $N\ge N_0$ and every group $g$, define
\[
    Q_{g,N}^{(1)}(\tau)
    =
    \bm X_g'\bm\beta_0^{(1)}
    +
    s\left(\tau-\frac12\right),
    \qquad
    \tau\in[0,1].
\]
Then $Q_{g,N}^{(1)}$ is strictly increasing and twice continuously differentiable, and its corresponding conditional mean satisfies
\[
    m_{g,N}^{(1)}
    =
    \int_0^1
    Q_{g,N}^{(1)}(\tau)\,d\tau
    =
    \bm X_g'\bm\beta_0^{(1)}.
\]

Take any nonzero vector $\bm d_0$.
Since $\bm\Sigma_X$ is positive definite,
$
    \bm d_0'\bm\Sigma_X\bm d_0
    =
    \sum_{g=1}^G
    \pi_g
    (
        \bm X_g'\bm d_0
    )^2
    >
    0.
$
Hence,
$
    \bm X_g'\bm d_0\neq0
$
for at least one group $g$.

Let
$
    \bm d
    =
    \varepsilon\bm d_0,
$
where $\varepsilon>0$ will be chosen sufficiently small, and define
$
    \bm\beta_0^{(2)}
    =
    \bm\beta_0^{(1)}
    +
    \bm d.
$
For each group $g$, let
$
    \lambda_g
    =
    \frac{\bm X_g'\bm d}{ \int_0^1 b_N(\tau)\,d\tau},
$
and, for every $N\ge N_0$, define
\[
    Q_{g,N}^{(2)}(\tau)
    =
    Q_{g,N}^{(1)}(\tau)
    +
    \lambda_g b_N(\tau).
\]

Because $G$ is fixed,
$
    \sup_{N\ge N_0}
    \sup_{\tau\in[0,1]}|b_N'(\tau)|
    <
    \infty,
$
and $ \int_0^1 b_N(\tau)\,d\tau \neq0$, we can choose $\varepsilon>0$ sufficiently small that
$
    \max_{1\le g\le G}
    |\lambda_g|
    \sup_{N\ge N_0}
    \sup_{\tau\in[0,1]}|b_N'(\tau)|
    <
    \frac{s}{2}.
$
It then follows that
\[
    \frac{d}{d\tau}
    Q_{g,N}^{(2)}(\tau)
    =
    s+\lambda_gb_N'(\tau)
    \ge
    \frac{s}{2}
    >
    0
\]
for every $g$, every $N\ge N_0$, and every $\tau\in[0,1]$.
Thus $Q_{g,N}^{(2)}$ is strictly increasing.
Moreover,
\[
    \sup_{N\ge N_0}
    \sup_{\tau\in(0,1)}
    \left|
        \frac{d^2}{d\tau^2}
        Q_{g,N}^{(2)}(\tau)
    \right|
    \le
    |\lambda_g|
    \sup_{N\ge N_0}
   \sup_{\tau\in[0,1]} |b_N''(\tau)|
    <
    \infty.
\]
Hence both $Q_{g,N}^{(1)}$ and $Q_{g,N}^{(2)}$ satisfy
Assumption~\ref{ass_smooth}(ii).

Because
$
    b_N(\tau_{k,N})=0
$
at every disclosed quantile index,
$
    Q_{g,N}^{(2)}(\tau_{k,N})
    =
    Q_{g,N}^{(1)}(\tau_{k,N})
$
for
$
    g=1,\ldots,G
$
and
$
    k=1,\ldots,K_N.
$
Thus, for every $N\ge N_0$, the two constructions generate exactly the same disclosed population quantiles.

On the other hand,
\begin{align*}
    m_{g,N}^{(2)}
    =
    \int_0^1
    Q_{g,N}^{(2)}(\tau)\,d\tau
    &=
    \int_0^1
    Q_{g,N}^{(1)}(\tau)\,d\tau
    +
    \lambda_g
    \int_0^1 b_N(\tau)\,d\tau
    \\
    &=
    \bm X_g'\bm\beta_0^{(1)}
    +
    \bm X_g'\bm d
    \\
    &=
    \bm X_g'\bm\beta_0^{(2)}.
\end{align*}
Therefore, for $\ell=1,2$,
$
    m_{g,N}^{(\ell)}
    =
    \bm X_g'\bm\beta_0^{(\ell)}
$
for
$
    g=1,\ldots,G
$
and
$
    N\ge N_0
$,
where
$
    \bm\beta_0^{(1)}
    \neq
    \bm\beta_0^{(2)}
$
since $\bm d\neq\bm 0$.

Finally, retain the same marginal distribution of $\bm X$ in the two constructions. 
For each $N\ge N_0$, $\ell=1,2$, and $\bm x\in\mathcal G_g$, define the conditional distribution of $Y$
given $\bm X=\bm x$ to be the distribution whose quantile function is $Q_{g,N}^{(\ell)}$.
Because the groups form a partition, this assignment is well defined and induces a valid joint distribution of $(Y,\bm X)$.

The two joint distributions are distinct. 
Indeed, positive definiteness of $\bm\Sigma_X$ implies that $\bm X_g'\bm d_0\neq0$ for at least one group $g$, and hence $\lambda_g\neq0$ for that group. This implies that
$
    Q_{g,N}^{(2)}
    \neq
    Q_{g,N}^{(1)},
$
since $b_N$ is nonzero.
Thus, for every $N\ge N_0$, there exist two distinct joint distributions with the same marginal distribution of $\bm X$ and the same disclosed conditional quantiles under
$\mathcal T_{K_N}$, but with
$
    \bm\beta_0^{(1)}
    \neq
    \bm\beta_0^{(2)}.
$
Hence, for every sufficiently large $N$, the regression coefficient is not point identified from the quantiles disclosed under $\mathcal T_{K_N}$.
\end{proof}

\clearpage
\def\thesection{}
\renewcommand{\thesubsection}{L-\arabic{subsection}}
\def\theequation{L.\arabic{equation}}
\def\theassumption{L.\arabic{assumption}}
\def\thelemma{L.\arabic{lemma}}
\def\thetheorem{L.\arabic{theorem}}
\def\theproposition{L.\arabic{proposition}}
\def\theexample{L.\arabic{example}}
\renewcommand{\thefigure}{L.\arabic{figure}}
\renewcommand{\thetable}{L.\arabic{table}}
\setcounter{equation}{0}
\setcounter{subsection}{0}
\setcounter{lemma}{0}
\setcounter{theorem}{0}
\setcounter{proposition}{0}
\setcounter{example}{0}
\setcounter{page}{1}
\section{Supplemental Material L: Derivation of the quantile-based variance estimator}

Section~\ref{subsec_variance} defines the reconstruction weights
$\alpha_{k,N}$ and the feasible variance estimator. The following
calculation explains the quantile covariance approximation used there.

The covariance structure of $\hat m_g$ follows from the joint covariance structure of the disclosed sample quantiles.
By linearity,
\[
    		\Var(\hat m_g)
    	=
    		\sum_{k=1}^{K_N}\sum_{l=1}^{K_N}
    		\alpha_{k,N}\alpha_{l,N}
   		 \Cov\bigl(\widehat Q_g(\tau_k), \widehat Q_g(\tau_l)\bigr).
\]
Using the Bahadur representation of sample quantiles
$
    		\widehat Q_g(\tau)
    	=
    		Q_g(\tau)
    		+
    		\frac{\tau - F_{n,g}(Q_g(\tau))}{f_g(Q_g(\tau))}
    		+
    		r_{g,n}(\tau),
$
where $r_{g,n}(\tau) = o_p(n_g^{-1/2})$
(see Lemma \ref{lem_bahadur_grid}),
we approximate the covariance by its leading linear terms:
\begin{align*}
    		\Cov(\widehat Q_g(\tau_k),\widehat Q_g(\tau_l))
    	&=
    		\Cov\bigl(\widehat Q_g(\tau_k) - Q_g(\tau_k), \widehat Q_g(\tau_l) - Q_g(\tau_l)\bigr)
    		\\
    	&\approx
    		\frac{1}{f_g(Q_g(\tau_k))f_g(Q_g(\tau_l))}
    		\Cov\bigl(F_{n,g}(Q_g(\tau_k)),F_{n,g}(Q_g(\tau_l))\bigr).
\end{align*}
Using the covariance structure of the empirical distribution function,
$
    		\Cov (F_{n,g}(x),F_{n,g}(y))
    	=
    		\frac{1}{n_g}
    		\{ F_g(\min(x,y)) - F_g(x)F_g(y) \},
$
and the identity $F_g(Q_g(\tau)) = \tau$, we obtain
\[
    		\Cov(\widehat Q_g(\tau_k),\widehat Q_g(\tau_l))
    	\approx
    		\frac{1}{n_g}
    		\frac{
        		\min(\tau_k,\tau_l)-\tau_k\tau_l
    		}{
        			f_g(Q_g(\tau_k))\,f_g(Q_g(\tau_l))
   		 }.
\]

Substituting the finite-difference estimates of $f_g(Q_g(\tau_k))$
gives $\widehat{\Var}(\hat m_g)$ in Section~\ref{subsec_variance}.
Under the conditions of Proposition~\ref{prop_var},
Lemma~\ref{lem_group_var} establishes
\[
    n_g\widehat{\Var}(\hat m_g)
    \xrightarrow{p}\Var(Y_i-m_g\mid\bm X_i\in g).
\]
Combining this result with $n_g/N\xrightarrow{p}\pi_g$ yields the
consistency of the regression variance estimator.

\end{document}

%% file: QOR_Replication/simulation/results_partial_qlm/table_region_coverage_even.tex
\begin{table}[H]
\centering
\begin{threeparttable}[H]
\centering
\caption{\label{tab:partial_identification_region_coverage_even} Coverage of the simultaneous confidence region under even grids}
\centering
{
\renewcommand{\arraystretch}{1.8}
\begin{tabular}[t]{cccccccccc}
\toprule
\multicolumn{1}{c}{ } & \multicolumn{3}{c}{Coverage} & \multicolumn{3}{c}{Length of $\beta_0$ interval} & \multicolumn{3}{c}{Length of $\beta_1$ interval} \\
\cmidrule(l{3pt}r{3pt}){2-4} \cmidrule(l{3pt}r{3pt}){5-7} \cmidrule(l{3pt}r{3pt}){8-10}
$N$ & $K = 3$ & $K = 9$ & $K = 19$ & $K = 3$ & $K = 9$ & $K = 19$ & $K = 3$ & $K = 9$ & $K = 19$\\
\midrule
1000 & 0.9378 & 0.9314 & 0.9296 & 0.6990 & 0.3928 & 0.2893 & 0.8680 & 0.5198 & 0.3982\\
5000 & 0.9464 & 0.9451 & 0.9454 & 0.5970 & 0.2910 & 0.1878 & 0.7079 & 0.3641 & 0.2461\\
10000 & 0.9511 & 0.9544 & 0.9540 & 0.5719 & 0.2665 & 0.1635 & 0.6685 & 0.3261 & 0.2092\\
\bottomrule
\end{tabular}
}
\begin{tablenotes}[flushleft]
\setstretch{2}
\item Notes: Coverage reports the empirical coverage probability of the simultaneous confidence region for the true identified set. The length columns report the average lengths of the projection intervals.
\end{tablenotes}
\end{threeparttable}
\end{table}

%% file: QOR_Replication/empirical_illustration/NAEP_data/table_naep_qonlyreg.tex
\begin{table}[!h]
\centering
\begin{threeparttable}
\caption{\label{tab:naep_qonlyreg}Quantile-only regression results using NAEP data}
\centering
{
\renewcommand{\arraystretch}{1.8}
\begin{tabular}[t]{lcccccc}
\toprule
\multicolumn{1}{c}{ } & \multicolumn{3}{c}{Point} & \multicolumn{3}{c}{Partial} \\
\cmidrule(l{3pt}r{3pt}){2-4} \cmidrule(l{3pt}r{3pt}){5-7}
Disclosure & Estimate & CI & Covered & Projection CI & Width & Covered\\
\midrule
Median & 29.7 & -- & -- & -- & -- & --\\
Quartiles & 29.3 & {}[28.6, 30.1] & No & {}[-104.7, 140.0] & 244.6 & Yes\\
Five percentiles & 28.7 & {}[28.0, 29.4] & Yes & {}[-37.7, 75.1] & 112.8 & Yes\\
\bottomrule
\end{tabular}
}
\begin{tablenotes}[flushleft]
\setstretch{2}
\item Notes: Benchmark estimate is 28.4. The benchmark estimate is obtained by regressing the published group mean scores on the non-EDS indicator and state fixed effects. “Covered” indicates whether the benchmark estimate
  based on the published mean scores is contained
  in the corresponding interval.
\end{tablenotes}
\end{threeparttable}
\end{table}

%% file: QOR_Replication/empirical_illustration/UK_election/output/paper_tables/table_main_results.tex
\begin{table}[!h]
\centering
\begin{threeparttable}
\caption{Labour majority and assessment duration: fixed-window qonlyreg results}
\label{tab:qonlyreg_main}
{
\renewcommand{\arraystretch}{1.8}
\begin{tabular}{cccccccc}
\toprule
$h$ & $|\mathcal G_h|$ & Co. & T / C & Est. (SE) & $U=150$ & $U=175$ & $U=200$ \\
\midrule
$\pm 3$ & 77 & 32 & 43 / 34 & -1.638 (0.065) & [-23.62, 17.38] & [-29.87, 23.63] & [-36.12, 29.88] \\
$\pm 4$ & 88 & 36 & 51 / 37 & -0.830 (0.061) & [-22.12, 17.39] & [-28.37, 23.64] & [-34.62, 29.89] \\
$\pm 5$ & 129 & 42 & 74 / 55 & -2.045 (0.050) & [-22.17, 5.03] & [-28.42, 11.28] & [-34.67, 17.53] \\
\bottomrule
\end{tabular}
}
\begin{tablenotes}[flushleft]
\setstretch{2}
\item Notes: $h$ is the window half-width in percentage points; $|\mathcal G_h|$ and Co.\ are the numbers of council-year groups and councils; T/C gives treated/control group counts. Est.\ (SE) reports the Labour-majority coefficient (standard error). Brackets give projections of the 95\% simultaneous confidence region under $Q_g(0)=0$ and upper support bound $U$ (days). The main specification uses $h=4$.
\end{tablenotes}
\end{threeparttable}
\end{table}

%% file: QOR_Replication/simulation/results_smooth_qlm/table_coverage_even.tex
\begin{table}[!h]
\centering
\caption{\label{tab:coverage_even}Coverage probability under evenly spaced grids}
\centering
\fontsize{9}{11}\selectfont
\begin{tabular}[t]{cccccccc}
\toprule
N & K = 3 & K = 5 & K = 7 & K = 10 & K = 15 & K = 20 & K = 50\\
\midrule
2000 & 0.926 & 0.928 & 0.934 & 0.938 & 0.941 & 0.944 & 0.951\\
20000 & 0.922 & 0.930 & 0.934 & 0.937 & 0.943 & 0.943 & 0.946\\
\bottomrule
\end{tabular}
\end{table}

%% file: QOR_Replication/simulation/results_compact_truncnorm_qlm/table_coverage_even.tex
\begin{table}[!h]
\centering
\caption{\label{tab:coverage_even_compact}Coverage probability under the asymmetric compact-support design and evenly spaced grids}
\centering
\fontsize{9}{11}\selectfont
\begin{tabular}[t]{cccccccc}
\toprule
N & K = 3 & K = 5 & K = 7 & K = 10 & K = 15 & K = 20 & K = 50\\
\midrule
2000 & 0.913 & 0.925 & 0.927 & 0.939 & 0.943 & 0.947 & 0.952\\
20000 & 0.848 & 0.878 & 0.890 & 0.910 & 0.920 & 0.931 & 0.943\\
\bottomrule
\end{tabular}
\end{table}

%% file: QOR_Replication/simulation/results_nonsmooth/table_coverage_even.tex
\begin{table}[!h]
\centering
\caption{\label{tab:coverage_even_nonsmooth}Coverage probability under evenly spaced grids}
\centering
\fontsize{9}{11}\selectfont
\begin{tabular}[t]{cccccccc}
\toprule
N & K = 3 & K = 5 & K = 7 & K = 10 & K = 15 & K = 20 & K = 50\\
\midrule
2000 & 0.882 & 0.905 & 0.917 & 0.926 & 0.936 & 0.939 & 0.951\\
20000 & 0.880 & 0.906 & 0.914 & 0.927 & 0.933 & 0.938 & 0.943\\
\bottomrule
\end{tabular}
\end{table}

%% file: QOR_Replication/simulation/results_partial_qlm/table_region_coverage_right_tail_coarse_mild.tex
\begin{table}
\centering
\begin{threeparttable}[!h]
\centering
\caption{\label{tab:partial_identification_region_coverage_right_tail_coarse_mild} Coverage of the simultaneous confidence region under mildly right-tail-coarse grids}
\centering
\begin{tabular}[t]{cccccccccc}
\toprule
\multicolumn{1}{c}{ } & \multicolumn{3}{c}{Coverage} & \multicolumn{3}{c}{Length of $\beta_0$ interval} & \multicolumn{3}{c}{Length of $\beta_1$ interval} \\
\cmidrule(l{3pt}r{3pt}){2-4} \cmidrule(l{3pt}r{3pt}){5-7} \cmidrule(l{3pt}r{3pt}){8-10}
$N$ & $K = 3$ & $K = 9$ & $K = 19$ & $K = 3$ & $K = 9$ & $K = 19$ & $K = 3$ & $K = 9$ & $K = 19$\\
\midrule
1000 & 0.9459 & 0.9389 & 0.9361 & 0.7577 & 0.4188 & 0.3023 & 0.9326 & 0.5498 & 0.4136\\
5000 & 0.9552 & 0.9474 & 0.9465 & 0.6554 & 0.3167 & 0.2009 & 0.7728 & 0.3930 & 0.2611\\
10000 & 0.9566 & 0.9553 & 0.9551 & 0.6298 & 0.2921 & 0.1765 & 0.7331 & 0.3548 & 0.2241\\
\bottomrule
\end{tabular}
\begin{tablenotes}[para,flushleft]
\footnotesize
\item \parbox{\linewidth}{Notes: Coverage reports the empirical coverage probability of the simultaneous confidence region for the true identified set. The length columns report the average lengths of the projection intervals.}
\end{tablenotes}
\end{threeparttable}
\end{table}

%% file: QOR_Replication/simulation/results_partial_qlm/table_region_coverage_right_tail_coarse_severe.tex
\begin{table}
\centering
\begin{threeparttable}[!h]
\centering
\caption{\label{tab:partial_identification_region_coverage_right_tail_coarse_severe} Coverage of the simultaneous confidence region under severely right-tail-coarse grids}
\centering
\begin{tabular}[t]{cccccccccc}
\toprule
\multicolumn{1}{c}{ } & \multicolumn{3}{c}{Coverage} & \multicolumn{3}{c}{Length of $\beta_0$ interval} & \multicolumn{3}{c}{Length of $\beta_1$ interval} \\
\cmidrule(l{3pt}r{3pt}){2-4} \cmidrule(l{3pt}r{3pt}){5-7} \cmidrule(l{3pt}r{3pt}){8-10}
$N$ & $K = 3$ & $K = 9$ & $K = 19$ & $K = 3$ & $K = 9$ & $K = 19$ & $K = 3$ & $K = 9$ & $K = 19$\\
\midrule
1000 & 0.9420 & 0.9409 & 0.9338 & 1.0606 & 0.5536 & 0.3703 & 1.2664 & 0.7035 & 0.4936\\
5000 & 0.9508 & 0.9504 & 0.9464 & 0.9662 & 0.4529 & 0.2704 & 1.1178 & 0.5463 & 0.3405\\
10000 & 0.9582 & 0.9557 & 0.9551 & 0.9414 & 0.4282 & 0.2459 & 1.0798 & 0.5075 & 0.3028\\
\bottomrule
\end{tabular}
\begin{tablenotes}[para,flushleft]
\footnotesize
\item \parbox{\linewidth}{Notes: Coverage reports the empirical coverage probability of the simultaneous confidence region for the true identified set. The length columns report the average lengths of the projection intervals.}
\end{tablenotes}
\end{threeparttable}
\end{table}

%% file: QOR_Replication/empirical_illustration/UK_election/output/paper_tables/table_strict_majority_sensitivity.tex
\begin{table}[!h]
\centering
\begin{threeparttable}
\caption{Sensitivity to defining treatment as strictly above 50 percent}
\label{tab:qonlyreg_strict_majority}
\small
\begin{tabular}{cccccccc}
\toprule
Window & $|\mathcal G_h|$ & Councils & T / C & Estimate (SE) & $U=150$ & $U=175$ & $U=200$ \\
\midrule
$\pm 3$ & 77 & 32 & 36 / 41 & -2.382 (0.066) & [-23.63, 17.38] & [-29.88, 23.63] & [-36.13, 29.88] \\
$\pm 4$ & 88 & 36 & 44 / 44 & -1.504 (0.061) & [-22.13, 17.41] & [-28.38, 23.66] & [-34.63, 29.91] \\
$\pm 5$ & 129 & 42 & 67 / 62 & -2.445 (0.049) & [-22.17, 5.04] & [-28.42, 11.29] & [-34.67, 17.54] \\
\bottomrule
\end{tabular}
\begin{tablenotes}[flushleft]
\footnotesize
\item Notes: The treatment is a Labour seat share strictly greater than 50 percent, so observations exactly at 50 percent are assigned to control. All other definitions match Table~\ref{tab:qonlyreg_main}.
\end{tablenotes}
\end{threeparttable}
\end{table}

%% file: QOR_Replication/empirical_illustration/UK_election/output/paper_tables/table_leave_one_council_out.tex
\begin{table}[!h]
\centering
\begin{threeparttable}
\caption{Leave-one-council-out sensitivity in the main window}
\label{tab:qonlyreg_leave_one_council_out}
\small
\begin{tabular}{cccccc}
\toprule
Excluded council & Side & Council-years & Assessments & Estimate & Change \\
\midrule
None (baseline) & -- & 0 & 0 & -0.830 & 0.000 \\
Bristol & treated & 5 & 22,290 & 0.107 & 0.936 \\
Kirklees & both across years & 8 & 36,721 & -0.024 & 0.806 \\
Croydon & control & 3 & 18,590 & -0.045 & 0.785 \\
Walsall & control & 2 & 9,355 & -1.405 & -0.575 \\
Nottinghamshire & control & 1 & 9,126 & -1.386 & -0.556 \\
\bottomrule
\end{tabular}
\begin{tablenotes}[flushleft]
\footnotesize
\item Notes: The first row is the baseline $\pm4$ estimate. The remaining rows show the five councils producing the largest absolute change when all of that council's observations are removed. Change is relative to the baseline. Across all 36 exclusions, 35 estimates remain negative. The complete council-level results are supplied in \texttt{table\_leave\_one\_council\_out\_full.csv}.
\end{tablenotes}
\end{threeparttable}
\end{table}

%% file: QOR_Replication/empirical_illustration/UK_election/output/paper_tables/table_year_fixed_effects_sensitivity.tex
\begin{table}[!h]
\centering
\begin{threeparttable}
\caption{Sensitivity to reporting-year fixed effects}
\label{tab:qonlyreg_year_fixed_effects}
\small
\begin{tabular}{cccccccc}
\toprule
Window & $|\mathcal G_h|$ & Councils & T / C & Estimate (SE) & $U=150$ & $U=175$ & $U=200$ \\
\midrule
$\pm 3$ & 77 & 32 & 43 / 34 & -2.434 (0.070) & [-26.89, 22.03] & [-33.14, 28.28] & [-39.39, 34.53] \\
$\pm 4$ & 88 & 36 & 51 / 37 & -1.698 (0.067) & [-25.09, 22.05] & [-31.34, 28.30] & [-37.59, 34.55] \\
$\pm 5$ & 129 & 42 & 74 / 55 & -3.361 (0.054) & [-25.14, 11.47] & [-31.39, 17.72] & [-37.64, 23.97] \\
\bottomrule
\end{tabular}
\begin{tablenotes}[flushleft]
\footnotesize
\item Notes: The treatment is a Labour seat share of at least 50 percent. Each model adds reporting-year indicators to the intercept and treatment indicator; 2017 is the omitted reporting year. The quantile reconstruction, fixed windows, assessment-count weights, and support assumptions match Table~\ref{tab:qonlyreg_main}. Parentheses contain qlm standard errors, and brackets report projections of the 95 percent simultaneous confidence region.
\end{tablenotes}
\end{threeparttable}
\end{table}

%% file: QOR_Replication/empirical_illustration/vitamin_d_meta_analysis/table_vitaminD_meta.tex
\begin{table}[!h]
\centering
\caption{\label{tab:vitd_main}Vitamin D levels and tuberculosis: an application to meta-analysis}
\centering
\begin{tabular}[t]{lcc}
\toprule
Method & Estimate & Interval\\
\midrule
Point estimator & -16.97 & {}[-21.92, -12.02]\\
Plug-in identified region &  & Empty set\\
Feasible confidence region &  & {}[-19.19, -12.01]\\
\bottomrule
\end{tabular}
\end{table}

%% file: bibliography.bib
@article{liu22_hai,
  author  = {Liu, X. and Spencer, A. and Long, Y. and Greenhalgh, C. and Steeg, S. and Verma, A.},
  title   = {A systematic review and meta-analysis of disease burden of healthcare-associated infections in {China}: an economic burden perspective from general hospitals},
  journal = {Journal of Hospital Infection},
  year    = {2022},
  volume  = {123},
  pages   = {1--11},
  doi     = {10.1016/j.jhin.2022.02.005}
}

@article{li17_hai,
  author  = {Li, Hao and Liu, Xinliang and Cui, Dan and Wang, Quan and Mao, Zongfu and Fang, Liang and Zhang, Furong and Yang, Ping and Wu, Huiling and Ren, Nili and He, Jianyun and Sun, Jing},
  title   = {Estimating the Direct Medical Economic Burden of Health Care-Associated Infections in Public Tertiary Hospitals in {Hubei Province, China}},
  journal = {Asia Pacific Journal of Public Health},
  year    = {2017},
  volume  = {29},
  number  = {5},
  pages   = {440--450},
  doi     = {10.1177/1010539517717366}
}

@article{farebrother79,
  author  = {Farebrother, Richard William},
  title   = {Estimation with Aggregated Data},
  journal = {Journal of Econometrics},
  year    = {1979},
  volume  = {10},
  number  = {1},
  pages   = {43--55}
}

@article{nicoletti12,
  author  = {Nicoletti, Cheti and Best, Nicky G.},
  title   = {Quantile Regression with Aggregated Data},
  journal = {Economics Letters},
  year    = {2012},
  volume  = {117},
  number  = {2},
  pages   = {401--404},
  doi     = {10.1016/j.econlet.2012.06.011}
}

@book{king97,
  author    = {King, Gary},
  title     = {A Solution to the Ecological Inference Problem:
               Reconstructing Individual Behavior from Aggregate Data},
  year      = {1997},
  publisher = {Princeton University Press},
  address   = {Princeton, NJ}
}

@article{kovalchik13,
  title={Aggregate-data estimation of an individual patient data linear random effects meta-analysis with a patient covariate-treatment interaction term},
  author={Kovalchik, Stephanie A},
  journal={Biostatistics},
  volume={14},
  number={2},
  pages={273--283},
  year={2013},
  publisher={Oxford University Press}
}

@techreport{mcculloch25,
  title={Sewers and urbanization in the developing world},
  author={McCulloch, Sean E and Schaelling, Matthew P and Turner, Matthew and Kitagawa, Toru},
  year={2025},
  institution={NBER Working Paper Series 33597}
}

@misc{dfe_working_together_2026,
  author       = {{Department for Education}},
  title        = {Working Together to Safeguard Children 2026:
                  A Guide to Multi-Agency Working to Help, Protect
                  and Promote the Welfare of Children},
  year         = {2026},
  month        = mar,
  howpublished = {Statutory guidance},
  publisher    = {{Department for Education}},
  note         = {\href{https://www.gov.uk/government/publications/working-together-to-safeguard-children--2}{Available at GOV.UK}. Accessed August 8, 2026}
}

@misc{dfe_cin_methodology,
  author       = {{Department for Education}},
  title        = {Children in Need: Methodology},
  year         = {2024},
  month        = oct,
  howpublished = {Explore Education Statistics},
  url          = {https://explore-education-statistics.service.gov.uk/methodology/children-in-need},
  note         = {Accessed August 8, 2026}
}

@article{alonso20,
  title={Political ideology and social services contracting: evidence from a regression discontinuity design},
  author={Alonso, Jos{\'e} M and Andrews, Rhys},
  journal={Public Administration Review},
  volume={80},
  number={5},
  pages={743--754},
  year={2020},
  publisher={Wiley Online Library}
}

@article{cai21,
  title={Estimating the sample mean and standard deviation from order statistics and sample size in meta-analysis},
  author={Cai, Siyu and Zhou, Jie and Pan, Jianxin},
  journal={Statistical methods in medical research},
  volume={30},
  number={12},
  pages={2701--2719},
  year={2021},
  publisher={SAGE Publications Sage UK: London, England}
}

@book{manski03,
  title={Partial identification of probability distributions},
  author={Manski, Charles F},
  year={2003},
  publisher={Springer}
}

@article{cowell82,
  title={The estimation and interpolation of inequality measures},
  author={Cowell, Frank A and Mehta, Fatemeh},
  journal={The Review of Economic Studies},
  volume={49},
  number={2},
  pages={273--290},
  year={1982},
  publisher={Wiley-Blackwell}
}

@article{hozo05,
  title={Estimating the mean and variance from the median, range, and the size of a sample},
  author={Hozo, Stela Pudar and Djulbegovic, Benjamin and Hozo, Iztok},
  journal={BMC medical research methodology},
  volume={5},
  number={1},
  pages={13},
  year={2005},
  publisher={Springer}
}

@article{molinari20,
  title={Microeconometrics with partial identification},
  author={Molinari, Francesca},
  journal={Handbook of econometrics},
  volume={7},
  pages={355--486},
  year={2020},
  publisher={Elsevier}
}

@article{bahadur66,
  title={A note on quantiles in large samples},
  author={Bahadur, R Raj},
  journal={The Annals of Mathematical Statistics},
  volume={37},
  number={3},
  pages={577--580},
  year={1966},
  publisher={JSTOR}
}

@article{nnoaham08,
  title={Low serum vitamin D levels and tuberculosis: a systematic review and meta-analysis},
  author={Nnoaham, Kelechi E and Clarke, Aileen},
  journal={International journal of epidemiology},
  volume={37},
  number={1},
  pages={113--119},
  year={2008},
  publisher={Oxford University Press}
}

@article{wan14,
  title={Estimating the sample mean and standard deviation from the sample size, median, range and/or interquartile range},
  author={Wan, Xiang and Wang, Wenqian and Liu, Jiming and Tong, Tiejun},
  journal={BMC medical research methodology},
  volume={14},
  number={1},
  pages={135},
  year={2014},
  publisher={Springer}
}

@article{luo18,
  title={Optimally estimating the sample mean from the sample size, median, mid-range, and/or mid-quartile range},
  author={Luo, Dehui and Wan, Xiang and Liu, Jiming and Tong, Tiejun},
  journal={Statistical methods in medical research},
  volume={27},
  number={6},
  pages={1785--1805},
  year={2018},
  publisher={SAGE Publications Sage UK: London, England}
}

@article{mcgrath20,
  title={Estimating the sample mean and standard deviation from commonly reported quantiles in meta-analysis},
  author={McGrath, Sean and Zhao, XiaoFei and Steele, Russell and Thombs, Brett D and Benedetti, Andrea and DEPRESsion Screening Data (DEPRESSD) Collaboration},
  journal={Statistical methods in medical research},
  volume={29},
  number={9},
  pages={2520--2537},
  year={2020},
  publisher={SAGE Publications Sage UK: London, England}
}

@article{kiefer67,
  title={On Bahadur's representation of sample quantiles},
  author={Kiefer, Jack},
  journal={The Annals of Mathematical Statistics},
  volume={38},
  number={5},
  pages={1323--1342},
  year={1967},
  publisher={JSTOR}
}

@book{vandervaart98,
  title={Asymptotic statistics},
  author={Van der Vaart, Aad W},
  year={1998},
  publisher={Cambridge university press}
}

@book{vandervaart96,
  title={Weak convergence and empirical processes: with applications to statistics},
  author={Van Der Vaart, Aad and Wellner, Jon A},
  year={1996},
  publisher={Springer-Verlag},
  address   = {New York},
  isbn      = {0-387-94640-3}  
}

@article{vandervaart11,
  title={A local maximal inequality under uniform entropy},
  author={Van Der Vaart, Aad and Wellner, Jon A},
  journal={Electronic Journal of Statistics},
  volume={5},
  number={2011},
  pages={192},
  year={2011}
}

@book{atkinson89,
  author    = {Atkinson, Kendall E.},
  title     = {An Introduction to Numerical Analysis},
  edition   = {Second},
  publisher = {John Wiley \& Sons},
  year      = {1989},
  address   = {New York},
  isbn      = {0-471-62489-6},
}

@article{dkw56,
  author = {Dvoretzky, Aryeh and Kiefer, Jack and Wolfowitz, Jacob},
  title = {Asymptotic minimax character of the sample distribution function and of the classical multinomial estimator},
  journal = {Annals of Mathematical Statistics},
  year = {1956},
  volume = {27},
  number = {3},
  pages = {642--669}
}

@article{massart90,
  author = {Massart, Pascal},
  title = {The tight constant in the Dvoretzky-Kiefer-Wolfowitz inequality},
  journal = {Annals of Probability},
  year = {1990},
  volume = {18},
  number = {3},
  pages = {1269--1283}
}
